\documentclass[11pt, oneside]{Thesis} 

\graphicspath{{Pictures/}} 

\usepackage[square, numbers, comma, sort&compress]{natbib} 
\hypersetup{urlcolor=blue, colorlinks=true} 

\usepackage{amsmath}
\usepackage{dsfont}
\usepackage{subfig}
\usepackage{graphicx}
\usepackage{listings}
\usepackage{yfonts}
\usepackage{simplewick}
\usepackage[svgnames]{xcolor}

\thesistitle{Numerical methods in the Conformal Bootstrap} 
\thesistype{Masters Thesis} 
\supervisor[mailto:miguelc@fc.up.pt]{Miguel \textsc{Costa}} 
\examiner{} 
\degree{Master of Science} 
\authors[mailto:alantunes@fc.up.pt]{António  \textsc{Antunes}} 
\addresses{} 
\subject{} 
\university[http://www.up.pt]{Universidade do Porto} 
\UNIVERSITY[http://www.up.pt]{UNIVERSIDADE DO PORTO} 
\department[http://dfa.fc.up.pt]{Departamento de Física e Astronomia} 
\DEPARTMENT[http://dfa.fc.up.pt]{DEPARTAMENTO DE FÍSICA E ASTRONOMIA} 
\faculty[http://www.fc.up.pt]{Faculdade de Ciências da Universidade do Porto} 
\FACULTY[http://www.fc.up.pt]{FACULDADE DE CIÊNCIAS DA UNIVERSIDADE DO PORTO} 

\begin{document}

\frontmatter 

\setstretch{1.3} 








\maketitle


\quotepage{Paul Dirac}{The fundamental laws necessary for the mathematical treatment of a large part of physics and the whole of chemistry are thus completely known, and the difficulty lies only in the fact that application of these laws leads to equations that are too complex to be solved.}


\addtotoc{Abstract} 

\abstract{
	
	\addtocontents{toc}{\vspace{1em}} 
	
	We present an accessible and easy to use approach to the Conformal Bootstrap. We begin by summarizing the main results of Conformal Field Theory that are useful in the Bootstrap program, for example the unitarity bounds and the Operator Product Expansion (OPE).
	 
	We proceed by deriving the bootstrap equation, which is the center-piece of the method, and show how to make it amenable to numerical methods, particularly Linear Programming.
	Using these techniques we attack 2D and 4D CFTs, reproducing general bounds on operator dimensions of the second lowest dimension operator ($\varepsilon$) as a function of the dimension of the lowest operator ($\phi$) . With a slight modification we also obtain bounds for OPE coefficients.
	
	As a generalization of the previous techniques, we attempt a more constraining bootstrap, which allows us to go beyond bounds. We develop the Extremal functional method, which we apply to obtain the spectrum and OPE coefficients of the 2D Ising model. We also explore the method of determinants, with which we can compute operator dimensions and OPE coefficients, using only the fusion rules of the theory.
	
	In order to apply the developed methods, we advocate and implement a new approach to obtain bounds on 2D CFTs with a finite central charge $c$, and replace the modular invariance of the Partition function on the Torus with a correlation function of twist operators.
	
	We finish with a conclusion where we recap the main results and discuss alternatives and future paths to be taken.

}
\cleardoublepage


\addtotoc{Resumo}

\abstract[title=Resumo,degree={Mestre de Ciência},connector=de]{
	
	\addtocontents{toc}{\vspace{1em}} 
	
Apresentamos, nesta tese, uma abordagem introdutória e acessível ao "Conformal Bootstrap". Começamos por recapitular os resultados base de Teoria de Campo Conforme que são importantes para o Bootstrap, como os limites de unitariedade e a "Operator Product Expansion"(OPE).

Procedemos, derivando a equação do Bootstrap que é a peça central desta abordagem, e mostramos como esta pode ser tornada tratável com o uso de métodos numéricos, nomeadamente, Programação linear. Usando estas técnicas, atacamos teorias em 2 e 4 dimensões, reproduzindo limites superiores às dimensões do segundo operador mais baixo da teoria ($\varepsilon$), em função do mais baixo ($\phi$) . Com uma ligeira modificação do método, conseguimos também obter constrangimentos para os coeficientes do OPE.

Como uma generalização das técnicas anteriores, tentamos um Boostrap mais restritivo, que nos permite ir para lá de limites superiores. Desenvolvemos o método do Funcional Extremo, que usamos para obter o espectro e os coeficientes do OPE do modelo de Ising 2D. Exploramos, também, o método dos determinantes, com o qual conseguimos novamente calcular dimensões e coeficientes do OPE, conhecendo apenas as regras de fusão da teoria.

Para aplicar os métodos estudados, motivamos e implementamos uma nova abordagem para obter limites superiores em teorias 2D com carga central finita $c$, substituindo a invariância modular da função de partição no toro, por funções de correlação de operadores com twist.

Terminamos com uma conclusão onde recapitulamos os resultados principais e discutimos caminhos alternativos e futuros aos desta tese.
}
\cleardoublepage


\setstretch{1.3} 

\acknowledgements{\addtocontents{toc}{\vspace{1em}} 

This thesis has been a very long journey, from no knowledge of Conformal Field Theory, to the ability of obtaining interesting results.
This journey, however, was not a lonely one. It was a journey with much discussion, ideas, cooperation, and friendship. And for it, there are many people who deserve a special word of gratitude.
 
First, I want to thank my supervisor, Miguel Costa, for having given me the chance to work with him on this very interesting topic. Without his guidance it would have been impossible to even begin understanding CFT, let alone the Conformal Bootstrap. The fact that he believed enough in my abilities to send me to the Bootstrap School in Brazil was a huge boost in my motivation and knowledge and I can't thank him enough for that. I look forward to continue my work with him in the future.
 Also, I had the good fortune of being able to discuss many problems with brilliant experienced people , in particular João Penedones, Tobias Hansen, Emilio Trevisani and Hongbin Chen. To them I am also grateful, in particular to Hongbin for his help with the Virasoro Blocks.
 There were also people who were constantly enduring by need of anoying them with what I had just learned. I am grateful to my friends and Colleagues: Zé Pedro Quintanilha, Pedro Paredes, Zé Ricardo, João Silva, Pedro Leal, Luís Ventura and Artur Amorim for their help in many ocasions, and specially to: Bruno Madureira, Maria Ramos, João Guerra, José Sá, Simão João and João Pires for their almost daily support.
 Before a thesis, however, a student needs to  climb the ladder of physics. In this process, a teacher is an invaluable help. I thank José Manuel Matos, for letting me appreciate the beauty of physics as soon as I did, and also to all the Professors of the Department of Physics and Mathematics, from whom I had the pleasure of learning much. A special thanks is due to Professors João Lopes dos Santos and Peter Gothen for their support and reference letters, and again to, Professor Miguel Costa, for all the reasons mentioned before, and many more.
 
 Before being a physics student, though, one needs to be a person. And in this, many important people in my life come in. I thank all my friends and family, including the Fechas family, for all the support and love they have provided me through this process, and all the others that preceeded it.
 Finally, I am grateful to the people who make my life what it is, and without whom there would be no point in doing all this:
 My parents, for unsurmountable support, love and undeserved patience, to whom I can not thank enough for everything they have given me,  for enduring endless days of stress and for making me feel loved even when I was too worried with work to be a good son.
 My girlfriend, Andreia, for helping me every single day, every single hour, for making me a better person, and for giving me a reason to be happy, and grateful everyday.
}


\tableofcontents 

\listoffigures 

\listoftables 


\setstretch{1.5} 

\listofabbreviations{rl} 
{
\textbf{LHS}  & \textbf{L}eft \textbf{H}and \textbf{S}ide \\
\textbf{RHS}  & \textbf{R}ight \textbf{H}and \textbf{S}ide  \\ 
\textbf{IR}   & \textbf{I}nfra-\textbf{R}ed \\
\textbf{UV}   & \textbf{U}ltra-\textbf{V}iolet \\
\textbf{CFT}  & \textbf{C}onformal \textbf{F}ield \textbf{T}heory \\
\textbf{QFT}  & \textbf{Q}uantum \textbf{F}ield \textbf{T}heory \\ 
\textbf{RG}  & \textbf{R}enormalization \textbf{G}roup \\ 
\textbf{SCT}  & \textbf{S}pecial \textbf{C}onformal \textbf{T}ransformation \\ 
\textbf{SOC}/\textbf{SOM} & \textbf{S}tate \textbf{O}perator \textbf{C}orrespondence/\textbf{M}ap \\ 
\textbf{OPE} & \textbf{O}perator \textbf{P}roduct \textbf{E}xpansion\\
\textbf{CB}  & \textbf{C}onformal \textbf{B}locks \\
\textbf{MM}  & \textbf{M}inimal \textbf{M}odels \\
\textbf{EFM}  & \textbf{E}xtremal \textbf{F}unctional \textbf{M}ethod\\
\textbf{LP(P)}  & \textbf{L}inear \textbf{P}rogramming (\textbf{P}roblem)\\
\hspace*{0.8in} & \hspace*{5in}
}




\listofsymbols{lll} 
{
	
$\mathcal{O}_{i}(x_{i})$ & Local operator inserted at $x_{i}$ & \\	
$f_{ijk}$ & Coefficient of the 3-pt function  $\langle\mathcal{O}_{i}\mathcal{O}_{j} \mathcal{O}_{k}\rangle$/ OPE coefficient $\mathcal{O}_{i}\mathcal{O}_{j} \sim f_{ijk}\mathcal{O}_{k} $ \\
$x_{ij}$ & the difference $x_{i} - x_{j}$  \\
$d$/$D$ & the number of spacetime dimensions  \\
$l$ & the spin label for a traceless symmetric representation of $SO(D)$ &  \\
$T^{\mu\nu}$ & the canonical stress energy tensor &  \\
$I$ & the Identity operator &  \\

& & \\ 

$\Delta$ & The full, quantum-corrected scaling dimension of an operator & \\

$\phi$ & A scalar operator, usually used as the one with lowest non-zero dimension & \\
$\varepsilon$ & A scalar operator, usually used as the second lowest non-zero dimension & \\
$\Lambda$ & A general cutoff, or a vector in the context of linear functionals & \\

& & \\

&The physical constants $c$, $k_ {B}$, $\hbar$ are set to 1. ~& \\
&The metric signature is Euclidean or $(-++\dots+)$, when applicable. ~& \\

}


%
%
%


\mainmatter 

\pagestyle{fancy}
\renewcommand{\chaptermark}[1]{\markboth{\thechapter.~\textsc{#1}}{}}
\fancyhead[LO,RE]{\leftmark}

\chapter{Fundamentals of CFT}
\label{Chapter2}

\section{Motivation}
Conformal Field Theories have been a subject of study in theoretical and mathematical physics for a few decades, but there are still many open problems and results to be explored. One of the main problems of modern physics is to understand strongly coupled behavior in QFT, but our usual perturbative techniques are by definition, only valid in the weak regime. Therefore, the usage of non-perturbative techniques is, perhaps, one of the most fundamental changes of paradigm in theoretical physics of the last years. In this program, Conformal field theories are very important, as, because of the strong restrictions imposed by Conformal symmetry (and more as we shall see in this thesis), many of the relevant objects are completely determined and we can try to learn about strongly coupled physics in this special case. Indeed, Conformal field theories are far from being trivial and there are many models of strongly interacting physical systems which possess Conformal invariance. Note that Conformal Symmetry is an extension of scale invariance. In particular, the following discussion will be centered around the importance of scale invariance but there are strong theorems which ensure that Unitary Scale invariant theories are in fact Conformal. Let us take a brief tour through the realm of application of CFTs. 
\subsection{CFTs in Condensed matter}
CFTs appear in many contexts of physical nature, notably critical phenomena/phase transitions in condensed matter physics. First, when we are interested in momentum scales much smaller than $\frac{1}{a}$, (where $a$ is the lattice spacing) a continuum description of the lattice becomes reasonable, and this scale disappears as we take the continuum limit.
 These field theories in condensed matter, are often interested in the behavior of correlation functions, which can have a characteristic length $\xi$, the correlation length. Close to a phase transition, this length becomes very large, as long range order kicks in, and the dependence on the scale is suppressed. The theory then, becomes scale invariant, and the CFT description becomes useful.
\subsection{CFTs as landmarks in the space of QFTs}
On a more field-theoretical context, Conformal theories are of central importance in a Wilsonian/effective field theory framework of Quantum field theory \cite{Peskin:1995ev}. When a field theory is quantized, the couplings in the Lagrangian, which are at the classical level, constants, become dependent on the energy scale at which the theory is behaving.

 This "running of the coupling constant" is interpreted in this context as the theory's reaction to high momentum degrees of freedom, which cause formal divergences in the usual regularization and renormalization schemes. In the Wilsonian picture, however, one introduces a cutoff scale $\Lambda$ which is to be seen as very large, and counts the contribution of the high-energy degrees of freedom only up to this scale. The effective field theory renormalization group, then, consists of integrating out the most energetic shells, and noticing that the result of doing this can be absorbed in the coupling constants, which change (after rescaling 
to the original momentum) as we perform successive integrations of lower and lower momentum. Considering the partition function: 
\begin{equation}
	Z[J]= \int_{|k|< \Lambda} \mathcal{D}\phi~ e^{\int[\mathcal{L}+J\phi]}
\end{equation}
The exact procedure is integrating a momentum shell $b\Lambda<|k|< \Lambda$ for $b = 1- \delta\ell $ for  a small $\delta\ell$. Indeed, after this hypothetical integration we would obtain:
\begin{equation}
	Z = \int_{|k|<b\Lambda} \mathcal{D}\phi \exp\left( -\int \mathcal{L}_{eff}\right) 
\end{equation}
Where the effective Lagrangian has absorbed the changes in the couplings, and we can now think of iterating this procedure.

 This defines a trajectory in the space of QFTs, known as a Renormalization group flow. When a theory is scale invariant, performing an integration of a very small shell does not alter the physical description, and so Scale invariant theories are fixed points in these RG trajectories. This description is made concrete with the introduction of two important quantities: the beta function, and the anomalous dimension $\gamma$.
 We have then the beta function associated to the coupling $\lambda$:
 \begin{equation}
 	\beta(\lambda)=\frac{\partial \lambda}{\partial \log(p/M)}
 \end{equation}
 
 Where we introduce a renormalization scale $M$. In, particular the beta function will vanish when the coupling does not depend on the momentum, i.e. for Scale invariant field theories.
 However, there is still the other quantity which measures the effect of renormalization, which does not in general vanish, the anomalous dimension: 
 \begin{equation}
 	\gamma_{\mathcal{O}} = M \frac{\partial}{\partial M} \log Z_{\mathcal{O}} 
 \end{equation}
 
 Where $Z_{\mathcal{O}}$ is the field-strength renormalization of operator $\mathcal{O}$. These quantities are balanced in the Callan-Symanzik equation, which still makes sense even when the beta function vanishes, in such a way that the anomalous dimensions for the operators $\mathcal{O}$ are non-trivial quantities even for CFTs. With this, we have then:
 
 \begin{equation}
 	\Delta_{\mathcal{O}} = \Delta_{DA} + \gamma_{\mathcal{O}}
 \end{equation}
 Where the scaling dimension has a DA (Dimensional analysis) part, and its anomalous correction. We understand the definition of $\Delta_{DA}$ in classical terms, but the true meaning of $\Delta_{\mathcal{O}}$ will be made rigorous in the next section.
 With this, we have established the quantities which balance the flow of Lagrangians through renormalization, and pointed to the importance of Scale invariant field theories.

\begin{figure}[htbp]
	\centering
	\includegraphics[width= 9 cm]{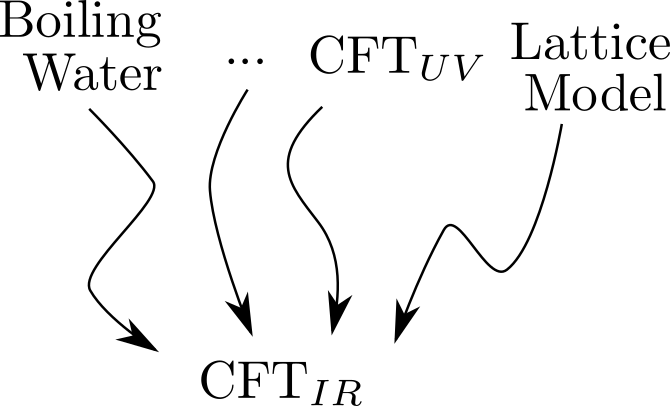}
	
	\caption[RG Flow]{Typical behavior of physical theories under RG flow. \cite{Simmons-duffin2016}}
	\label{fig:RGF}
\end{figure}

  Because of this, it is typical for theories to flow asymptotically (either at very low energies: IR, or very high energies: UV) to Conformal field theories. Indeed, the famous phenomenon of universality is a consequence of many different theories (other field theories, lattice theories, perhaps even more exotic theories) flowing to the same scale invariant fixed point in the IR. This, of course guarantees that they possess the same critical exponents, correlation functions, etc. because they are being described by the same effective Conformal field theory. Therefore one would expect that the typical behavior in RG flow is what is drawn in Figure \ref{fig:RGF}.

CFTs are the ends of RG flow, so we should study them in detail to understand what a QFT really is.

\subsection{AdS/CFT Correspondence}
AdS/CFT is a conjectured (but accepted)  duality which relates the description of Conformal Field Theories in $d$ dimensions with theories of gravity/String theories in an asymptotically AdS background of dimension $d+1$ \cite{Penedones:2016voo}. The CFT is interpreted as living on the boundary of AdS. This duality has been one of the hottest topics in theoretical physics in the last decade, and it has found many phenomenological applications. The basic example is the Duality between $\mathcal{N} = 4$ Super Yang-Mills theory, which is a Conformal field theory (in fact a SuperConformal field theory), and a Type II B String theory on a $ AdS_{5} \times S_{5}$ background spacetime. We will see an interesting interpretation of a result under this Gauge/Gravity duality in Chapter \ref{twist}.

\section{Basics of Conformal Field Theory}
Assuming a working knowledge of Quantum Field Theory, there are many physical consequences that can be extracted by demanding global Conformal invariance as an extension to the Poincaré Symmetry of usual QFT. We will briefly develop the structural ideas of CFT in order to understand the Conformal Bootstrap which is the main objective of this thesis. The following discussion is centered in $d\geq3$ and is mostly based on \cite{Simmons-duffin2016,Qualls2015,Rychkov2016}.
\subsection{Conformal Symmetry and the Conformal Algebra}

	Symmetries/Conserved Charges are associated with Killing vectors ($\epsilon = \epsilon^{\mu}\partial_{\mu}$) of the background spacetime. Poincaré/Euclidean invariance ensures a conserved energy-momentum tensor current $\partial_{\mu}T^{\mu \nu}=0$ (local) in flat spacetime, with the Killing equation and the associate vector field solutions (translations and rotations/Lorentz boosts): \begin{equation}
	\partial_{\mu}\epsilon_{\nu} +\partial_{\nu}\epsilon_{\mu}=0 ~,~ p_{\mu}= \partial_{\mu} ~,~ m_{\mu\nu}= x_{\nu}\partial_{\mu}-x_{\mu}\partial_{\nu}
	\end{equation}
	If we allow the assumption of a Traceless energy-momentum tensor: \begin{equation}
	T^{\mu}_{~ \mu}(x)=0
	\end{equation}
	We may obtain further symmetries (related to the insensitivity of the EM Tensor to Conformal changes of the metric which is equivalent to tracelessness), namely a charge:
	\begin{equation}
		Q_{\epsilon} = \int dS_{\mu} \epsilon_{\nu} T^{\mu \nu}
		\label{eqn:Charge}
	\end{equation}

	 will have an associated conserved current:
	\begin{equation}
		0=\partial_{\mu} J^{\mu}=\partial_{\mu}(\epsilon_{\nu} T^{\mu \nu}) = \frac{1}{2} (\partial_{\mu}\epsilon_{\nu} +\partial_{\nu}\epsilon_{\mu})T^{\mu \nu}
	\end{equation}

	With the previous condition (where we used symmetry and conservation of $T^{\mu \nu}$ ) and a traceless EM tensor, one now has the Conformal killing equations:
	\begin{equation}
	\label{eq:Ckil}
	\partial_{\mu}\epsilon_{\nu} +\partial_{\nu}\epsilon_{\mu}= c(x)\delta_{\mu\nu}
	\end{equation}
	Where $c(x)$ is a generic function, and we have further solutions: \begin{equation}
	d= x^{\mu}\partial_{\mu} ~,~ k_{\mu} = 2 x_{\mu}(x.\partial) - x^{2}\partial_{\mu}
	\end{equation}
	These correspond to Dila(ta)tions (which are related to scale invariance) and Special Conformal Transformations which can be shown to be Inversions followed by translation followed by another inversion. 
	We can summarize the transformations in a small table:
	 \begin{center}\vspace{1cm}
	 	\begin{tabular}{l l l l }
	 		\toprule
	 		Transformation & Generator & Change in coordinates  \\
	 		\midrule
	 		Translation & $p^{\mu}$ & $x'^{\mu}= x^{\mu}+ a^{\mu}$ \\
	 	    Rotation & $m^{\mu\nu}$ &$x'^{\mu} = M^{\mu}_{~ \nu} x^{\nu} $ \\
	 		Dilation & d & $x'^{\mu} = \alpha x^{\mu} $\\
	 		SCT & $k^{\mu}$ & $x'^{\mu}=\frac{x^{\mu}- x^{2}b^{\mu}}{1- 2(b\cdot x) + b^{2}x^{2}}$ \\
	 		
	 	\bottomrule
	 	\end{tabular}
	 	\captionof{table}{List of Conformal transformations, their generators and finite form}
	 \end{center}\vspace{1cm}
	 
	 From this we can count $D$ translations, $\frac{D(D-1)}{2}$ rotations, 1 dilation and $D$ SCTs. This gives a total of $\frac{(D+2)(D+1)}{2}$ generators, which incidentally coincides with the number of generators of rotations of $D+2$ dimensional space.

	These vector fields can be used to construct conserved charges, as in equation \ref{eqn:Charge}, which obey the Conformal algebra: \begin{align} \nonumber &[D,K_\mu]=-iK_\mu ~,~ [D,P_\mu]=iP_\mu  ~,~ [K_\mu,P_\nu]=2i\eta_{\mu\nu}D-2iM_{\mu\nu} \,,\\
\nonumber	&[K_\mu, M_{\nu\rho}] = i ( \eta_{\mu\nu} K_{\rho} - \eta_{\mu \rho} K_\nu )  ~,~ [P_\rho,M_{\mu\nu}] = i(\eta_{\rho\mu}P_\nu - \eta_{\rho\nu}P_\mu) \,, \\
	&[M_{\mu\nu},M_{\rho\sigma}] = i (\eta_{\nu\rho}M_{\mu\sigma} + \eta_{\mu\sigma}M_{\nu\rho} - \eta_{\mu\rho}M_{\nu\sigma} - \eta_{\nu\sigma}M_{\mu\rho})\,, \end{align}
	The first line of commutators is the most interesting one, because it establishes $K$ and $P$ as lowering and raising operators for the Dilation operator, whose eigenvalues are the Conformal dimensions.
	Notice that the last line of the Conformal algebra is exactly the Poincaré algebra, so it is indeed clear that introducing Conformal invariance is a way to extend the usual global symmetry of spacetime.
	Also, note that 	the Conformal algebra in $d$ euclidean dimensions is isomorphic to SO($d+1,1$). So, one can build Conformal kinematics from Poincaré kinematics in $d+1,1$. This is the idea of the embedding space formalism. 
	In fact, it is a simple but painful exercise to check that defining a set of generators $L_{AB}$ such that $A= -1,0,1,\dots,d$, and identifying:
	 \begin{align}
	\nonumber &L_{\mu,\nu}=M_{\mu\nu}\\
	\nonumber &L_{-1,0} = D \\
	 \nonumber &L_{0,\mu} = \frac{1}{2}(P_{\mu} + K_{\mu})      \\ 
	 &L_{-1,\mu} = \frac{1}{2}(P_{\mu} - K_{\mu}) 
	 \end{align}
	 That the commutation relations of SO$(d+1,1)$ are obeyed. Notice that the $d+2$ $A$ indexes have one time-like component, unlike the euclidean signature we are generally working in, from the d $\mu$ components.
	
	\subsection{Representations of the Conformal algebra}
	 The Conformal Algebra also allows us to classify operators in representations. We define primary operators, and characterize them by the quantum number $\Delta$, the Conformal dimension, the representation $S$ it transforms in, under rotation (its spin), and the fact that it is annihilated by $K$ (in Representation theory this correspond to a lowest weight state):
	 \begin{equation}
	 [D,\mathcal{O}(0)]= \Delta \mathcal{O}(0) ~,~ [M_{\mu\nu},\mathcal{O}]= \mathcal{S}_{\mu\nu}\mathcal{O}(0) ~,~
	 [K_{\mu},\mathcal{O}(0)]=0
	 \end{equation}
	 We also define descendants, which are obtained by acting with the momentum operator successively, and increase their dimension relative to the primary: 
	 \begin{equation}
	 P_{\mu_{1}\dots \mu_{n}}\mathcal{O} \rightarrow \Delta + n
	 \end{equation}
	 The full set of a primary and all its descendants is known as a Conformal family, associated to the primary of dimension $\Delta$ and spin $l$.
	 All operators in a CFT are combinations of Primaries and Descendants. A proof of this fact can be seen, for example, in \cite{Simmons-duffin2016}.
	 \subsection{Conformal kinematics}
	 Symmetry is a very strong tool to determine basic correlation functions in CFT. For example, translation and rotation invariance sets $\langle \mathcal{O}_{1}(x_{1}) \mathcal{O}_{2}(x_{2}) \rangle = f(|x_{1}-x_{2}|) $. Furthermore (and this shows the importance of scale invariance), one needs, under scalings: $\langle \mathcal{O}_{1}(x_{1}) \mathcal{O}_{2}(x_{2}) \rangle = \lambda^{\Delta_{1}+\Delta_{2}} \langle \mathcal{O}_{1}(\lambda x_{1}) \mathcal{O}_{2}(\lambda x_{2}) \rangle$. This homogeneity condition can be solved by:
	 \begin{equation}
	 	\langle \mathcal{O}_{1}(x_{1}) \mathcal{O}_{2}(x_{2}) \rangle =
	 	\frac{C}{|x_{1}-x_{2}|^{\Delta_{1}+\Delta_{2}}}
	 \end{equation}
     For a general Conformal transformation, where we have a local scale factor $\Omega(x)$, we demand  $ \langle \mathcal{O}_{1}(x_{1}) \mathcal{O}_{2}(x_{2}) \rangle = \Omega(x_{1}')^{\Delta_{1}} \Omega(x_{2}')^{\Delta_{2}} \langle \mathcal{O}_{1} (x_{1}') \mathcal{O}_{2}(x_{2}') \rangle $. This condition is already satisfied for translations, rotations and dilatations, which means we only need to check SCTs, but as we mentioned earlier, these can be built with inversions. It is trivial to check the transformation property of distances under the aforementioned actions:
     \begin{equation}
     	(x-y)^{2} = \frac{(x'-y')^{2}}{\Omega(x') \Omega(y')}
     \end{equation}
     This fact, together with our demand for the transformation of the correlation function, fixes even further the 2-pt function.
	 Imposing the full Conformal invariance, then, gives us, for scalars:
	 \begin{equation}
	 \langle \mathcal{O}_{1}(x_{1}) \mathcal{O}_{2}(x_{2}) \rangle = \frac{C \delta_{\Delta_{1}\Delta_{2}}}{|x_{1}-x_{2}|^{2\Delta_{1}}}
	 \end{equation}
	 Which means that 2-pt functions are diagonal on the space of operators.
	 We can choose an orthonormal basis of operators and absorb $C$.
	 
	  The 3-pt function is still fixed by symmetry, with the same arguments as before:
	 \begin{equation}
	 \langle \mathcal{O}_{1}(x_{1}) \mathcal{O}_{2}(x_{2}) \mathcal{O}_{3}(x_{3}) \rangle =\frac{f_{123}}{|x_{12}|^{\Delta_{1}+\Delta_{2}-\Delta_{3}} |x_{23}|^{\Delta_{2}+\Delta_{3}-\Delta_{1}} |x_{31}|^{\Delta_{3}+\Delta_{1}-\Delta_{2}}}
	 \end{equation}
	 But the overall constant $f_{ijk}$ must be determined some other way.
	 
	 Up to this point it looks like we will completely solve all CFTs just be symmetry, but of course, it is not that simple.
	 The difficulty emerges when we begin studying the 4-pt  correlator.
	 There are two Conformally independent combinations of 4 points:\begin{equation}
	 u=\frac{x_{12}^{2}x_{34}^{2}}{x_{13}^{2}x_{24}^{2}} ~,~ v= \frac{x_{23}^{2}x_{14}^{2}}{x_{13}^{2}x_{24}^{2}}
	 \end{equation}
	 So the 4-pt function is dependent on a function of $u$ and $v$:
	 \begin{equation}
	 \langle \phi(x_{1}) \phi(x_{2}) \phi(x_{3}) \phi(x_{4}) \rangle = \frac{g(u,v)}{x_{12}^{2\Delta_{\phi}} x_{34}^{2\Delta_{\phi}}}
	 \end{equation}
	 The LHS of this equation is manifestly invariant under exchanges of the $x_{i}$ and this will lead to consistency conditions which will be useful for the Conformal Bootstrap. 
	 
	 The kinematics for operators which transform in non-trivial representations of SO($d$) are fixed in a similar fashion, although we have to deal with possibly multiple tensor structures.
	 
\begin{figure}[htbp]
	 	\centering
	 	\includegraphics[width= 9 cm]{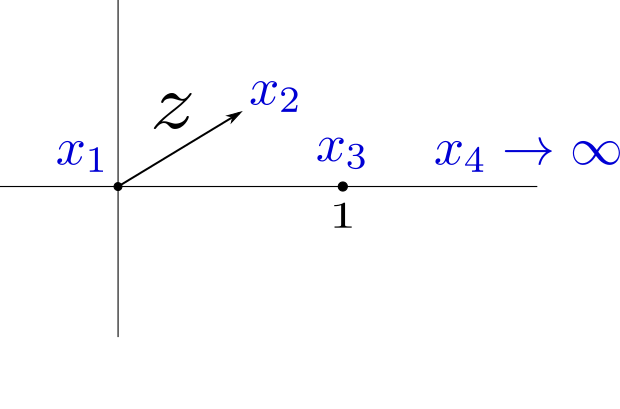}
	 	
	 	\caption[A sketch of $z$, $ \bar{z}$ coordinates]{A choice of a configuration of 4 points using Conformal transformations which defines the $z$ variable.}
	 	\label{fig:zcoord}
	 \end{figure}

	 It is also important to notice another way of describing the 4-points (up to Conformal transformations) with two degrees of freedom. Let us consider, then, 4 points $x_{1},x_{2},x_{3},x_{4}$. By using a Special Conformal transformation, we can invert one of the points, say $x_{4}$, translate it to the origin, and the invert it back to infinity. By performing a translation, we can move a second point ($x_{1}$) to the origin, without affecting infinity. Then by dilations, we can bring $x_{3}$ to distance one of the origin, and rotate it (obviously without moving $x_{1},x_{4}$) to $(1,0,\dots,0)$. Now, we have three of the points in a line. Finally we can perform rotations around this line ($x_{3} =(1,0,\dots,0)$ ; $x_{1} =(0,0,\dots,0)$ ; $ x_{4} =(\infty,0,\dots,0) $) and set the final point $x_{2}$ to the $xy$ plane, with coordinates $(x,y,0,\dots,0)$. Again, this means that we can describe any set of 4 points up to Conformal transformations with just two numbers. Note that this argument was independent of the number of spatial dimensions. We show a sketch of these coordinates in Figure  \ref{fig:zcoord}.
	 
	 It is common to express this $x,y$ coordinates in terms of complex coordinates $z= x+iy$ and $\bar{z}=x-iy$. With this choice we can easily compute the cross-ratios and obtain:
	 \begin{equation}
	 	u= z \bar{z} ~;~ v=(1-z)(1-\bar{z})
	 \end{equation}
	 
	\section{Dynamics: Radial Quantization}
\begin{figure}[htbp]
			\centering
			\includegraphics[width= 8 cm]{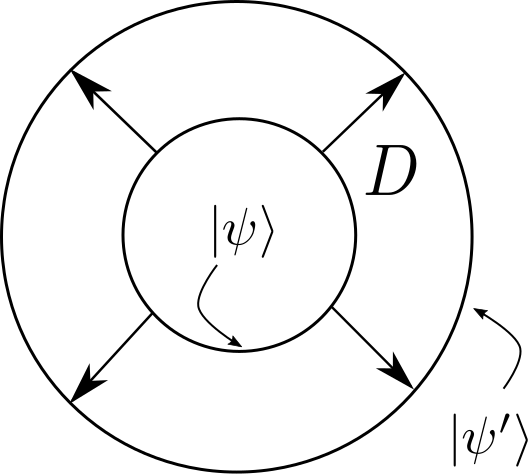}
			
			\caption[Radial Quantization]{Sketch of the radial Hilbert spaces and of the action of the Dilation operator}
			\label{fig:RQ}
		\end{figure}

		Up to now we have implicitly assumed some path integral description, but we can explicitly quantize the theory by taking a radial foliation and defining Hilbert spaces on each sphere. This is useful because quantization in terms of commutators is better computationally, and by choosing the radial direction we are respecting the euclidean symmetry of our space, instead of unnaturally choosing a time direction (canonical quantization) along which we foliate. We propagate states through spheres, as in Figure \ref{fig:RQ}, using the $D$ operator which plays the role of a Hamiltonian.  
		
		Using this formulation we have a natural way to define states from operators and vice-versa. We can define (summing over all possible boundary values on a sphere) a state associated to the insertion of an operator $\mathcal{O}$: \begin{equation}
		|\mathcal{O}\rangle \equiv \int \mathcal{D}\phi_{b} |\phi_{b}\rangle\langle \phi_{b}| \mathcal{O}(x)|0\rangle
		\end{equation}
		With the coefficient defined by the bulk path integral with the appropriate boundary conditions: 
		\begin{equation}
		\langle \phi_{b}| \mathcal{O}(x)|0\rangle = \int_{
			\phi(1, \mathbf{n})= \phi_{b}(\mathbf{n}) }^{r<1} \mathcal{D}\phi(r,\mathbf{n}) \mathcal{O}(x) e^{-S[\phi]}
		\end{equation}
		Reciprocally, we can also obtain an operator given a state, or more precisely build a correlation function by gluing states in the path integral. So we have:
		\begin{equation}
		\mathcal{O}(0) \leftrightarrow \mathcal{O}(0)|0\rangle \equiv |\mathcal{O}\rangle
		\end{equation}
		
		More intuitively, any state can be evolved with the dilatation operator all the way back to origin, so the state becomes defined locally close to $x=0$. This is precisely what we mean by a local operator.
		
		\subsection{Radial Quantization on the cylinder}

		By doing a Weyl rescaling of the Metric (to which CFT's are insensitive, up to factors), we can write correlation functions on the cylinder, reinterpreting radial quantization: \begin{equation}
		ds^{2}_{\mathbb{R}^{d}}= dr^{2} + r^{2}ds^{2}_{S^{d-1}}= e^{2\tau}(d\tau^{2}+ds^{2}_{S^{d-1}})
		\end{equation}
		Where $r=e^{\tau}$ and $\tau$ goes from $-\infty$ to $+\infty$ and therefore can be interpreted as a time direction in radial quantization.
		Sometimes its useful to compute correlation functions on the cylinder, which are Conformally equivalent to flat Correlation functions.
		
		Also, this time direction gives an easy way of defining the hermitian conjugate, relating $\langle in|$ and $|out\rangle$ states, and giving a definition for a scalar product and unitarity in the usual way, by sending $\tau$ to $-\tau$.
		Back in flat space, this gives us rules to take hermitian conjugates of charges, using the transformation of operators under the Conformal rescaling of the metric, and the elementary rule on the cylinder. 
		The main consequences of this are:
		\begin{equation}
			D^{\dagger} = D ~,~ P_{\mu}^{\dagger} = K_{\mu}
		\end{equation}
		\section{Reflection positivity and Unitarity Bounds}
		Many reasonable QFTs are unitary, this means that probability is preserved quantum mechanically and that states should have positive norms: $\langle\phi|\phi\rangle \ge 0$. To impose this condition, we can Wick rotate to Euclidean time, and obtain the condition of reflection positivity:
		\begin{equation}
		\langle 0 | \mathcal{O}_{t_{En}} \dots \mathcal{O}_{t_{E1}}\mathcal{O}_{-t_{E1}} \dots \mathcal{O}_{-t_{En}}|0\rangle \geq 0
		\end{equation} 
		Where we inserted operators on the vacuum at negative time to define a general state.
		Imposing this on matrix elements of the form:
		\begin{equation}
		\langle \mathcal{O}| P_{\mu}^{\dagger} P_{\mu}|\mathcal{O}\rangle
		\end{equation}
		We can use the properties of the adjoint, the commutation relations of the Conformal algebra, and for $|\mathcal{O}\rangle$ a scalar primary state ($K_{\mu}|\mathcal{O}\rangle = 0$, $M_{\mu\nu}|\mathcal{O}\rangle = 0$), we get:
		\begin{equation}
			\langle \mathcal{O}| [K_{\mu} ,P_{\mu}]|\mathcal{O}\rangle = \langle \mathcal{O}|(2D \delta_{\mu\nu} -2M_{\mu\nu})|\mathcal{O}\rangle = 2 \Delta \delta_{\mu\nu} \langle \mathcal{O}|\mathcal{O}\rangle
		\end{equation}
		Choosing $\mu$ to be any component, and the fact that $\langle \mathcal{O}| P_{\mu}^{\dagger} P_{\mu}|\mathcal{O}\rangle= |P_{\mu}|\mathcal{O}\rangle|^{2}$, we can obtain a weak version of the unitarity bound, that justifies our intuition that we should allow for lowest weight states:
		\begin{equation}
			\Delta \ge 0
		\end{equation} 
		One can obtain the celebrated unitarity bounds, by doing similar calculation in more complicated matrix elements: 
		\begin{align}
		& \Delta_{l=0} \geq \mathsf{Max}\left( \frac{d-2}{2},0\right)  \\
		& \Delta_l \geq l + d - 2
		\end{align}
	These are our first examples of bounds on operator dimensions determined from consistency conditions and symmetry, which is the main philosophy of the Conformal Bootstrap. We will discuss this with more detail in the next chapter.
	\section{Operator Product Expansion}
		If we insert two operators and perform the path integral, we have (because every state is a linear combination of primaries and descendants):\begin{equation}
		\mathcal{O}_{i}(x)\mathcal{O}_{j}(0) |0\rangle = \sum_{k} C_{ijk}(x,P)\mathcal{O}_{k}(x)|0\rangle 
		\end{equation}
		Where the sum is over all primaries of the theory, and $C_{ijk}$ is a differential operator which generates all descendants and is fixed by Conformal symmetry. Then by the State-Operator map, we have (OPE):
		\begin{equation}
		\mathcal{O}_{i}(x_{1})\mathcal{O}_{j}(x_{2}) = \sum_{k} C_{ijk}(x_{12},\partial_{2})\mathcal{O}_{k}(x_{2}) 
		\end{equation}
		The expansion converges within a ball where there are no other operator insertions.
	\subsection{N-point functions and the OPE}
		After one fixes $C_{ijk}$, which depend only on $\Delta$'s and $f_{ijk}$'s (for example, by comparing the expression for a 3-pt function, with and without using the OPE, this will give us that the $C_{ijk}$ are exactly proportional to $f_{ijk}$), the OPE can be used to compute any correlation function of the theory! \begin{equation}
		\langle \mathcal{O}_{1}(x_{1})\mathcal{O}_{2}(x_{2}) \dots \mathcal{O}_{n}(x_{n})\rangle = \sum_{k}C_{12k}(x_{12},\partial_{2})\langle \mathcal{O}_{k}(x_{2}) \dots \mathcal{O}_{n}(x_{n}) \rangle
		\end{equation}
		This expansion can be applied successively, to reduce the problem to computing one-point functions, which are trivial in flat space:
		\begin{equation}
		\langle \mathcal{O} \rangle = \delta_{\mathcal{O}I}
		\end{equation}
		This is just stating that the vacuum of the theory is stable ($I$ denotes the identity operator).
		In general we will write the OPE with the coefficients $f_{ijk}$ factored out (as we argued that happens naturally), so that $C_{ijk}$ will depend only on scaling dimensions and the number of dimensions of spacetime.
		 
\chapter{The Conformal Bootstrap}
\label{Boot}

\section{The Bootstrap Philosophy}

So far, we have established the main tools of Conformal Field Theory in the Global Conformal symmetry sense (this is typically referred to in the literature as CFT in $D \geq 3$ but it still makes sense to discuss $2D$ CFTs in this framework, we just are not considering the most general set of locally Conformal transformations, and we are escaping the formalism of complex analytic/Virasoro algebra CFT as seen in Appendix \ref{appendix2d}). There is, implicitly, a line of thought that we are following that will lead us to the core of this thesis' work:
\begin{enumerate}
	\item{Focus on the CFT itself, not some UV Lagrangian, or microscopic realization of the theory.}
	\item{Exploit everything that Conformal Symmetry fixes kinematically.}
	\item{Impose further consistency conditions.}
	\item{Constrain, or in the best case scenario, solve, the CFT using the previous data.}
\end{enumerate}

This philosophy is intrinsically non-perturbative, in the sense that it studies properties of a CFT regardless of the value of any coupling constant/interaction strength, which means, of course, that there is no expansion in a small parameter and the usual Feynman Diagrams are not needed.
So far, we are essentially in stage 2 of our list, but, before we move on to phase 3 (which is the actual Conformal Bootstrap), we will need to develop a few more tools; we will begin by defining the Conformal Blocks \cite{Simmons-duffin2016}. 

\section{Conformal Blocks}
	Consider a 4-pt function of identical scalars and take the OPE between $1\leftrightarrow 2$ and $3 \leftrightarrow 4$: 
\begin{align}\nonumber &\langle \phi(x_{1}) \phi(x_{2}) \phi(x_{3}) \phi(x_{4}) \rangle = \sum_{\mathcal{O},\mathcal{O}'}f_{\phi \phi \mathcal{O}}f_{\phi \phi \mathcal{O}'}C_{a}(x_{12},\partial_{2})C_{b}(x_{34},\partial_{4})\langle \mathcal{O}^{a}(x_{2}) \mathcal{O}'^{b}(x_{4}) \rangle\\
& = \sum_{\mathcal{O}}f_{\phi \phi \mathcal{O}}^{2}C_{a}(x_{12},\partial_{2})C_{b}(x_{34},\partial_{4})\frac{I^{ab}(x_{24})}{x_{24}^{2\Delta_{\phi}}} 
 \end{align}
 Where we used the fact that the 2-Pt functions are diagonal as determined in \ref{Chapter2}, from the first to the second line and the $a,b$ indexes denote possible tensor structures when spinning operators are involved.
 
 Note that the 4-pt function is now expressed as a sum over all primary operators $\mathcal{O}$ whose Conformal Families contribute to the $\phi \times \phi$ OPE. Each of these elements will be named a Conformal Block, indexed by the dimension and spin of its respective primary operator, and it contains all the information about the associated Conformal family.

	So, we identify the structure of the 4-pt function as discussed in the previous chapter, where we had written the coordinate dependence of the correlator as $g(u,v)$, with $g$ a generic, undetermined function, but know we have expanded it in terms of simpler quantities, in an organized way.
	 \begin{equation}
		g(u,v)= \sum_{\mathcal{O}} f_{\phi\phi \mathcal{O}}^{2} g_{\Delta_{\mathcal{O}},l_{\mathcal{O}}}
	\end{equation}
	Where we define the Global Conformal Block:
	\begin{equation}
		g_{\Delta_{\mathcal{O}},l_{\mathcal{O}}} = x_{12}^{2\Delta_{\phi}} x_{34}^{2\Delta_{\phi}}  C_{a}(x_{12},\partial_{2})C_{b}(x_{34},\partial_{4})\frac{I^{ab}(x_{24})}{x_{24}^{2\Delta_{\phi}}} 
	\end{equation}
	Where only traceless symmetric tensors of even spin participate in the sum in this case of identical scalars (this can be seen from the 3-pt function with 2 identical scalars and one spinning operator, and will be used extensively when we consider the spectrum in the $\phi\phi$ OPE).
	Now thinking of projecting the 4-pt function onto the contribution of a Conformal primary, we can define:
	\begin{equation}
		|\mathcal{O}| = \sum_{\alpha,\beta =\mathcal{O}, \partial\mathcal{O},\dots} \frac{|\alpha\rangle \langle\beta|}{\langle \alpha|\beta\rangle}
	\end{equation}
	We have, of course, $\sum |\mathcal{O}| = I$, which means we can insert it in the middle of the four-point function, and obtain a sum of terms, all of which are of the form: 
	\begin{equation}
	\label{eq:proj}
		\langle 0| \phi(x_{3})\phi(x_{4})|\mathcal{O}|\phi(x_{1})\phi(x_{2})|0\rangle =  \frac{f_{\phi\phi\mathcal{O}}^{2}g_{\Delta_{\mathcal{O}},l_{\mathcal{O}}}(u,v)}{x_{12}^{2\Delta_{\phi}} x_{34}^{2\Delta_{\phi}}}
	\end{equation}
	This is easy to obtain by taking the OPE on the LHS.
	
	The Conformal Blocks can be obtained explicitly from the Conformal Casimir equation. As the Conformal algebra is isomorphic to SO($D+1,1$) we can use the Casimir operator:
	\begin{equation}
		C= -\frac{1}{2} L_{ab}L^{ab}
	\end{equation}
	As a Casimir, it has the same eigenvalue ($\lambda_{\Delta,l} =\Delta(\Delta-d) +l(l+d-2) $) on every state in a given representation.
	
	So acting with $C$ as a differential operator on equation \ref{eq:proj}, and using the fact that $|\mathcal{O}|$ will be an eigenvector on both sides it is found that:
	\begin{equation}
		\mathcal{D}g_{\Delta,l}(u,v)  = \lambda_{\Delta,l} g_{\Delta,l}(u,v)
	\end{equation}
	Where $\mathcal{D}$ is the Casimir operator written as a differential operator on $z,\bar{z}$ variables:
	\begin{equation}
		\mathcal{D} =2(z^{2}(1-z)\partial_{z}^{2}-z^{2}\partial_{z}) + 2(\bar{z}^{2}(1-\bar{z})\partial_{\bar{z}}^{2}-\bar{z}^{2}\partial_{\bar{z}})  + 2(d-2)\frac{z\bar{z}}{z-\bar{z}}((1-z)\partial_{z} - (1-\bar{z})\partial_{\bar{z}})
	\end{equation}

	Alternatively, they can also be obtained by a power expansion \cite{Hogervorst:2013sma,Costa:2016xah} in terms of cylinder radial quantization, or by recursion relations \cite{Penedones:2015aga}, which are based on the fact that states with zero norm make the normalization blow up, so there should be poles in the CB when we think of it as a function of $\Delta$. The residues are then proportional to the global Block itself and a recursion relation is obtained.

	For example, in $2d$, and only regarding Global Conformal Symmetry, the Casimir equation has an exact solution \cite{Dolan:2000ut,Dolan:2003hv}:
	\begin{equation}
	\label{eqn: 2dblock}
	g_{\Delta,l}(u,v) = k_{\Delta + l}(z) k_{\Delta -l}(\bar{z}) + k_{\Delta - l}(z) k_{\Delta + l}(\bar{z})
	\end{equation}
	With:
	\begin{equation}
	k_{a}(x)= x^{\frac{a}{2}}~ _{2}F_{1}(\frac{a}{2},\frac{a}{2},a;x)
	\end{equation}
	
	Where $_{2}F_{1}(a,b,c;z)$ is a Hypergeometric function.
	
	\section{Crossing Symmetry: The Conformal Bootstrap}
	
	\subsection{Imposing Crossing Symmetry}
		When we calculated the 4-pt correlator we chose the 1-2; 3-4 OPE channel, however, we could have taken the OPE between different pairs of points! This is exactly the crossing symmetry which is manifest in the four point function:
		\begin{equation}
			\langle \phi(x_{1}) \phi(x_{2}) \phi(x_{3}) \phi(x_{4}) \rangle = \langle \phi(x_{1}) \phi(x_{4}) \phi(x_{2}) \phi(x_{3}) \rangle 
		\end{equation} 
		This is obvious, because in the path integral formulation, the fields $\phi$ are just functions, and the order of their product is irrelevant. Note, however that one should be careful about convergence of the OPE when selecting different pairs of points, by making sure that one can build a sphere containing only the two points (if necessary do a Conformal transformation to an equivalent configuration where it is easier to visualize this procedure).	
		
		Indeed, using a Wick contraction bracket to denote the OPE, the crossing symmetry equation, becomes:

		\[
			\contraction{\langle}{\phi(x_{1})}{}{\phi(x_{2})}
			\contraction{\langle \phi(x_{1}) \phi(x_{2})}{\phi(x_{3})}{}{\phi(x_{4})}
			\langle \phi(x_{1}) \phi(x_{2}) \phi(x_{3}) \phi(x_{4}) \rangle
			=
				\contraction[2ex]{\langle}{\phi(x_{1})}{\phi(x_{2}) \phi(x_{3})}{\phi(x_{4})}
				\contraction{\langle \phi(x_{1}) }{\phi(x_{2})}{}{\phi(x_{3})}
				\langle \phi(x_{1}) \phi(x_{2}) \phi(x_{3}) \phi(x_{4}) \rangle
				\tag{2.6-A}
				\label{eq:2.6-A}
		\]
			 \begin{figure}[htbp]
			\centering
			\includegraphics[width= 11 cm]{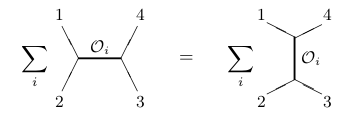}
			
			\caption[Crossing Symmetry]{Schematic representation of Crossing symmetry in terms of $s$ and $t$ OPE channels.}
			\label{fig:Cross}
		\end{figure} 
		Schematically one can think of the OPE as a vertex of interaction between two $\phi$ operators and the whole set $\sum\mathcal{O}$ of operators that participate in the OPE. In this way, one sees the external scalar fields "exchanging" $\mathcal{O}$'s. Note, of course, that this is \textbf{not} a Feynman diagram or a sketch of a scattering event. It is merely a way to visualize the OPE channels, in a recognizable $s$ or $t$ channel from usual interactions in QFT. Note this schematic description in Figure \ref{fig:Cross}.
		
		Now that we are equipped with the CBs of identical scalars we can make our crossing symmetry equation explicit. It is clear from the definitions of $u$ and $v$ that permuting the $x_{i}$ will alter the aspect of functions of these cross ratios.
		Indeed, doing $2 \leftrightarrow 4$ we get: 
		\begin{equation}
			 \frac{x_{12}^{2}x_{34}^{2}}{x_{13}^{2}x_{24}^{2}} \leftrightarrow \frac{x_{23}^{2}x_{14}^{2}}{x_{13}^{2}x_{24}^{2}} ~,~ u\leftrightarrow v
		\end{equation}
		Then, we see, for the full four point function:
		\begin{equation}
			\frac{g(u,v)}{x_{12}^{2\Delta_{\phi}}x_{34}^{2\Delta_{\phi}}} \rightarrow \frac{g(v,u)}{x_{12}^{2\Delta_{\phi}}x_{34}^{2\Delta_{\phi}}} \frac{x_{12}^{2\Delta_{\phi}}x_{34}^{2\Delta_{\phi}}}{x_{14}^{2\Delta_{\phi}}x_{32}^{2\Delta_{\phi}}}
		\end{equation}
		
		And demanding both sides to match we read:
		\begin{equation}
			g(u,v) = g(v,u) \left( \frac{u}{v}\right) ^{\Delta_{\phi}}
		\end{equation}
		
		Where we noticed $\frac{x_{12}^{2\Delta_{\phi}}x_{34}^{2\Delta_{\phi}}}{x_{14}^{2\Delta_{\phi}}x_{32}^{2\Delta_{\phi}}} =\left( \frac{u}{v}\right) ^{\Delta_{\phi}} $
		
	 Noting the appropriate replacements for the Conformal Block expansion in both channels we get:
		\begin{equation}
		\label{eqn:Bootstrap} \sum_{\mathcal{O}} f_{\phi\phi\mathcal{O}}^{2}\left(  v^{\Delta_{\phi}}g_{\Delta,l}(u,v) - u^{\Delta_{\phi}} g_{\Delta,l}(v,u)\right)=0 
		\end{equation}
Notice that the sum over operators $\mathcal{O}$ is, in fact, a sum over dimensions and spins $\Delta$ and $l$, as a primary operator is identified by the eigenvalues of the commuting charges defining a representation of the Conformal Algebra ($D$ and $M_{\mu\nu}$).
 Now, with equation \ref{eqn:Bootstrap}, we have a non-trivial constraint on all Conformal Field Theories: the Bootstrap/crossing equation. This deserves a small reflection on what defines a CFT.
 \subsection{What is a CFT?}
 Up to now, we were essentially defining a CFT by the existence of local operators called primaries (the Conformal family is completely determined by symmetry), with dimension $\Delta_{i}$, along with correlation functions, which we saw can be reduced from n-points all the way down to one point. However, it is crucial to notice that the OPE, and the 3-pt function depend on the OPE coefficients $f_{ijk}$, so we also need these numbers to define a CFT.
 
  Naively, we could just state that given numbers ${\Delta_{i},f_{ijk}}$, which are called the CFT data, we can define a CFT without ambiguity. However this is not true. A random set of numbers will \textbf{not} define a CFT, because there exists a highly non-trivial constraint on the relationship between the spectrum $\Delta_{i}$ and the OPE behavior $f_{ijk}$, which is, of course, the Crossing Symmetry equation \ref{eqn:Bootstrap}. 
 
 It is natural to ask the question if these conditions are enough to have a fully consistent theory. It is generally believed \cite{Simmons-duffin2016,Rychkov2016} that these constraints almost completely fix a theory, however, one can, in principle, impose further conditions by demanding that the CFT is well defined in topologically non-trivial spaces. For example, requiring that a 2D CFT makes sense in a Torus, by imposing appropriate periodicity conditions on the partition function. This is known as modular invariance, and we will discuss it briefly in Chapter \ref{twist}.
 
 Now we are just starting to see the power of the Bootstrap, we have developed a much richer definition of a CFT without touching any CFT in particular!
 
 \section{Attacking the Crossing equation}
 \label{summingvectors}
 
 Notice that the Bootstrap equation \ref{eqn:Bootstrap} is written as a statement for every value of the cross-ratios $u$ and $v$, where the unknowns are the spectrum and the OPE coefficients. It can, then, be seen as a functional equation. Some basic analytical results can be obtained by performing an expansion in $u$, or considering certain special limits, but our main focus will be a numerical approach.
 Before that, one can write the equation in a more geometrical fashion:
  \begin{equation}
  \label{eqn: VecBoot}
  \sum_{\Delta,l}p_{\Delta,l}  \overrightarrow{F_{\Delta,l}^{\Delta_{\phi}}} = 0
  \end{equation}
 Note that the coefficient $p_{\Delta,l}$  is positive, as we did the identification $p_{\Delta,l} =f_{\phi\phi\mathcal{O}}^{2} $ where the operator $\mathcal{O}$ has dimension $\Delta$ and spin $l$; Also we define:  \begin{equation} \label{eqn: 2.9} \overrightarrow{F_{\Delta,l}^{\Delta_{\phi}}} = v^{\Delta_{\phi}}g_{\Delta,l}(u,v) - u^{\Delta_{\phi}} g_{\Delta,l}(v,u)\end{equation} Which is a crossing symmetric function, associated to an operator $ \mathcal{O} \leftrightarrow (\Delta,l)$ for a fixed dimension of the external scalar $\Delta_{\phi}$.

 The idea is to see the space of functions of $u$ and $v$ as an infinite dimensional vector space, and perhaps look at finite dimensional subspaces of the equation and obtain constraints from that.
 One can think of truncating the space by writing the equation for particular values of $u$ and $v$, demanding that the derivatives match on a certain point, or other more exotic statements.
 
 A good way to visualize this is to see equation \ref{eqn: VecBoot} as a sum of vectors with positive coefficients equal to zero. If one states this, for example in a 3-dimensional subspace, one could think of many vectors from the origin (corresponding to all operators in the sum) adding up to zero. This idea will be fruitfully exploited in the next section.

  \section{Algorithm to bound dimensions from the Bootstrap equations}
  \label{sec: 2.5}
  Armed with our geometrical intuition, one might consider the following problem: Given a spectrum for the CFT (in fact a spectrum contributing to the scalar-scalar OPE), will we always find a choice for the $p_{\Delta,l}$ such that the vectors add up to zero, and the theory obeys crossing? 
  
  If, for some (poor) choice of spectrum, all of the vectors were on a single half-space, that is, we could find a separating plane, through the origin, such that all the vectors are on one side, then, it is clear that adding them together with positive coefficients could never lead to a zero result on the Bootstrap equation. This would declare that our guess of spectrum was inconsistent, and we could exclude that candidate theory!

  In a more algorithmic fashion, one then can try to use the following procedure:
  
  \textbf{Algorithm}
  	\begin{enumerate}
  		\item{Assume some $\Delta,l$ spectrum for operators in the $\phi \times \phi$ OPE.}
  		\item{Look for a linear functional $\alpha$ such that it is non-negative on all vectors $F$ for the chosen spectrum, and strictly positive on at least one, which we choose to be the Identity Operator.}
  		\item{By applying $\alpha$ to both sides of the Bootstrap equation one gets $0>0$, a contradiction, which excludes our hypothesis for the spectrum.}
  	\end{enumerate}
  	
  The idea of the linear functional $\alpha$ is just a generalization of the separating plane mentioned earlier. In fact, what we are doing implicitly with the separating plane, is introducing a vector (the normal to the plane) and demanding that the scalar product with all the vectors in the Bootstrap sum be non-negative. This is what it means for all the vectors to be on the same side of the plane. However, the concept of a linear functional can be looked at as something more abstract than taking dot products, but this will suffice for our applications. Before exploring the possibilities for this $\alpha$, let us briefly recap step 3 with more care.
  Consider applying $\alpha$ to both sides of equation \ref{eqn: VecBoot}:
  \begin{equation}
  	\alpha \left( \sum_{\Delta,l}p_{\Delta,l}  \overrightarrow{F_{\Delta,l}^{\Delta_{\phi}}}\right) = \alpha(0) 
  \end{equation}
  
  By linearity of $\alpha$:
  
  \begin{equation}
  \sum_{\Delta,l}p_{\Delta,l}   \alpha \left(\overrightarrow{F_{\Delta,l}^{\Delta_{\phi}}}\right) = 0 
  \end{equation}
  
  If we can satisfy condition 2, we can write: 
  \begin{equation}
  \sum_{\Delta,l} \overbrace{\underbrace{p_{\Delta,l}}_{\ge 0}   \underbrace{\alpha \left(\overrightarrow{F_{\Delta,l}^{\Delta_{\phi}}}\right)}_{\ge 0}}^{\ge 0} = 0 
  \end{equation}
  The overall parcels we are summing are non-negative, and furthermore, for the identity operator we demanded $\alpha(F_{I}) > 0$ and we have $f_{\phi\phi I}=1>0$ because inserting the identity in a 3-pt correlator gives $\langle\phi\phi I \rangle=\langle \phi \phi \rangle$, such that $f_{\phi\phi I}$ corresponds to the normalization of the 2-pt function which we can choose to be one; i.e., the term corresponding to $I$ is strictly positive. This means we have $0 > 0$, a contradiction, and step 3 follows.
  
  Unfortunately, we are being quite naive in step 2. Not only is the space of linear functionals over functions of $u$ and $v$ enormous, but the vector space of functions is also infinite. As we mentioned before, one can truncate the Bootstrap equation, in the sense of considering a finite dimensional subspace of the space of functions (considering particular values, derivatives at a point, etc.).
  
   However, one still has, in principle, infinitely many $\Delta$s and $l$s, even though our linear functionals act on a finite dimensional space.
   In fact, we need to find linear functionals such that an infinite number of constraints are satisfied:
  
  \begin{equation}
  \label{eqn: const}
  	\alpha \left(\overrightarrow{F_{\Delta,l}^{\Delta_{\phi}}}\right)  \ge 0
  \end{equation}
  
  From this analysis, we have implicitly formulated two approaches to our problem of solving crossing symmetry. In the context of Linear Programming, they are usually called the primal and dual formulations of a feasibility problem (we are interested in checking if a set of inequalities define a non-empty region or not.)
  First we have, the Primal formulation:
  \begin{equation}
  	\exists p_{\Delta,l}: ~   \sum_{\Delta,l}p_{\Delta,l}  \overrightarrow{F_{\Delta,l}^{\Delta_{\phi}}} = 0 ;~ p_{I}=1 
  \end{equation}
  If we find such OPE coefficients we check that our choice of spectrum is consistent with crossing.
  On the other hand our description of a linear functional $\alpha$ suggests the Dual formulation:
  \begin{equation}
  	\exists \mathbf{\Lambda}\in \mathbb{R}^{N}: ~  \mathbf{\Lambda} \cdot \mathbf{F}_{\Delta,l}^{\Delta_{\phi}} \ge 0 ~ \forall (\Delta,l)\in S'
  \end{equation}
  Where $S$ denotes the spectrum, and we also require $\mathbf{\Lambda} \cdot  \mathbf{F}_{I} \ge \epsilon$ where the greater or equal sign can be replaced by an equality, setting a normalization (Note that $\alpha(\mathbf{F}) = \mathbf{\Lambda}\cdot\mathbf{F}$). Finding a solution to this problem excludes the spectrum $S'$, and as we advocated previously, this is the LPP we will pursue when doing numerical Bootstrap: the Dual problem.
  
  To actually solve it, there are a few alternatives to explore. One can try to find specific functionals by hand, or trial and error, and obtain sporadic information about the CFT, or find a systematic/computational way of exploring the space of functionals and identifying those who give non-trivial constraints.
  On the other hand, we still have to deal with the infinity of constraints, and there are also a few choices here. One can use a Semi-definite programming approach, which is able to treat the dimension $\Delta$ as continuous, but has to introduce a cutoff in the spin $l$. This is the technique most commonly used in the recent literature but it is based on a more sophisticated approach in terms of the numerics, requiring a rather too deep incursion into the computational techniques, which is against the spirit of accessibility and easy use of this thesis. 
  
  Alternatively, one can additionally discretize the dimension and introduce a cutoff $\Delta_{max}$. This will give a finite number of inequalities, which is treatable with more standard computational techniques, namely Linear Programming, as will be seen in the next chapter. This technique was used in the earlier days of the Bootstrap literature but it is quite convenient for our purposes of obtaining results with the simplest and most accessible techniques, allowing for results to be obtained without any external software, other than a traditional scientific programming package, like \textit{SciPy} for Python users, or a standard general/symbolic calculation software like \textit{Mathematica}.
  Also, there are many modern numerical Bootstrap papers implemented in terms of Linear Programming \cite{ElShowk:2012hu,Paulos:2014vya}.
\chapter{Bootstrap Bounds}
\label{Bounds}

\section{Guessing a 2D subspace}

Before making our approach systematic and getting to the core of our results, we first proceed by one of the techniques mentioned in section \ref{sec: 2.5}. 
We consider a specific finite-dimensional subspace of the vector space of functions of $u$ and $v$: $ \mathfrak{F}_{u,v}$. We consider two pairs of combinations of the values assumed by the conformal Blocks in their crossing symmetric form. For the first component of our subspace, we use $(u,v)= (\frac{1}{2},\frac{3}{5}) ~;~ (\frac{1}{2},\frac{1}{3}) $ and for the second one: $ (u,v)= (\frac{1}{2},\frac{3}{5}) ~;~ (\frac{1}{3},\frac{1}{4}) $ this is defined for every operator $(\Delta,l)$ participating in the OPE of  $\phi \times \phi$ and we first choose $\phi$ to be a scalar of a  2D theory with $\Delta_{\phi} = \frac{1}{8}$. (Note that we chose a 2d subspace of $ \mathfrak{F}_{u,v}$, for simplicity, but this has nothing to do with the choice of a CFT in two spacetime dimensions, these two twos are not to be confused).

\begin{figure}[htbp]
 	\centering
 	\includegraphics[width= 8 cm]{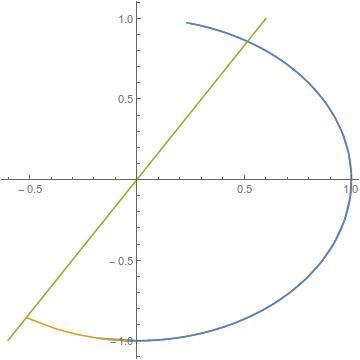}
 	
 	\caption[2D Ising bound]{Aspect of the 2d vectors and the separating line, for the case $\Delta_{\phi} = \frac{1}{8}$ (2D Ising)}
 	\label{fig:2dis}
 \end{figure}

To be precise, we consider the function (essentially we normalize by the Identity): \begin{equation}
\label{eq:H}
	H_{\Delta,l}^{\Delta_{\phi}}=\frac{F_{\Delta,l}^{\Delta_{\phi}}}{u^{\Delta_{\phi}}-v^{\Delta_{\phi}}}
\end{equation} and take our vectors to be: \begin{equation}
	\left( H(\frac{1}{2},\frac{3}{5})-H(\frac{1}{2},\frac{1}{3}),H(\frac{1}{2},\frac{3}{5})-H(\frac{1}{3},\frac{1}{4})\right) 
\end{equation}
Notice that the function $ F_{\Delta,l}^{\Delta_{\phi}}$ is defined in terms of the Conformal Blocks (see eq. \ref{eqn: 2.9}), which are in this case, the CBs of the two dimensional global conformal algebra: Eqs. \ref{eqn: 2dblock}. If we plot these vectors (normalized) as points in the plane with the coordinates being the components of the vector, we obtain Figure \ref{fig:2dis}. 

 Each curved line has a color which corresponds to a different spin (blue is $l=0$ and yellow is $l=2$). For a fixed spin, as we start the dimension $\Delta$ at $0$ and slowly increase it, we get a curve in this 2d-space. For the spin 2 (and similarly for higher spins which are not represented for the sake of clarity), the vector begins at the intersection with the straight green line, and slowly drifts to the point $(0,-1)$. However, for the case of the scalar, the vector starts near $(0,-1)$, rotates anti clockwise crossing the green line, then comes back down, and stabilizes near $(0,-1)$ for large $\Delta$. The green line shows that it is not possible to separate all the vectors into one side of the plane. In fact, it is clear that there must exist operators in the theory which correspond to points which are above the green line, otherwise, we could not add all the vectors together with positive coefficients and get zero. If we are precise, we find that the values of $\Delta$ for which the points are above the green line are:
 \begin{equation}
 	\Delta \in [0.161, 1.035] 
 \end{equation}
 This means we can make our first claim based on the Bootstrap.
 
 \textbf{Claim}:
 
 \textit{For a 2D CFT with a scalar of dimension $\Delta_{\phi}=\frac{1}{8}$ there exists another scalar, say $\varepsilon$, with dimension $\Delta_{\varepsilon} \in [0.161, 1.035]$}
 
 We can make our claim more rigorous by using the algorithm described in section \ref{sec: 2.5}.
 First we assume that our spectrum \textbf{does not} contain any operators with $l=0$ and  $\Delta \in [0.161, 1.035] $.
 
  Next, we need to find a linear functional on this 2d subspace. We can take $\alpha$ to be the projection of our vectors in the direction of the vector $\vec{a}$. We choose $\vec{a}$ to be the unit normal vector in the direction orthogonal to the green line, pointing downward. We then take the projection onto this vector:
	\begin{equation}
	\alpha = \vec{a}\cdot\left( H(\frac{1}{2},\frac{3}{5})-H(\frac{1}{2},\frac{1}{3}),H(\frac{1}{2},\frac{3}{5})-H(\frac{1}{3},\frac{1}{4})\right) 
	\end{equation}
	
	Because we did not include the mentioned interval in the spectrum, all the vectors are to the right of the green line, so the scalar product is non-negative for every operator in the spectrum (in fact, its clear that it is strictly positive for most of them). Therefore, we found a linear functional that satisfies all the constraints. Our spectrum is therefore excluded, and there must exist a scalar operator $\varepsilon$ with $\Delta_{\varepsilon} \in [0.161, 1.035]$. This proves our claim. 
	
	Note that this is a rigorous, non-perturbative claim, about the full quantum-corrected dimension $\Delta$. In fact a theory with a scalar of dimension $\frac{1}{8}$ is well known, it is the Ising Model. Indeed, the only other Virasoro primary in the theory is a scalar $\varepsilon$ of dimension $\Delta_{\varepsilon}=1$ which is really close ($3.5\%$) to the upper bound that we obtained.
	  \begin{figure}[htbp]
	  	\centering
	  	\includegraphics[width= 9 cm]{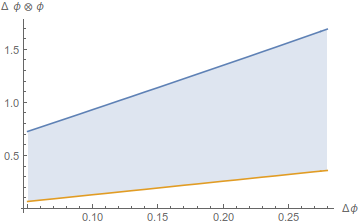}
	  	
	  	\caption[2D bound]{Upper and lower bound for $\Delta_{\varepsilon}$ given a  general $\Delta_{\phi}$ }
	  	\label{fig:2dbound}
	  \end{figure} 
	
	We can make a similar construction while choosing different values for $\Delta_{\phi}$. For each choice of $\phi$ a similar behavior is observed, in the sense that it is always possible to find a straight line that isolates a set of vectors corresponding to scalars in a small interval of dimensions. We can then use the same proof as before choosing the appropriate normal vector, and generalize the result.
	
	This will give us a constraint on the general space of CFTs, as parametrized by the dimension of the two first scalars in the theory as we can note in Figure \ref{fig:2dbound}.
	 This creates a strip of allowed theories in the space of Unitary 2D CFTs, just by considering a (criteriously chosen) 2 dimensional subspace of an infinite dimensional equation. This is highly suggestive of the power of the Bootstrap, and one begins to wonder what happens as we choose larger subspaces.

   \section{CB Derivatives}
   \label{sec: CBD}
In the previous section, we did not have to look for values of the function $F_{\Delta,l}$ in which the geometric construction worked. We were lucky enough to be inspired by \cite{Simmons-duffin2016}. Of course, this is good to prove the concept of the Conformal Bootstrap, but it is very far from being a systematic way to gain information. For example, the same choice of subspace is useless in providing information when one goes to 4D CFT and uses the appropriate Conformal Blocks.

A more sustainable way of finding a good linear functional to apply our algorithm, would be to choose a family of $\alpha$'s, parametrized by a choice of one or more numbers, to be tuned in order to find an $\alpha$ compatible with the constraints \ref{eqn: const}. Here is the crucial argument: if one cannot find parameters such that all the requirements are satisfied, this does \textbf{not} mean, that the Bootstrap equation does not contain information about our physics. That could be the case sometimes, of course, but, in principle, it just means that we are not looking in a big enough space of linear functionals, or maybe, we are not choosing the right kind of functional to obtain information. However, if parameters are found such that we are able to apply phase 2 of our algorithm, we can indeed obtain a result from the Bootstrap.

   The most common way to do this, is to consider functionals that are a finite (with a cutoff at order $\Lambda$) linear combination of derivatives (with respect to $z$, $m$ times and to $\bar{z}$, $n$ times) of our infinite vectors $\overrightarrow{F_{\Delta,l}^{\Delta_{\phi}}}(z,\bar{z})$ taken at a point which is quite symmetric: $z=\bar{z}=\frac{1}{2}$ (it is symmetric in the sense that  derivatives $(m,n)$ are the same as $(n,m)$; It is known as the crossing symmetric or self-dual point). Of course, the parameters that we can tune to move through the space of functionals are the constant coefficients ($a_{m,n}$) multiplying the linear combinations. Therefore, we can write the general form of our linear functional:
   	\begin{equation}
   	\alpha = \sum_{m,n}^{\Lambda} a_{m,n} \partial_{z}^{m} \partial_{\bar{z}}^{n}|_{z=\bar{z}=\frac{1}{2}}
   	\end{equation}
   	Now, we need to find a way to make the computer scan the space of $a_{m,n}$'s such that $\alpha(F_{\Delta,l}) \geq 0$ for all $\Delta,l$ in the spectrum:
	\begin{equation}
	\alpha(F_{\Delta,l}) = \sum_{m,n}^{\Lambda} a_{m,n} \partial_{z}^{m} \partial_{\bar{z}}^{n}F_{\Delta,l}(z,\bar{z})|_{z=\bar{z}=\frac{1}{2}} \ge 0
	\end{equation}
	
	As we mentioned in \ref{sec: 2.5}, our approach will be to treat the spectrum as discrete, and introduce a maximum dimension for the operators $\Delta_{max}$. Additionally, we must  introduce a maximum spin, $l_{max}$ after which we cutoff all constraints. Of course, if we find an $\alpha$ which satisfies \ref{eqn: const}, we should a posteriori check that it also works for large $\Delta,l$.
	
	Taking the expression \ref{eqn: 2dblock}, and the definition \ref{eqn: 2.9}, we have an explicit formula to apply our linear functional for 2D CFTs. In fact, we computed the required derivatives, directly through symbolic computation in \textit{Mathematica}. Furthermore, we took our initial cutoff in the derivate order to be $\Lambda=8$, and by direct calculation one checks that the non-zero terms correspond to $(m,n) \in \{(1, 0)$ , $ (2, 1)$ , $(3, 0)$ , $(3, 2)$ , $(4, 1)$ , $(4, 3)$ , $(5, 0)$ , $ (5, 2)$, $(7, 0)$ , $ (6,1)\} $, so we chose a ten dimensional linear functional.
	
	For the sake of concreteness, we usually take the spacing between dimensions to be $\delta\Delta=0.1$, and for operators of lower dimension we typically refine our grid to $\delta\Delta=0.001$. Furthermore, the cutoffs are chosen to be $\Delta_{max}=80$ and $l_{max}=24$. The choice of cutoff is easy to justify, as suppressing the demand that the constraints for these values of $(\Delta,l)$ be satisfied does not substantially alter  the results.
	 The choice of spacing is taken in order to be sufficiently fast for calculations in an ordinary laptop, but refined enough that a found $\alpha$ does not fail to solve the constraints for intermediate values of the dimension (if it does so, one can ask for further constraints to be satisfied in the troublesome areas).
	
	Now that everything is in place, we can put our algorithm to work. First, we fix the dimension of the lowest operator of the theory to be $\Delta_{\phi}$, and take it as the external operator in the correlator.
	We assume a spectrum, which contains the identity operator ($\Delta_{I}=0$), the scalar $\phi$, which we choose as defining the theory, then a second scalar operator $\varepsilon$ with a dimension to determine, and then all possible operators with dimension $\Delta \ge \Delta_{\varepsilon}$ that respect the unitarity bound. Taking the system to be symmetric under inversion $\phi \rightarrow -\phi$, the possible operators appearing in the OPE, will be the identity, $\varepsilon$ and everything after that, so our constraints are:
	
	\begin{equation}
		\alpha(F_{I}) > 0 ~;~ \alpha(F_{\Delta,l}) \ge 0 ~\mathnormal{for}~ \Delta \geq \Delta_{\epsilon}
	\end{equation}
	
	This translates as a finite (but very big) number of linear inequalities, with the following aspect:
	
	\begin{equation}
		a_{1,0} \partial_{z}^{1} F_{\Delta,l}(z,\bar{z})|_{z=\bar{z}=\frac{1}{2}} + a_{2,1} \partial_{z}^{2} \partial_{\bar{z}}^{1}F_{\Delta,l}(z,\bar{z})|_{z=\bar{z}=\frac{1}{2}} + \dots + a_{6,1} \partial_{z}^{6} \partial_{\bar{z}}^{1}F_{\Delta,l}(z,\bar{z})|_{z=\bar{z}=\frac{1}{2}} \ge 0
	\end{equation}
	
	Where there are 10 coefficients, associated to our choice of taking derivatives up to the eighth order. Of course, to satisfy hypothesis 2 of the Algorithm, the action of $\alpha$ on one of the operators should be strictly positive. The only natural operator to choose is the unit operator, because it is important in all field theories, and we will see that it has a very special weight in the Bootstrap equations. We then have our only strict inequality:
	
	\begin{equation}
		a_{1,0} \partial_{z}^{1} F_{I}(z,\bar{z})|_{z=\bar{z}=\frac{1}{2}} + a_{2,1} \partial_{z}^{2} \partial_{\bar{z}}^{1}F_{I}(z,\bar{z})|_{z=\bar{z}=\frac{1}{2}} + \dots + a_{6,1} \partial_{z}^{6} \partial_{\bar{z}}^{1}F_{I}(z,\bar{z})|_{z=\bar{z}=\frac{1}{2}} > 0
	\end{equation}
	
	In our case we generally have at least $(80)/0.1 + (80-2)/0.1 + \dots + (80-24)/0.1 \approx 9000$ inequalities, and 10 unknowns. Most of the time we actually have a bit more than that, because we increase the number of operators chosen (make the discretization finer) for small values of the dimension $\Delta$.
	
	\subsection{Using CB Derivatives to find the optimal bound}
	\label{optimal}
	
	Before approaching this problem of many linear inequalities, we briefly discuss the way we can use this to bound the dimension of the operator $\varepsilon$, $\Delta_{\varepsilon}$. Suppose we pick our spectrum with a very large gap, that is $\Delta_{\varepsilon,1} \gg \Delta_{\phi}$. In general this choice is inconsistent, so we expect that the equations have a solution. If this happens, then our algorithm allow us to exclude this choice of spectrum, that means that $\Delta_{\varepsilon} \ne \Delta_{\varepsilon,1}$. We claim that in fact, if the equations have a solution, the result is stronger, in particular, it means that $ \Delta_{\varepsilon} < \Delta_{\varepsilon,1}$. Suppose then, that we found a solution to the system, with a gap $\Delta_{\varepsilon,1}$, excluding this choice. Next, assume a second spectrum with gap $\Delta_{\varepsilon,2} > \Delta_{\varepsilon,1}$. Then all the equations associated to this spectrum were already included in the previous one, because we assumed all possible operators above $ \Delta_{\varepsilon,1}$ to exist. Because of this, the first solution will work for any spectrum with $\Delta_{\varepsilon,2} > \Delta_{\varepsilon,1}$, proving our claim.
	
	Now suppose we choose $\Delta_{\varepsilon}$ to be very close to $\Delta_{\phi}$, which is physically reasonable (Liouville theory for example, see \cite{Ribault:2015sxa}). The Bootstrap cannot exclude theories that we know exist, otherwise they would not be consistent. Then, we expect the system of equations to be unsolvable, which means that our algorithm cannot work, and we can not exclude our choice of spectrum.
	
	This suggests that there exists a method to find the best possible value, an optimal bound for $\Delta_{\varepsilon}$. Choose a high value for $\Delta_{\varepsilon}$ such that the equations are solvable, and we exclude everything above, then lower it successively (maybe using a binary search for speed), until the equations stop having a solution after lowering the dimension by the minimum possible amount. This will give, by construction, the best possible bound for this choice of linear functional.
	
	\subsection{Solving many linear inequalities}
	\label{sec: LinearP}
	
\begin{figure}[htbp]
 	\centering
	 	\includegraphics[width= 9 cm]{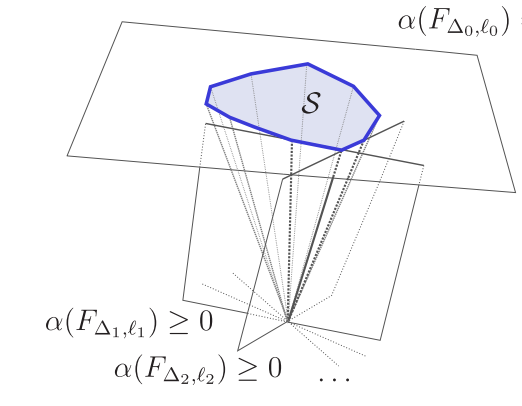}
	 	
	 	\caption[A cone in the parameter space of our functional]{Each positivity condition defines a side of our polytope cone. The Identity contribution sets a base. \cite{Poland2012} }
      \label{fig:polycone}
	 \end{figure}

	Now we have a way to optimize our bound, but we still haven't discussed how to verify if the inequalities actually have a solution or not. In this framework we can have yet another geometrical interpretation that will help us understand if our problem is solvable. Each inequality defines a half-space through the origin, and as we take into account more and more inequalities, we must consider the intersection of these half-spaces. This cuts a cone in our 10-dimensional vector space, but because of our discretization, our cone can be thought to have a "base" which is a very complicated polytope. In the limit where the constraints are taken to be continuous, this base becomes a smooth "curve":  this reminds us of the Semi-definite approach, and our discretization in then a polygonal approximation to the methods of \cite{Poland2012}. We see a sketch in Figure \ref{fig:polycone}.

 Essentially, any inequality we consider should define a half-space which intersects the cone, and if all of them have a common point (a solution), then in fact there are infinitely many solutions, because multiplying the solution vector by a positive constant will still respect the sign of the linear combinations. Otherwise, the solution space would be just the point at the origin, but this is excluded by the demand of the strict inequality for the Identity Operator. (In figure \ref{fig:polycone}, the Identity contribution is normalized to one, so we see plane that does not contain the origin; this makes the previous argument clearer.)
 Alternatively, one can choose a strict normalization $\alpha(F_{I}) = 1$ instead of an open one,  $\alpha(F_{I}) > 0$ and therefore have a unique solution. The numerics are sometimes sensitive to these choices, so one should use whatever stabilizes the problem.
 
 The most common tool to attack a problem of many linear inequalities, is usually related to finding the minimum of a linear function on the set defined by the inequalities (our constraints). These problems are usually known as \textbf{Linear Programming} problems, and there are many algorithms available to solve them. The most common ones are in a family called "Simplex Algorithms", but in our case, where we have many more inequalities than unknowns, the most appropriate choice in the "Interior Point" Method. Both of them are implemented in \textit{Mathematica} and are easy to use.
 
 There are however a few subtleties that make the use of these algorithms in our Bootstrap Program not completely trivial:
 
 \begin{enumerate}
 	\item{There is no natural quantity to minimize.}
 	\item{Our unknowns $a_{m,n}$ are not naturally bounded, there is no reason they should be positive.}
 	\item{We have a strict inequality.}
 \end{enumerate}
 We now describe our strategies to attack these problems:

   \begin{enumerate}
   	\item{We can choose to minimize the first variable $a_{1,0}$, or $-\alpha(F_{I})$. The second choice is natural in the bounds to OPE coefficients, which we will discuss later, and the first is the simplest conceivable linear function. (In this context, a linear function is of the form $c \cdot x$ where $c$ is a constant vector, and $x=(a_{1,0},\dots, a_{6,1})$ is the vector of unknowns; we can also take $c= 0$ and simply find a feasible solution) }
   	\item{Because if we find a solution any positive multiple of it will still work, all we need to do is to allow the $a_{m,n}$ to be negative (usually the algorithms take them as positive). We then make the additional (weak) constraints $ a_{m,n} > -0.1 $. This just ensures that negative values are allowed.}
   	\item{We define a small value $ \delta  $, such that we replace the strict inequality, with $ \alpha(F_{I}) \ge \delta $. Numerically it does not make sense to demand a strict inequality, so we write everything in these terms ($\ge$). As we mentioned before, we can normalize everything with an equality, but this may imply that we need to tune condition 2.}
   	
   \end{enumerate}
   
   Because the zero vector is really close to being a solution (it only violates the strict inequality), there are some occasional artificial solutions, with all $a_{m,n}$ really close to zero. When these solutions are checked they are sometimes false, so the introduction of the "machine $\delta$" defined in 3, is important to drive the Interior Point method away from these "solutions"; of course the normalization $\alpha(F_{I}) = 1$ is also a good alternative to solve this problem.
   
   With all the machinery set in place, we can begin obtaining serious results.
   
  \section{Bounds in 2D Globally Conformal CFT }
  
  As seen in \ref{Boot}, the Global Conformal Blocks for identical scalars in a 2D CFT, have an explicit formula obtained by Dolan and Osborn \cite{Dolan:2000ut} \cite{Dolan:2003hv}:
  \begin{align}
  	&	g_{\Delta,l}^{2D}(u,v) =z^{\frac{\Delta + l}{2}}~ _{2}F_{1}(\frac{\Delta + l}{2},\frac{\Delta + l}{2},\Delta + l;z) ~  \bar{z}^{\frac{\Delta - l}{2}}~ _{2}F_{1}(\frac{\Delta - l}{2},\frac{\Delta - l}{2},\Delta - l;\bar{z})\\
    &  	 + 	z^{\frac{\Delta - l}{2}}~ _{2}F_{1}(\frac{\Delta - l}{2},\frac{\Delta - l}{2},\Delta - l;z) ~ \bar{z}^{\frac{\Delta + l}{2}}~ _{2}F_{1}(\frac{\Delta + l}{2},\frac{\Delta + l}{2},\Delta + l;\bar{z}) \nonumber
  \end{align}
  
  Note that in general, 2D Conformal Symmetry is extended to Virasoro Symmetry, which has infinitely many generators. in particular, it has a central generator (which commutes with every other) and is known as the central charge. In most reviews of 2D CFT, the theories are characterized by the representations of the Virasoro algebra under which the operators transform, their Conformal dimensions ($h=\frac{\Delta+l}{2} ~;~ \bar{h}= \frac{\Delta -l}{2}$) and by the value of this central charge $c$. It seems then, that we are neglecting a lot of information about 2D theories, and that we will only be able to make a very broad guess of where allowed theories live, as global Conformal symmetry is far from being the full set of information about 2D Conformal physics.
  
  However, in spite of this massive omission, we will proceed with the program of \ref{sec: CBD}. We take our spectrum to respect the unitarity bounds, that is $\Delta > 0$ for scalars, and $\Delta > l + d -2 = l$ for operators with Spin. Also we assume that $\phi$ is an even operator, so that it does not appear in its own OPE. Therefore, we use the identity operator, a second scalar of theory $\varepsilon$ whose dimension we bound, and every other possible operator which respects $\Delta > \Delta_{\varepsilon}$ and $\Delta > l$.
  
  The derivatives were computed directly with \textit{Mathematica}, and we used the numerical parameters mentioned in the previous section (indeed, we made the discretization for the operator dimensions much smaller $\delta\Delta= 0.001$ for operators in the range $\Delta \in [0.,2.]$, which is the expected region for the values of $\Delta_{\varepsilon}$ when $\Delta_{\phi}$ is between $0$ and $0.2$).
  
  In Particular, for comparison with the first bound, in section \ref{Bounds}, we obtained the optimal Bound for the 2D Ising Model (Unitary theory with $\Delta_{\phi}=\frac{1}{8}$). The result is:
  \begin{equation}
  	\Delta_{\varepsilon} < 1.004
  \end{equation}
  
  This is interesting for a few reasons:
  \begin{itemize}
  	\item{It doesn't seem to have been a big improvement from our original 2D subspace in \ref{Bounds} ($\Delta_{\varepsilon} < 1.035$).}
  	\item{It appears that as we increase the dimension of the subspace we take (now we are in a 10 parameter space), the bound seems to converge to $\Delta_{\varepsilon}\approx1$}
  \end{itemize}
  
  It seems like we are not getting very far, but we must remember that the 2D Ising Model has an exact solution, and in fact we know that $\Delta_{\varepsilon}=1$. Then it is clear why the Bootstrap can't get the bound any lower than $1$: There exists a physical Conformal theory, which of course is consistent with our requirements of symmetry, unitarity and Crossing invariance. Because we are only getting upper bounds, eventually our constraints hit a physical "wall", which is a real theory. 
  With this in mind, our result is a lot more impressive, our bound is within $0.4\%$ of being saturated by a real theory.
  
  Our intuition is now clear, as we increase the number of parameters for our linear functional, our upper bounds will go down, until they converge around some value, and we won't see significant changes, but we can obtain higher precision results. Of course, with our approach we are technically only obtaining upper bounds, but slicker techniques as in \cite{Poland2012}, allow the joint constraints of more correlation functions, which in some cases can carve out islands in the space of CFTs, giving concrete values for dimensions, up to some precision (in general this is only possible when giving additional information about a specific theory, like the number of relevant operators or the existence of a gap for spinning operators,  there is no substantial improvement with using multiple correlators in the general carving out process that we present in this thesis, if nothing else is assumed about the spectrum).
  
  As before, the procedure can be extended to any value of $\Delta_{\phi}$, and we can actually carve out the map of 2D Conformal theories, as in \ref{fig:2dkink}:
  
  \begin{figure}[htbp]
  	\centering
  	\includegraphics[width= 9 cm]{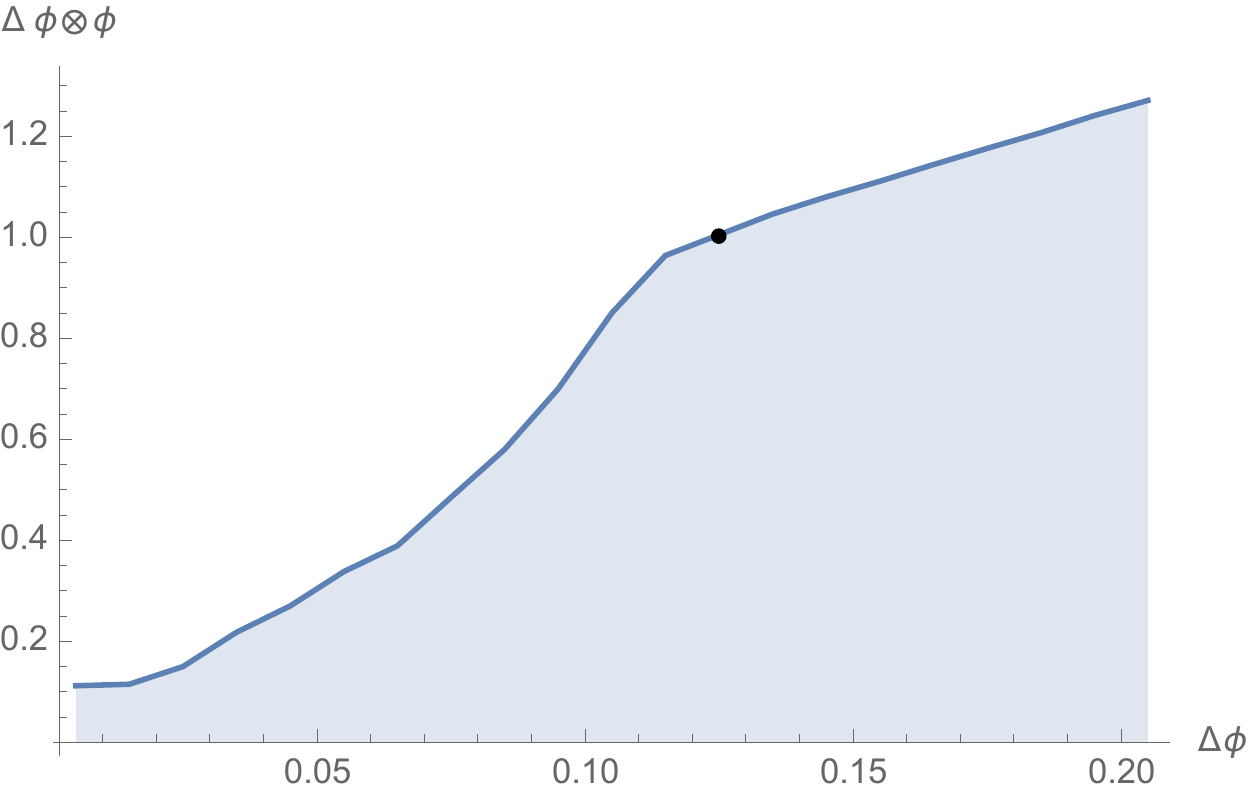}
  	
  	\caption[2D kink in CFT space]{A carved out map of 2D CFTs, in terms of the dimensions for the first two scalar primaries $\phi$ and $\varepsilon \sim \phi \times \phi$, using CB derivatives}
  	\label{fig:2dkink}
  \end{figure} 
  
  The result is quite remarkable as there exists a distinct feature in the plot, a kink around $(\Delta_{\phi},\Delta_{\varepsilon})\approx (0.12,1.)$, which is close to the known value for the dimensions of the Ising model (black dot). In fact for a sufficiently large number of derivatives, the kink does converge to the Ising Model \cite{Kos2014,Simmons-duffin2016}. To the right of this, there is an approximately linear behavior, and when we consider the well known minimal-models, we can check that they are in the allowed region, and in fact they are close to saturating our bound. It is known, from the yellow book, for example, that for the minimal models $\mathcal{M}_{m,m+1}$ ($m>3 ; m\in \mathbb{N}$) the dimensions are exactly:
  \begin{equation}
  	\Delta_{\phi}= \frac{1}{2} - \frac{3}{2(m+1)}  ~;~ \Delta_{\varepsilon} = 2 - \frac{4}{m+1}
  \end{equation}
  
\begin{figure}[htbp]
   	\centering
   	\includegraphics[width= 9 cm]{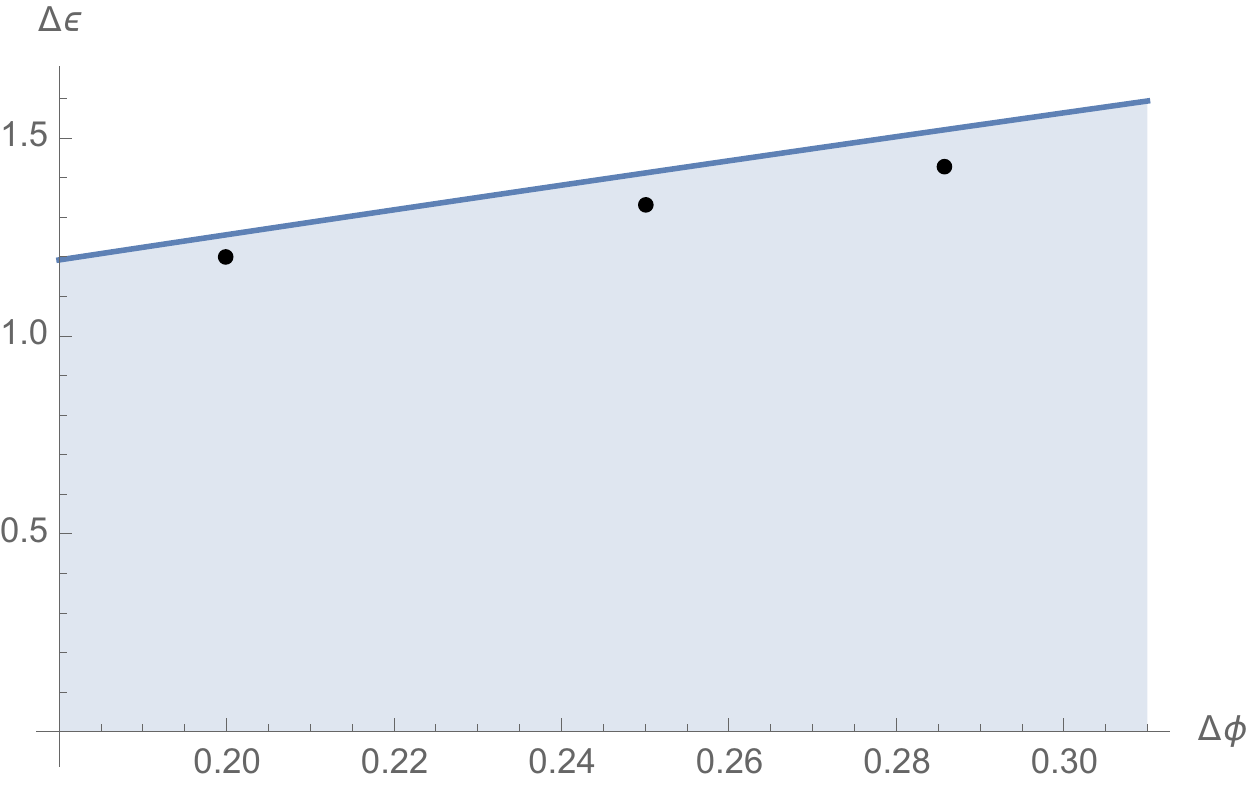}
   	
   	\caption[Minimal models in the CFT space]{A visual check that the dimensions of the Minimal Models are inside our allowed region}
   	\label{fig:2dmin}
   \end{figure}

  In particular, $m=3$ is the Ising model and this gives  $(\Delta_{\phi},\Delta_{\varepsilon})= (\frac{1}{8},1)$, which is indeed in the allowed region. Furthermore, $m= 4,5,6$ have dimensions $(\Delta_{\phi},\Delta_{\varepsilon}) = (\frac{1}{5},\frac{6}{5});(\frac{1}{4},\frac{4}{3});(\frac{2}{7},\frac{10}{7})$, respectively. If we consider dimensions that go a bit higher, we can actually verify this in the plot \ref{fig:2dmin}, where the allowed region is in blue, and the MM are the black dots.
  
   The plot is strong support to our claim that we can't improve our bounds much more with just this technique, as we have the physical minimal models as walls that hold our bounds.
   
   This brings us back to the kink we mentioned earlier. We explored the right side of it quite thoroughly, and noticed that the bounds can probably be improved to obtain higher precision, but the results won't change dramatically. However, to the left of it, there is a steep descent in the dimensions of operators, which means that we are getting smaller gaps. In this region, it is clear that there are no relevant theories which prohibit our bounds from going down, and in fact we obtain much stronger results than in our first approach with just 2 dimensional functionals.
   
\begin{figure}[htbp]
   	\centering
   	\includegraphics[width= 9 cm]{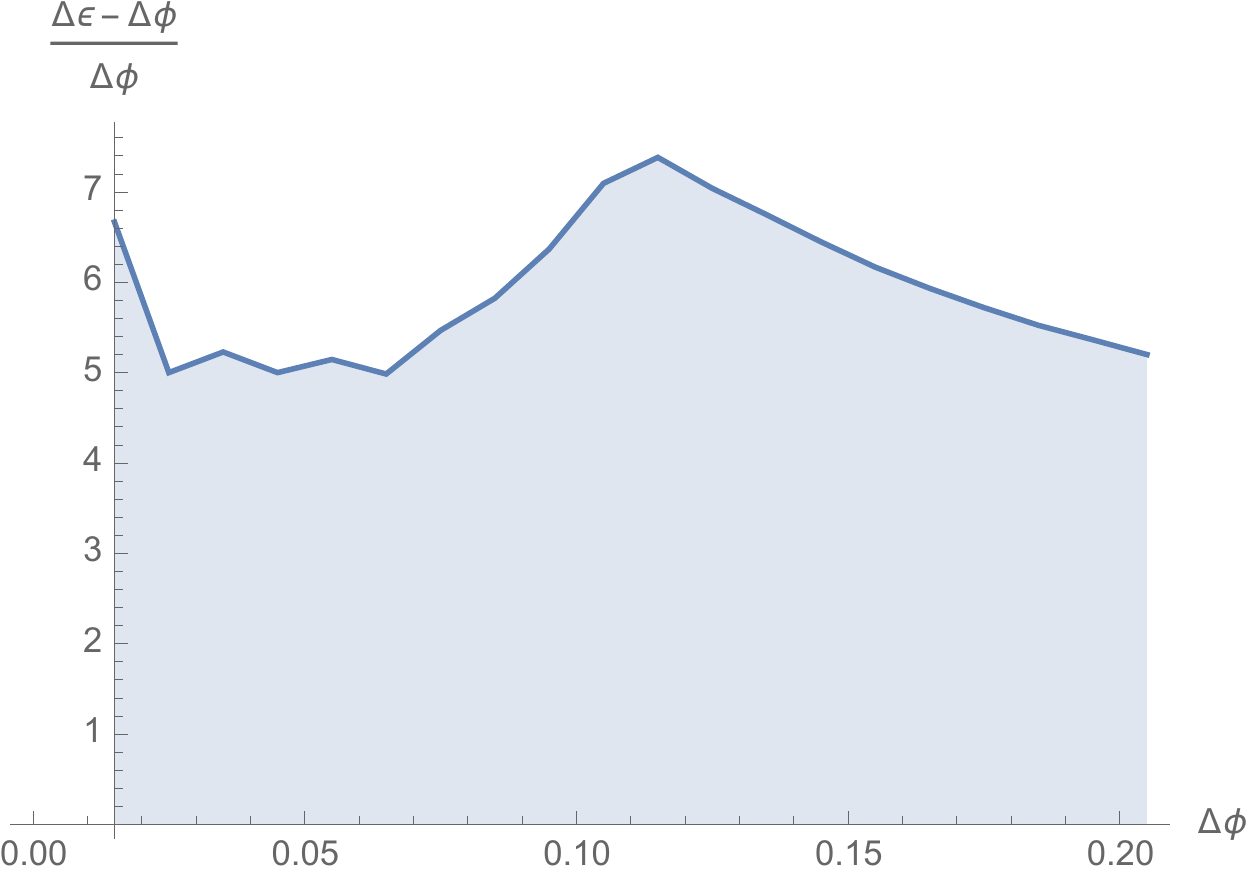}
   	
   	\caption[Relative gap bounds for 2D CFT]{Relative gap in 2D CFTs taken from the previous plots}
   	\label{fig:relgap}
   \end{figure}

   It is also interesting to notice the size of the difference $\Delta_{\varepsilon} - \Delta_{\phi}$ relative to the gap from $\phi$ to the identity: $\frac{\Delta_{\varepsilon} - \Delta_{\phi}}{\Delta_{\phi} - \Delta_{I}} = \frac{\Delta_{\varepsilon} - \Delta_{\phi}}{\Delta_{\phi}} \equiv \xi $ this gives another view on the shape of the spectrum and we can use Figure \ref{fig:relgap} for reference.
   
   The kink close to the Ising model is still prominent in terms of this new scaled variable, and it stresses our ability to notice important features of the space of CFTs, without having introduced any data about a specific theory in the Bootstrap. Also notice that for large values of $\Delta_{\phi}$, $\xi$ is smoothly decreasing. This is consistent with the minimal models, which satisfy $\xi_{m}= 3 + \frac{4}{m-2}$.
   It is also interesting to notice that the scaled variable changes less abruptly, with all theories satisfying $\Delta_{\phi} \in [0.02,0.2] $ sitting approximately in $\xi \in [5,7]$
   
   The main conclusion is that one cannot define a consistent 2D unitary CFT with arbitrarily high gaps between the dimensions of the operators.
   Very roughly, we can say that: 
   
   \begin{equation}
   	\frac{\Delta_{\varepsilon} - \Delta_{\phi}}{\Delta_{\phi}}\lesssim 7
   \end{equation}
   
   \section{Bounding 4D CFTs }
   
   So far, we have done some non-trivial checks on the consistency of well-known 2D CFTs, and we have also obtained some general bounds. However, because of the infinite symmetry of the Virasoro Algebra, the ability to completely solve CFTs, independently of the Bootstrap Method makes our approach feel more like a check, than a truly new way to obtain physical results. Indeed, there are many examples, in particular the afore mentioned Minimal Models, in which the intricacies of the infinite Conformal symmetry allow the existence of relatively simple spectra, in terms of multiplicities of representations, in which the dimensions of operators and even OPE coefficients are completely determined.
   
   When we make the jump to $D\geq3$, however, this scenario changes dramatically \cite{Rattazzi2008}.
    In particular, the Conformal group is finite dimensional, and our introductory discussion in \ref{Chapter2} contains, albeit briefly, all the results that can be imposed from Ward Identities associated to the Global Conformal symmetry. One of the main reasons for this added difficulty, is the fact that there are infinitely many primary operators in a CFT with $D \geq3$. This is a standard fact, whose proof can be seen in \cite{Simmons-duffin2016}.
   
   Thankfully, our discussion regarding the symmetries of Conformal Blocks being the same as the 4-pt function still holds, and we only need to focus on the Conformal Casimir equation, which Dolan and Osborn solved exactly for 4-dimensions as well. Now the CB contain the whole information about the symmetry of the theory, and we need not worry about the Virasoro Algebra and central charge.
   
   This time, the Global Blocks read \cite{Dolan:2000ut,Dolan:2003hv}:
   
   \begin{equation}
   	g_{\Delta,l}^{4D}(u,v) = \frac{z\bar{z}}{z - \bar{z}} k_{\Delta + l}(z) k_{\Delta - l -2}(\bar{z}) - k_{\Delta - l -2}(z)  k_{\Delta + l}(\bar{z}) 
   \end{equation}

   Where we recall that: $	k_{a}(y)= y^{\frac{a}{2}} ~_{2}F_{1}(\frac{a}{2},\frac{a}{2},a,y)$.
   Notice that there is a formal divergence for $z = \bar{z}$, so one should be careful when using symbolic calculus packages to compute derivatives around our favorite point $z=\bar{z}=\frac{1}{2}$. In Particular, \textit{Mathematica}'s \textbf{Series Coefficient} function deals with this problem automatically. 
   
   The advantages of the Bootstrap Method become clear now. Once we have set up our program to compute the Conformal Block derivatives and solve the Linear inequalities in the framework described in \ref{sec: LinearP}, the procedure is essentially automatic. All we have to do is adjust our assumptions on the spectrum.
   This time, the unitarity bound gives us $\Delta \ge \frac{d-2}{2}= 1$ for scalars (this is the dimension of a free scalar) and $\Delta > l+d-2 = l+2 $ for spinning operators. We take the identity operator ($\Delta=0; l=0$) and consider that the scalar primary of lowest dimension $\phi$ does not participate in his OPE to respect parity, and again $\varepsilon$ is the second lowest operator whose dimension we will bound. Once more, we consider every other operator with $\Delta > \Delta_{\varepsilon} $ and $\Delta > l+2$, up to our numerical cutoff.
   
   Applying our optimal bound technique, we obtain the plot in Figure \ref{fig:4DB}:
   
  \begin{figure}[htbp]
  	\centering
  	\includegraphics[width= 9 cm]{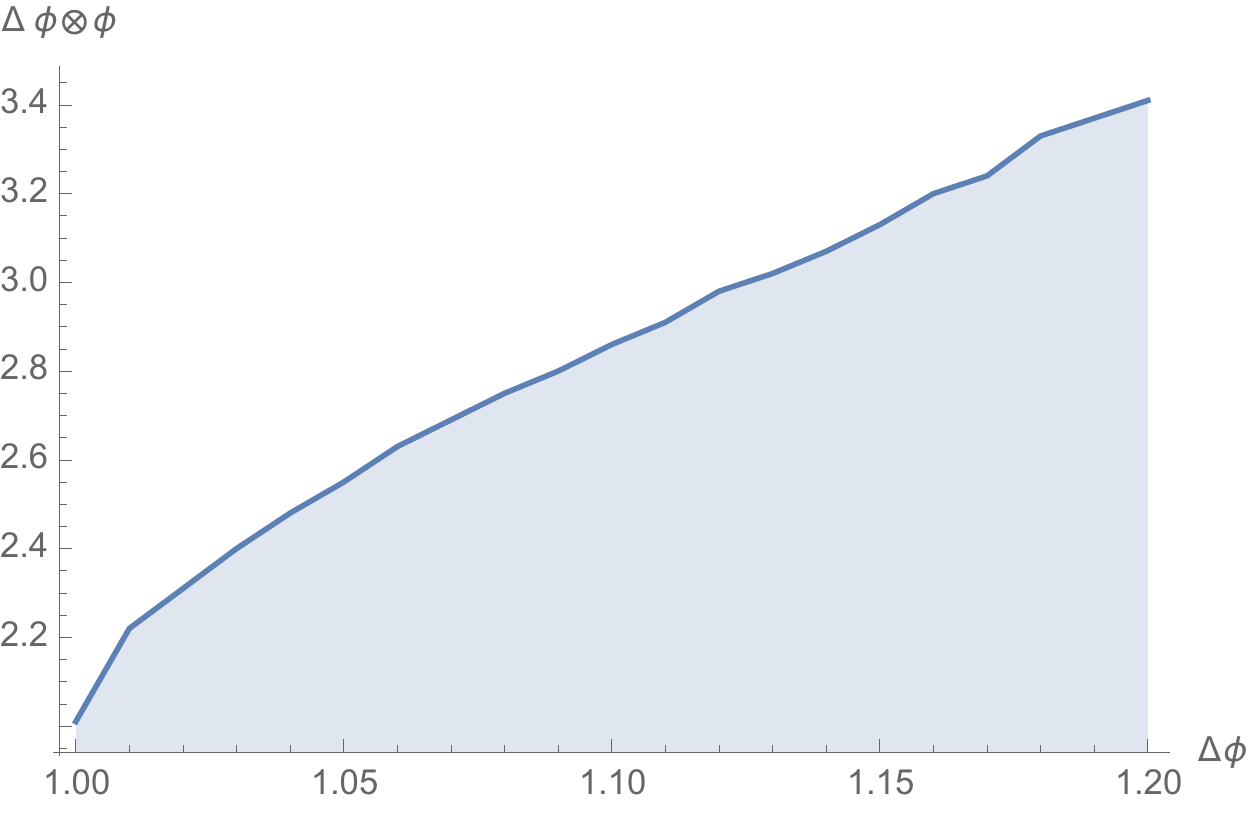}
  	
  	\caption[Operator dimensions in 4D CFT]{Upper bound for the dimension of $\varepsilon$ in terms of $\Delta_{\phi}$ in 4D CFT}
  	\label{fig:4DB}
  \end{figure} 
  
  In this case, the behavior is a lot smoother, and there are no noticeable features that point to the existence of any theories of special interest. At least from the point of view of the Ising Model, this is to be expected, as it becomes a mean field theory in 4D, so its operator dimensions should be trivial, which essentially means they should coincide with a free scalar field and its square, with dimensions $(\Delta_{\phi}, \Delta_{\varepsilon}) = (1,2)$ which corresponds to the origin of \ref{fig:4DB} and is, as one would hope, in the allowed region.
  
  \subsection{Bounding OPE coefficients in 4D CFT }
  So far we have been interested in one part of the CFT data, the values of the Conformal dimensions, or the low-lying spectrum of our theory.
  However, as we discussed previously, to describe a consistent CFT, the compatibility of the OPE is crucial, as we used it to construct the Conformal Blocks. In particular the OPE coefficients, or the 3-pt function coefficients, define which operators contribute to correlators, in the CB expansion, and fix the form of all higher n-point functions.
  
  If we think about the OPE in the schematic language of \ref{fig:Cross} we can think of the OPE coefficient as a "coupling" between the two operators being fused and the one appearing in the OPE, but of course, $f_{ijk}$ merely measures the correlation between $\mathcal{O}_{i},\mathcal{O}_{j},\mathcal{O}_{k}$, and to some extent how important a particular operator is in the OPE of the other two. Note that, as we argued before, the OPE is convergent within some radius, so large dimension operators can't have arbitrarily high OPE coefficients, as this would spoil OPE convergence.
  
  Having reminded ourselves of the importance of the $f_{ijk}$, let us look at the Bootstrap equation on $\langle \phi \phi \phi \phi \rangle$ with different eyes, with particular attention to OPE coefficients $f_{\phi \phi \tilde{\mathcal{O}}} \equiv \lambda_{\tilde{\mathcal{O}}}$.
  Our crossing symmetry equation \ref{eqn: 2.9} does not privilege any operator, but since we are interested in the relationship of the external $\phi$ with $\tilde{\mathcal{O}}$, let us write it isolated \cite{Caracciolo2010}:
  
  	\begin{equation}
  	\lambda_{\tilde{\mathcal{O}}}^{2}\alpha(F_{\tilde{\Delta},\tilde{l}})= -\alpha(F_{I}) - \sum_{\mathcal{O} \neq \tilde{\mathcal{O}}} \lambda_{\mathcal{O}}^{2} \alpha(F_{\Delta,l})
  	\end{equation}
  	
  Also, notice that we separated the contribution of the identity operator and of course, $f_{\phi\phi I}$=1, by normalization of 2-pt functions. Now, we want to isolate our OPE coefficient, so we make a different constraint than the ones we usually make. We force:
  
  \begin{equation}
  	\alpha(F_{\tilde{\Delta},\tilde{l}})=1
  \end{equation}
  
  Which means we can read-off the OPE coefficient:
  
  	\begin{equation}
  	\lambda_{\tilde{\mathcal{O}}}^{2}= -\alpha(F_{I}) - \sum_{\mathcal{O} \neq \tilde{\mathcal{O}}} \lambda_{\mathcal{O}}^{2} \alpha(F_{\Delta,l})
  	\end{equation}
  	
 Now, of course most of the the quantities on the RHS are unknown, but if we make the usual constraints: $ \alpha(F_{\Delta,l})\ge 0$, we can obtain an upper bound for the OPE coefficient:
 \begin{equation}
 	 \lambda_{\tilde{\mathcal{O}}}^{2}= -\alpha(F_{I}) - \sum_{\mathcal{O} \neq \tilde{\mathcal{O}}}
 	 \underbrace{ \lambda_{\mathcal{O}}^{2} \alpha(F_{\Delta,l})}_{ \ge 0} \le - \alpha(F_{I})
 \end{equation}	
 That means, once again by considering every possible operator participating in the OPE of $\phi \times \phi $, and demanding positivity, we get the bound: 
 
 \begin{equation}
 	\lambda_{\tilde{\mathcal{O}}}^{2} \le - \alpha(F_{I}) 
 \end{equation}
 
 Now because we have an inequality, and we became familiar with the linear programming techniques, we have an obvious way to make our bound as good as possible: minimize the quantity $ - \alpha(F_{I}) $. This is essentially using the fact that we can choose any $\alpha$ and apply it to the crossing equation, and it will still be true, so we take the one that gives us the best possible result. Notice, that unlike when we bounded operator dimensions, there is no iterative procedure to find the optimal bound, by construction, the bound we obtain is optimal.
 
\begin{figure}[htbp]
   	\centering
   	\includegraphics[width=11 cm]{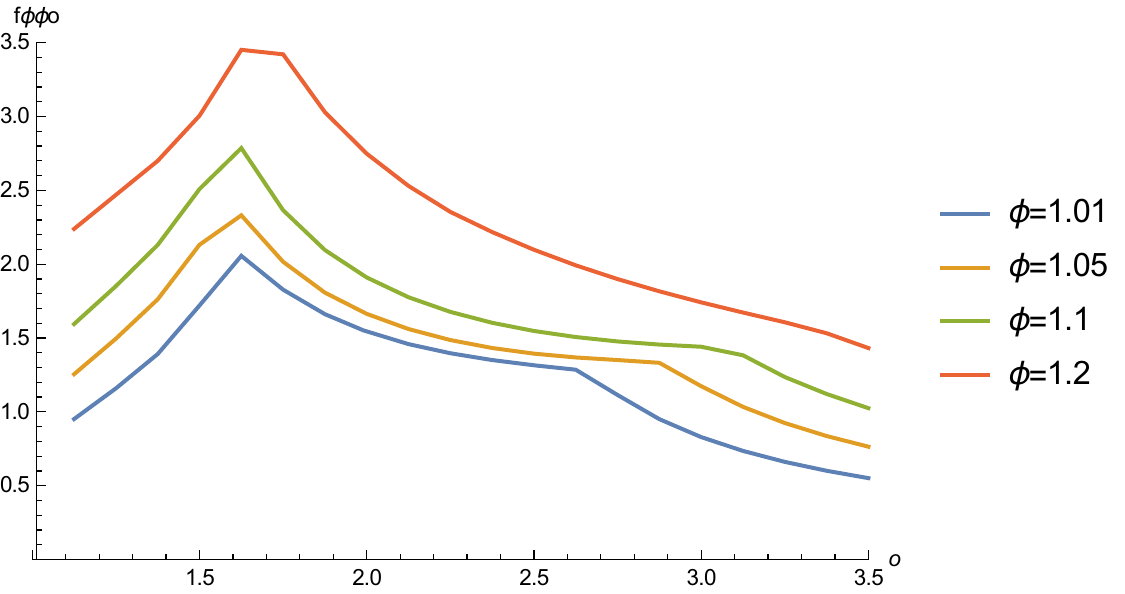}
   	
   	\caption[Bounds on OPE coefficients in 4D CFT]{Upper bound for the OPE coefficient $f_{\phi \phi \mathcal{O}}$, as we change the dimension of $\mathcal{O}$ and choose a few values for $\Delta_{\phi}$}
   	\label{fig:OPE4D}
   \end{figure}

 Now, it is easy to implement this procedure for various choices of $\mathcal{O}$ and $\phi$ as seen in Figure \ref{fig:OPE4D}.

   Note that our physical intuition is checked, as all OPE coefficients become smaller as we increase the dimension of the intermediate/exchanged operator, to guarantee OPE convergence. It is also noticeable that the upper value on OPE coefficients tends to increase with the external dimension $\Delta_{\phi}$ when we keep $\Delta_{\mathcal{O}}$ fixed, this makes sense, because we know that the bound on $\Delta_{\varepsilon}$ always grows with $\Delta_{\phi}$ so, low lying operators have the tendency to appear later, and thus need a larger OPE coefficient, to balance the infinite tower of high spin and dimension (which should exist even for smaller $\Delta_{\phi})$.
   
   There is also a very clear peak, slightly to the left of $\Delta_{\mathcal{O}}$, which intuitively should correspond to the first operator in the OPE, that is $\varepsilon = \phi^{2}$, however it corresponds to values smaller than those calculated from the dimension bound. This is not inconsistent, it merely reflects the fact that when we minimize the OPE coefficient, we cannot assume the next operator to be exactly at the extremal value, so we have extra constraints which arise from all possible operators starting at $\Delta_{\phi}$.

\chapter{Beyond bounds: more restrictive Bootstrap}
\label{Beyond}

\section{The extremal functional method}

So far, the Bootstrap approach has given us quite a bit of information about the structure of CFTs. However, it does not seem to be restrictive enough to actually solve a given theory.

By smartly refining our previous methods, we will show that, for certain extremal situations \cite{El-Showk:2016mxr,ElShowk:2012hu}, the Conformal Bootstrap can actually fix the dimensions of several operators, even for higher spin, and subsequently fix the OPE coefficients $f_{\phi\phi \mathcal{O}}$ of the 2 external operator, with the operator we choose.

\subsection{A sharper look at the optimal bound}
Recalling the procedure \ref{optimal}, we refine our description of the crossing equation as a sum of vectors in \ref{summingvectors}:
The white region in our exclusion plot corresponds to forbidden theories, that is, choices of spectrum that cannot satisfy crossing $  \sum_{\Delta,l}p_{\Delta,l}  \overrightarrow{F_{\Delta,l}^{\Delta_{\phi}}} = 0$, again, this means that there is no choice of OPE coefficients that adds the F-vectors to zero, or, more geometrically, that our set of vectors are on a single half-space. In general, let us consider this picture, when the theory is strongly excluded (or the linear programming problem is solved), when the theory is allowed (no solution for LPP), and in the optimal case where  decreasing $\Delta_{gap}$ by our minimum interval $\delta\Delta$, switches the LPP from solvable to unsolvable. Schematically we represent the case of a two dimensional subspace in Figure \ref{fig:vectors}:
 \begin{figure}[htbp]
 	\centering
 	\includegraphics[width= 11 cm]{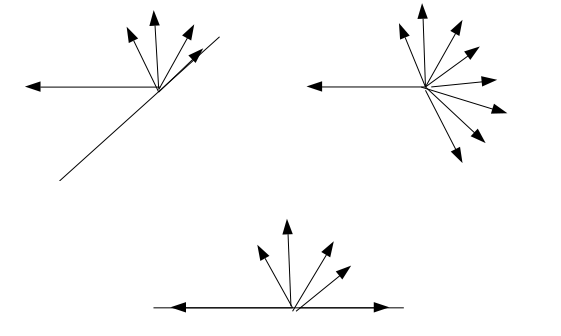}
 	
 	\caption[2d vectors for hypothetical spectrums ]{Sketch of 2d vectors $F_{\Delta,l }$ for a forbidden theory (top left), an allowed one (top right) and the extremal/optimal case (bottom)}
 	\label{fig:vectors}
 \end{figure} 
 
 For the forbidden case (top left), it is clear that they cannot be added to zero, because for any given vector, its negative is not part of the spectrum, or dually, we have a straight line that separates all the vectors on a half-space. For the allowed CFT, we can find no such line, or, in the primal formulation, we can think of positive coefficients that make the sum zero.
 The limiting, extremal case is quite interesting, as we can find a separating line, and of course the vectors above it can never participate as there are no vectors below the line to cancel them. However, it is possible that vectors lying on the line itself exist, and satisfy crossing!

 This is suggesting that the vectors that have a zero dot product with the vector $\mathbf{\Lambda}$ orthogonal to the line, actually exist and define the spectrum of theory. In this case, our spectrum naturally becomes sparse, as we expect from many well-known CFTs. This is a consequence of the fact that most of the vectors included in the sum are ruled out, but a few special ones lying on the $\Lambda$ plane, remain. To some degree, our CFT gets truncated, and a few operators are chosen out, defining a unique solution to crossing (as we said before, it is impossible to add any other vector that is not on the plane, as it cannot be canceled, by extremality, so this proves uniqueness). A unique solution to crossing is the Bootstrap way of saying "a unique CFT"!
 
 From the point of view of the linear functionals $\alpha$, this means that the zeros of the functional are the operators existing in the spectrum, so we can now think of plotting the following functions:
 \begin{equation}
 	\sigma_{l}(\Delta) \equiv \alpha(F_{l}(\Delta))
 \end{equation}
 where we think of $\Delta$ as a variable and not a parameter or lable.
 Therefore, for $\Delta_{\varepsilon} = \Delta^{*} $, that is, for an extremal spectrum, we have that:
 \begin{equation}
 		\sigma_{l_\mathcal{O}}(\Delta_{\mathcal{O}}) = 0
 \end{equation}
 Means that operator $\mathcal{O}$ is part of the spectrum.
 
 Of course, this procedure is approximate, for many reasons, namely the fact that the optimal bound is obtained numerically, with a precision that depends on the discretization, and more seriously on the number of derivatives taken, and the fact that the zeros of the functional suffer from the same dependencies.
 It has been checked however, that as we increase the number of derivatives the values obtained for the spectrum tend to converge rather quickly, specially for low-lying operators. This also means that as we increase the strength of our constraints by computing many derivatives, we will obtain more and more operators \cite{ElShowk:2012hu}. 
 
 Once we have a small number of operators composing our spectrum, we can still use the crossing equation for even more information. Typically, the number of operators will be smaller than the number of equations, which leaves the system undetermined under normal conditions. However, by restricting the Bootstrap equation to a subspace of dimension equal to the number of zeros found, we can solve it, and obtain an estimate for the OPE coefficients.
 Indeed, we have, for $k$ zeros of the function 	$\sigma_{l_\mathcal{O}}(\Delta_{\mathcal{O}})$, a truncated expression for the crossing symmetry equation: \begin{equation}
 \begin{pmatrix}
 \partial_{z}^{1} \partial_{\bar{z}}^{0}F_{\Delta_{1},l_{1}}(z,\bar{z})|_{z=\bar{z}=\frac{1}{2}} & \dots &\partial_{z}^{1} \partial_{\bar{z}}^{0}F_{\Delta_{k},l_{k}(z,\bar{z}})|_{z=\bar{z}=\frac{1}{2}}  \\
 \vdots & \ddots & \vdots \\ 
 \partial_{z}^{m} \partial_{\bar{z}}^{n}F_{\Delta_{1},l_{1}}(z,\bar{z})|_{z=\bar{z}=\frac{1}{2}} & \dots &\partial_{z}^{m} \partial_{\bar{z}}^{n}F_{\Delta_{k},l_{k}}(z,\bar{z})|_{z=\bar{z}=\frac{1}{2}} 
 \end{pmatrix}
 \cdot
 \begin{pmatrix}
 f_{\phi\phi \mathcal{O}_{1}}  \\
 \vdots \\
 f_{\phi\phi \mathcal{O}_{k}} 
 \end{pmatrix}
 = -  \begin{pmatrix}
  \partial_{z}^{1} \partial_{\bar{z}}^{0}F_{I}(z,\bar{z})|_{z=\bar{z}=\frac{1}{2}}  \\
 \vdots \\
 \partial_{z}^{m} \partial_{\bar{z}}^{n}F_{I}(z,\bar{z})|_{z=\bar{z}=\frac{1}{2}}  
 \end{pmatrix}
 \end{equation}

  where we have taken the pair $(m,n)$ to be the one such that we have exactly $k$ equations in total, and the RHS is the contribution of the identity, whose OPE coefficients we previously argued to be known and equal to 1.
  Of course, it is not clear that we get the same result regardless of the set of equations we choose. A natural choice, however is to pick the ones associated to lower dimensional derivatives, and increase their number, as we increase the number of operators. The convergence of the OPE coefficients obtained via this method is also discussed in detail in \cite{ElShowk:2012hu}.
  
  \subsection{Using the EFM to solve the 2D Ising Model}
  For obvious reasons, we will apply the previously described method to a theory which we will prove to be everyone's favorite 2D system, the critical Ising model. We will make the following two assumptions:
  \begin{enumerate}
  	\item{The theory is extremal}
  	\item{The theory has $\Delta_{\phi}= \frac{1}{8}$}
  \end{enumerate}
  
  From this, we will check, using the previous sections arguments, that the only solution to the crossing equation under these assumptions is the 2D Ising model.
  Recall that from the previous chapter, we had obtained $\Delta^* = 1.004$ for $\Delta_{\phi} = 0.125$.
   \begin{figure}[htbp]
   	\centering
   	\includegraphics[width= 9 cm]{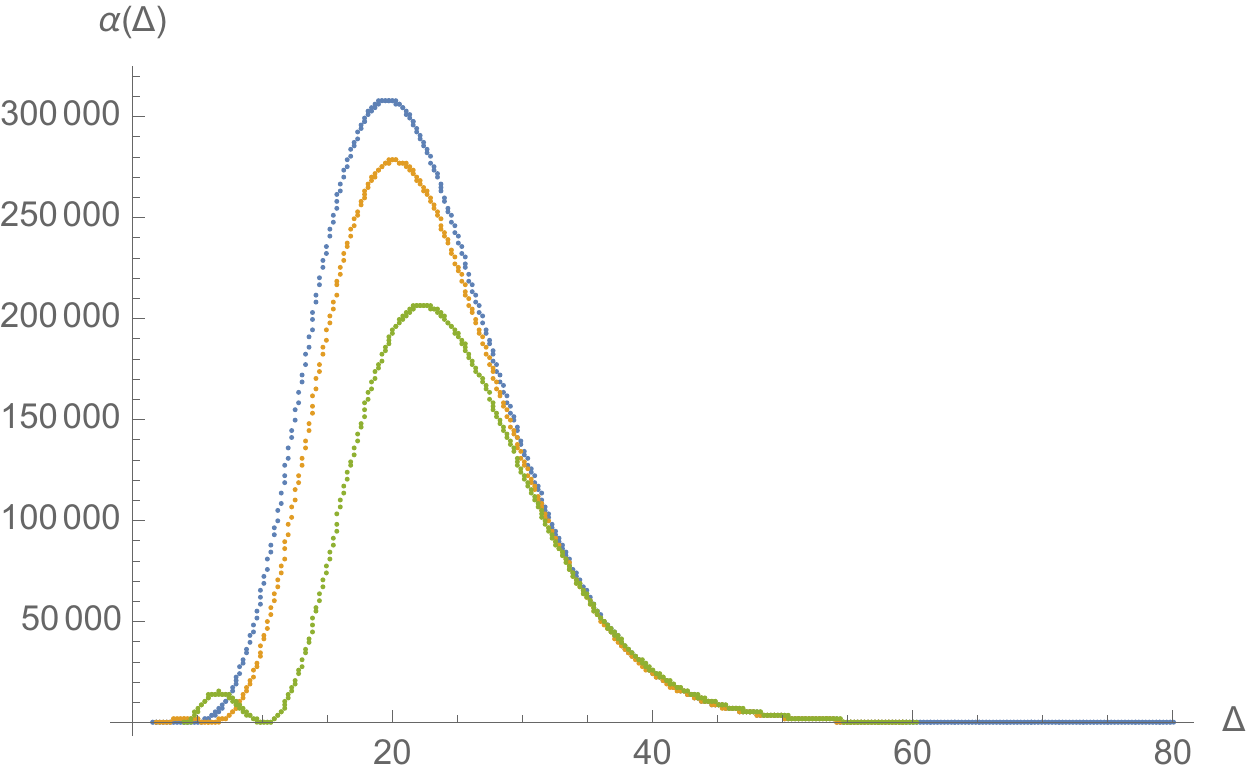}
   	
   	\caption[Plot of the functional $\alpha(F(\Delta))$ in the non-extremal case]{Plot of functional $\alpha(F(\Delta))$ for several spins($l=0$ blue, $l=2$ yellow, $l=4$ green). Some dips are noticeable, but all the values are strictly positive (in fact strictly greater than $1$ as we demanded in the LPP) }
   	\label{fig:functional}
   \end{figure}
  
  First of all, just for concreteness, let us plot the functional for when the gap is comfortably above the critical values, and check that it is indeed positive definite. Taking $\Delta_{gap} = 1.5$ and making the extra strict requirement $\alpha(F_{l}(\Delta)) \ge 1$ , we are able to solve the LPP and plot $ \alpha(F_{l}(\Delta))$ for the first values of $l$ in figure \ref{fig:functional}.

\begin{figure}[htbp]
     	\centering
     	\includegraphics[width= 9 cm]{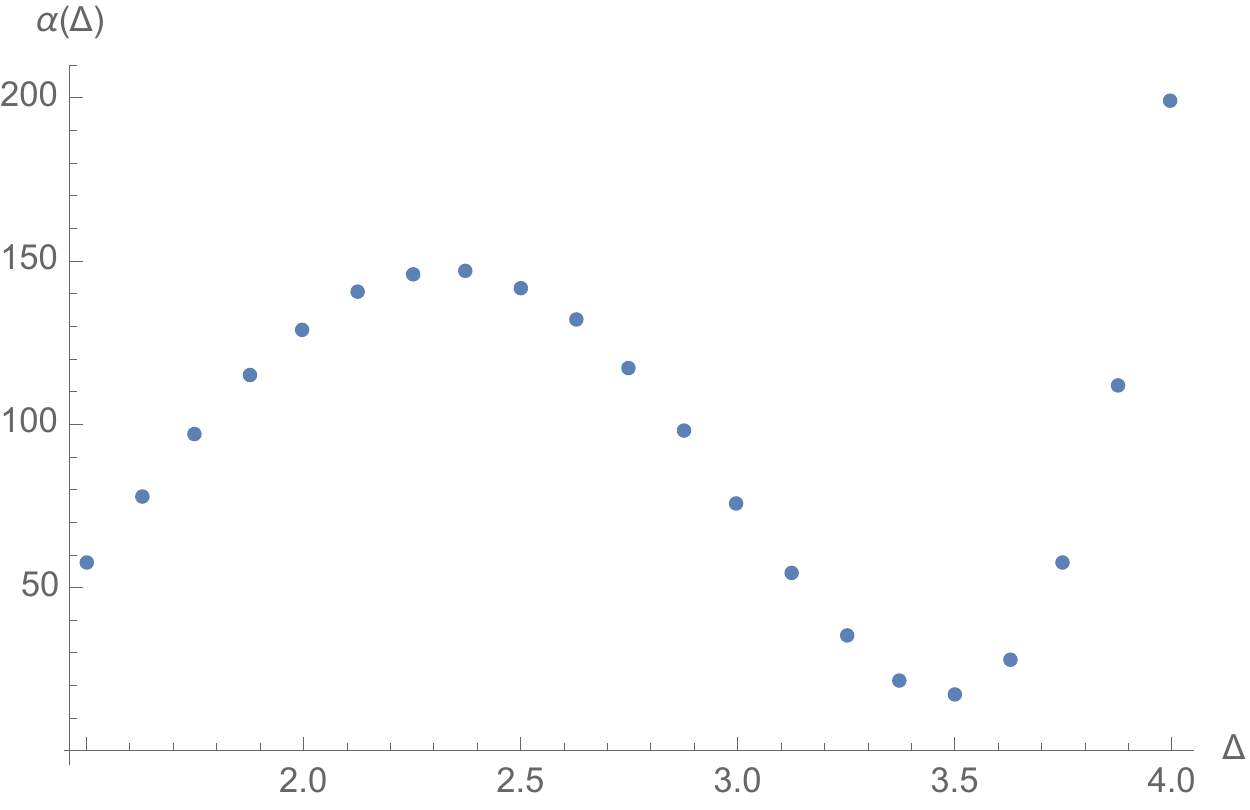}
     	
     	\caption[A dip in $\alpha(F_{l=0}(\Delta))$ in the non-extremal case]{A dip in the scalar non-extremal functional, close to $\Delta = 4$. Later on this will correspond to a zero of the extremal functional. }
     	\label{fig:funcdip}
     \end{figure}

\begin{figure}[htbp]
     	\centering
     	\includegraphics[width= 9 cm]{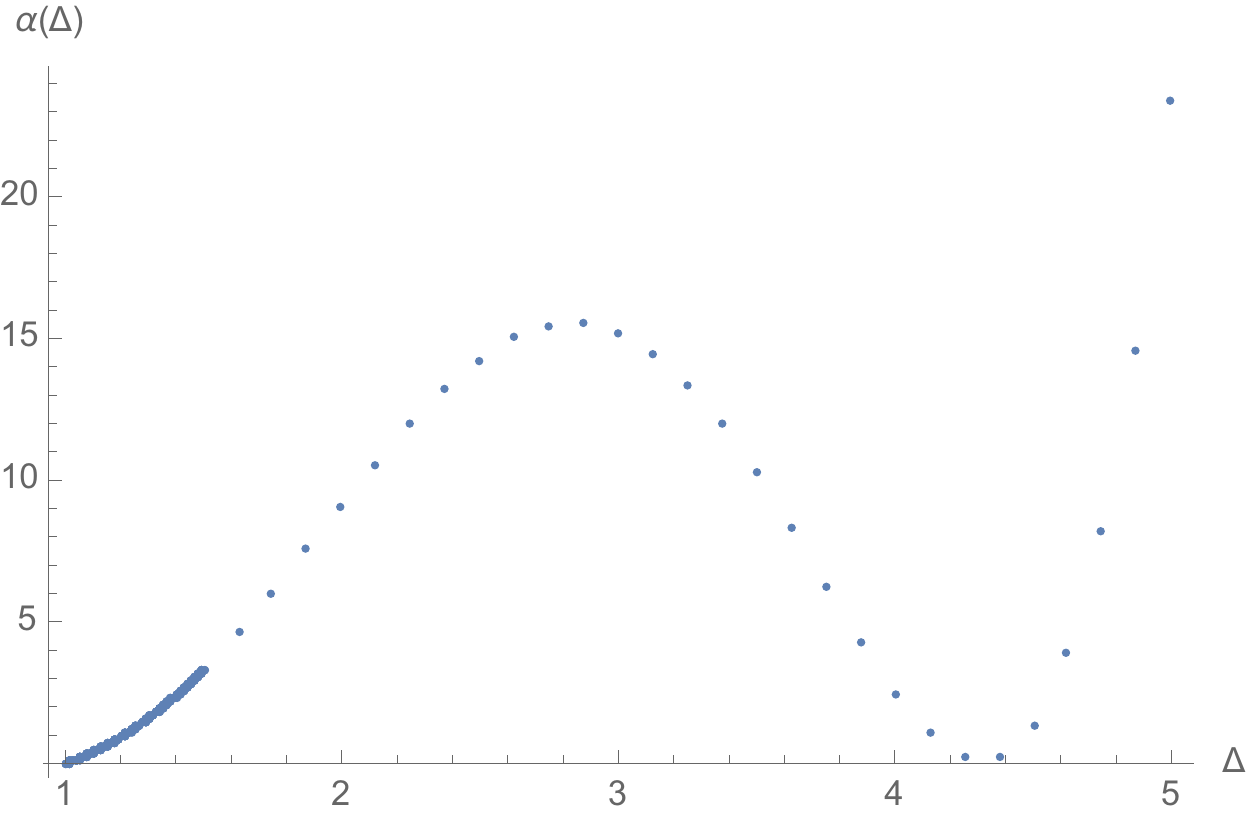}
     	
     	\caption[Zeroes for the scalar Extremal Functional]{The zeroes of the scalar extremal functional, at $\Delta_{1,0} =1.004$ and $\Delta_{2,0} = 4.3$ }
     	\label{fig:extrsca}
     \end{figure}

    From the previous figure, our conclusions that we had actually excluded this value for $\Delta_{gap}$ become evident, as we see the strictly positive value for $\alpha(F(\Delta)) = \Lambda \cdot F(\Delta )$.
    Interestingly, we noticed that for the scalar sector, there is a really low dip (see figure \ref{fig:funcdip}), which, however, stays strictly positive. This is already suggestive that when we approach the extremal gap, this dip will become one of the aforementioned zeros, which are the physical operators (We denote them by $\Delta_{k,l}$, with  $k$ an integer index).
    Now, we can use this idea to probe the extremal case, and check if our discussion of the existence of null vectors actually works. For example, (and from now on setting $\Delta_{gap}= \Delta^{*}= 1.004$), noticing the value for the scalar sector, we have  Figure \ref{fig:extrsca}.
    
     It is clear that our scalar Operator at $\Delta^{*}$ is annihilated by the linear functional and is indeed the second lowest dimension scalar of the theory. As predicted before, we get a second non-trivial operator, with dimension $\Delta \approx 4.3$ 
     
\begin{figure}[htbp]
     	\centering
     	\includegraphics[width= 9 cm]{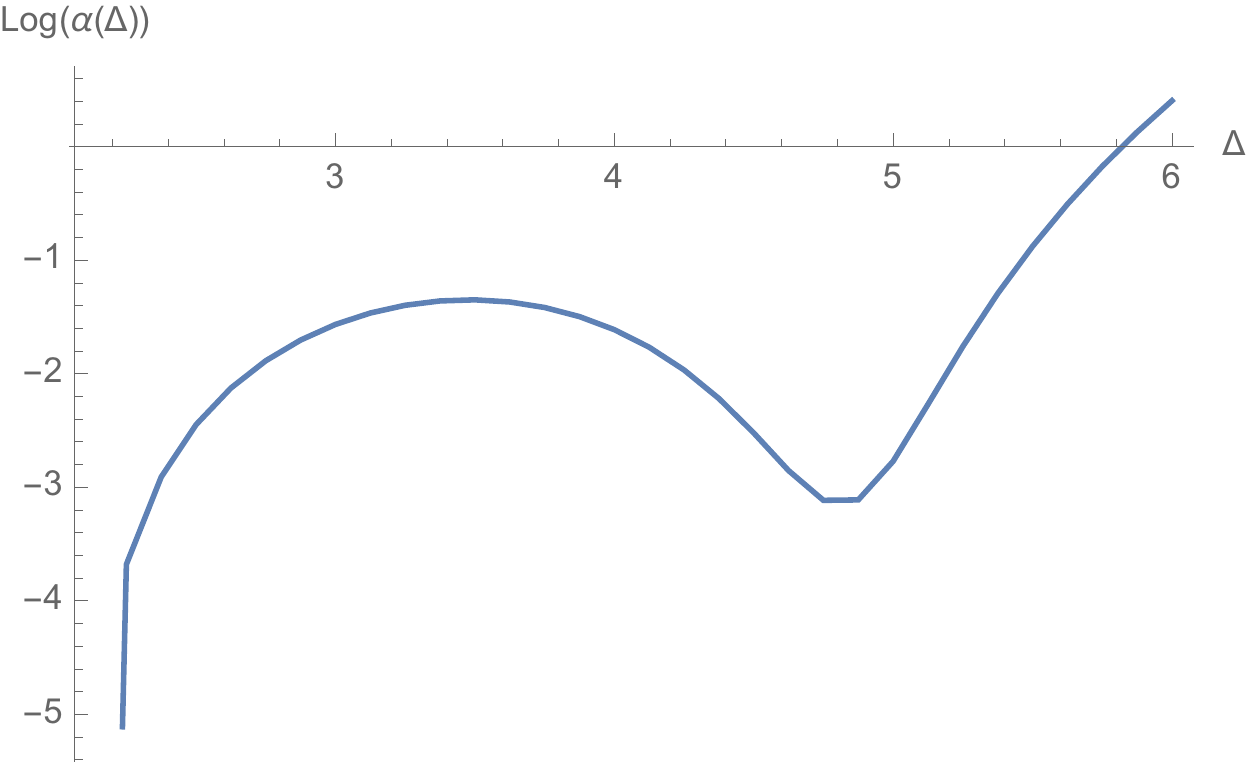}
     	
     	\caption[Zeroes for the spin 2 Extremal Functional]{The zeroes of the scalar extremal functional, at $\Delta_{1,2} =2.0$ and $\Delta_{2,2} = 4.8$ }
     	\label{fig:extr2}
     \end{figure}

     Now, we can follow this procedure to the higher spin sector, and find non-trivial spinning operators. Of course we expect the existence of the Stress-Energy Tensor, as this is a fundamental entity of any CFT (in 2 dimensions $\Delta_{T}=2$, $l_{T}=2$ ). Indeed we get for spin 2, the log-plot (as is more common in the EFM literature) of Fig. \ref{fig:extr2}.
     As expected, $\sigma_{2}(\Delta)$ has a distinct zero at $\Delta=2$, corresponding to $T_{\mu\nu}$ and there is also a (softer numerically, should improve with more derivatives) zero at $\Delta \approx 4.8$.
     
     Finally, for completeness, we include the spin 4 sector in Figure \ref{fig:extr4}: 
     \begin{figure}[htbp]
     	\centering
     	\includegraphics[width= 9 cm]{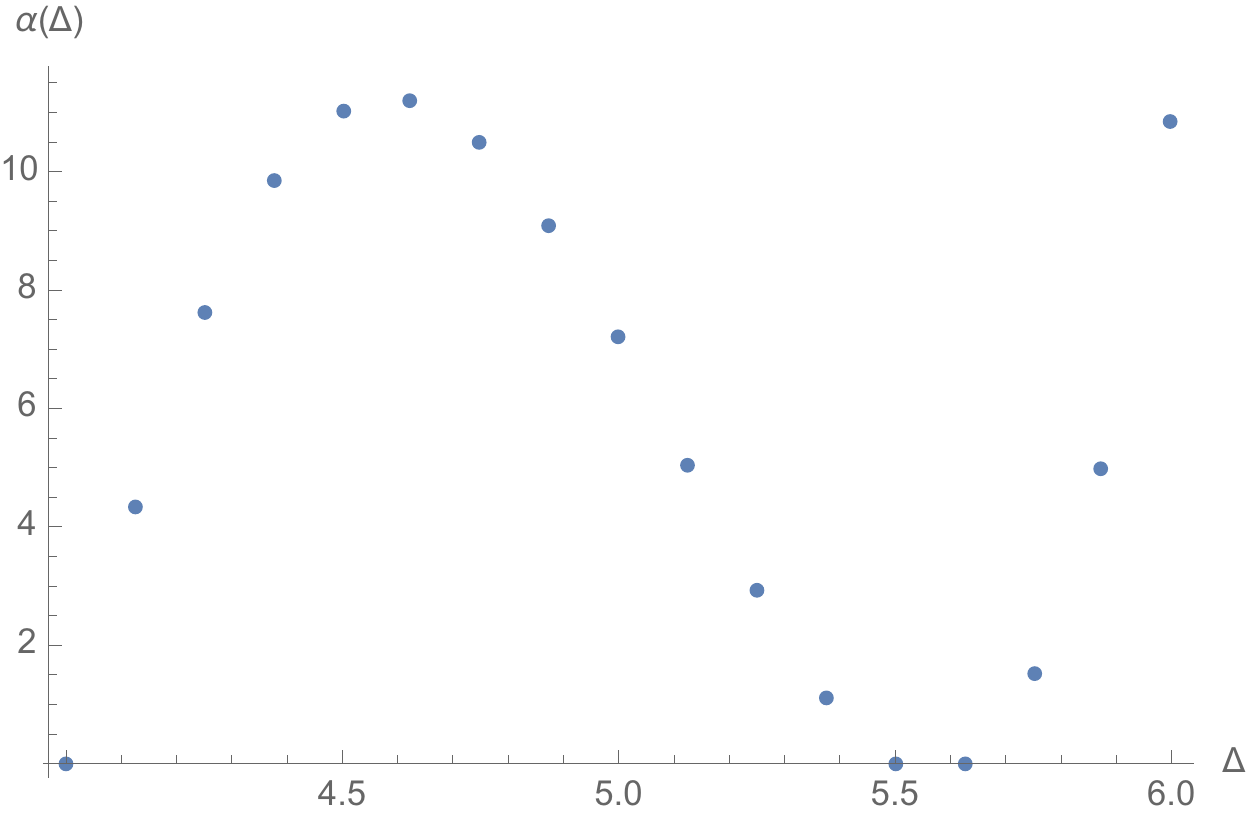}
     	
     	\caption[Zeroes for the spin 4 Extremal Functional]{The zeroes of the scalar extremal functional, at $\Delta_{1,4} =4.0$ and $\Delta_{2,4} = 5.55$ }
     	\label{fig:extr4}
     \end{figure}
     
     Which continues the trend of including an operator that satisfies the unitarity bound ($\Delta = l+ d-2 = l$), in this case $\Delta =4$, and a non-trivial spin 4 operator with $\Delta = 5.55$.
     If we go to higher spins, we stop having non-trivial zeros, but at spin 6, we still have the unitarity bound operator ($\Delta=6,l=6$).
     This gives us a set of seven zeros of the functional, or seven physical operators, which we summarize on a table, and compare to the known values from the exact solution of the critical 2D Ising model ($\Delta_{th}$)
     
    \begin{center}\vspace{1cm}
    	\begin{tabular}{l l l l l l}
    		\toprule
    		~ & $l=0$ & $l=2$ & $l=4$ & $l=6$ \\
    		\midrule
    		$\Delta_{1,l}$ & 1.004 &2. &4. &6. \\
    		$\Delta_{th}$ & 1 &2 &4 &6 \\
    		$\Delta_{2,l}$ & 4.3 & 4.8 &5.55 &~ \\
    		$\Delta_{th}$ & 4 &6 &5 &~ \\
    		
    		\bottomrule
    	\end{tabular}
    	\captionof{table}{Operator dimensions obtained for $\Delta_{\phi}=0.125$ via EFM}
    \end{center}\vspace{1cm}
    
    At this number of derivatives $N=10$ which is incredibly small compared to the specialized literature, we still get results which are correct within $20\%$, in the worst possible case. As mentioned earlier, increasing the number of derivatives improves these values, but we still get a qualitative check of the structure of the spectrum!
    Now, as discussed previously, we can take these seven approximate operators, and get an estimate for the OPE coefficients, by truncating the Bootstrap equation, only summing the Conformal Blocks of the 7 operators we determined, and taking the contribution of the identity as the non-homogeneous part (The OPE coefficients for $I$ are already fixed to one).
    Therefore, we have a set of 7 unknowns, but 10 linear equations, so we truncate them and choose only the first 7, as we discussed.
    This gives the Equations:
    \begin{align}
	\nonumber  & f_{\phi \phi \mathcal{O}_{1,0}}^{2} \partial_{z}^{1}\partial_{\bar{z}}^{0}F_{1.004,0}(z,\bar{z})|_{z=\bar{z}=\frac{1}{2}}  +  \dots + f_{\phi \phi \mathcal{O}_{1,6}}^{2} \partial_{z}^{1}\partial_{\bar{z}}^{0}F_{6,6}(z,\bar{z})|_{z=\bar{z}=\frac{1}{2}}  = -\partial_{z}^{1} \partial_{\bar{z}}^{0}F_{I}(z,\bar{z})|_{z=\bar{z}=\frac{1}{2}}  \\ 
	 \nonumber &\dots\\ 
	 & 	f_{\phi \phi \mathcal{O}_{1,0}}^{2} \partial_{z}^{5}\partial_{\bar{z}}^{0}F_{1.004,0}(z,\bar{z})|_{z=\bar{z}=\frac{1}{2}}  +  \dots + f_{\phi \phi \mathcal{O}_{1,6}}^{2} \partial_{z}^{5}\partial_{\bar{z}}^{0}F_{6,6}(z,\bar{z})|_{z=\bar{z}=\frac{1}{2}}  = -\partial_{z}^{5} \partial_{\bar{z}}^{0}F_{I}(z,\bar{z})|_{z=\bar{z}=\frac{1}{2}} 
    \end{align}

    Where $\partial_{z}^{5}\partial_{\bar{z}}^{0}$ is the last of the seven derivatives we chose. This is now a set of 7 linear equation on the 7 unknowns $f_{\phi\phi \mathcal{O}_{i}}^{2}$. Using a linear solver, we find the OPE coefficients, and compare them to the known results (We compare to the OPE coefficients of the actual operator whose dimensions are slightly off from the ones we obtained).
    
     \begin{center}\vspace{1cm}
     	\begin{tabular}{l l l l l l l l l l l}
     		\toprule
     		$f_{\phi\phi(\Delta,l)}$ & $(1.004,0)$ & $(4.3,0)$ & $(2,2)$ & $(4.8,2)$ & $(4,4)$ & $(5.55,4)$ & $(6,6)$\\
     		\midrule
     		Obtained & 0.499 & 0.0147 & 0.178 & 0. &0.019 &0.0056 & 0.0057 \\
	     	Exact Solution & 0.5 & 0.0156 & 0.177 & 0.0026 &0.021 &0.0055 & 0.0037\\
      		\bottomrule
     	\end{tabular}
     	\captionof{table}{OPE coefficients for the 2D Ising model solved with the approximate operator dimensions}
     \end{center}\vspace{1cm}
     
     It is remarkable that even though we already had errors of the order of $10\%$ in the operator dimensions, we still get quantitatively useful results, namely for those operators with smaller errors in the determination of $\Delta$ ($\varepsilon$, $T_{\mu\nu}$, $\dots$). This seems to be because of the convergence of the OPE (forgetting about operators with large dimensions does not change the balance of the crossing equation too much), and because of the fact that we have a sparse spectrum. 
     Only a few operators exist, and they must balance each other on the crossing equation, which means that small nudges on the dimension gives small nudges to the OPE coefficients, in such a way that the dominant operators for the convergence, remain dominant.

     It also neat to check that the central charge, $c = \frac{\Delta_{\phi}^{2}}{f_{\phi \phi T}^{2}} \approx 0.493 $  is rather close to the value for the Ising model $c=\frac{1}{2}$. The central charge determines the Virasoro algebra for the 2D CFT in question, but in this case, we obtained information about it using only the global part of the Conformal Algebra.
     This justifies our initial statement that taking $\Delta_{\phi} = \frac{1}{8}$ and looking for the extremal theory defines the 2D Critical Ising model! (We just defined a whole theory from kinematical and consistency statements, including extremality, using the single dynamical input $\Delta_{\phi} = \frac{1}{8}$)
     
     \section{The method of Determinants}
     After our approach with the EFM, we have used the Bootstrap to actually solve a CFT. However, our procedure was dependent on the fact that it was on the edge of the allowed region, or what we called an Extremal CFT. This is interesting, but limited, because many CFTs may not be extremal, and also unintuitive, because we obtained a solution to a theory we were not necessarily thinking about, and there was no physical understanding of the condition of extremality (generality is great, but it would also be interesting to get results about theories we are particularly interested in).
     
     In the second part of this Chapter, we present a method, advocated by Gliozzi \cite{Gliozzi:2013ysa,Gliozzi:2014jsa}, in which one obtains an approximate spectrum and OPE coefficients, but for a theory of which we are generally aware of its fusion properties. By this, we mean that we know the structure of the OPE:
     \begin{equation}
     	\phi \times \phi \sim \sum_{i}N_{i} \mathcal{O}_{\Delta_{i},l_{i}}
     \end{equation}
  Where we label the operators $\mathcal{O}$ being exchanged by their quantum numbers ($\Delta_{i},l_{i}$), and the number $N_{i}$ accounts for possible degeneracy of these quantum numbers. No knowledge of OPE coefficients is implicit (only the fact that these are the only non-zero ones).
  
  \subsection{Truncating the OPE and the Rouché-Capelli Theorem}
  Having chosen a theory, and knowing its fusion rule, we get to one of the main assumptions of this method: Because the OPE converges, truncating it after a finite number of operators is a reasonable approximation, which gets better the more operators we keep. This distinguishes the present idea from our previous methods where we made very basic assumptions about the spectrum, and included every possible operator in the crossing sum rule.
  As in all of our previous Bootstrap methods, however, we think of the crossing equation as being turned into a finite number of equations, by considering derivatives around the crossing symmetric point. First, we write the crossing equation in a more convenient way (Isolate and Divide by the Identity contribution):
  \begin{equation}
  	 \sum_{\mathcal{O}\ne I} f_{\phi\phi\mathcal{O}}^{2}\left( \frac{ v^{\Delta_{\phi}}g_{\Delta,l}(u,v) - u^{\Delta_{\phi}} g_{\Delta,l}(v,u)}{u^{\Delta_{\phi}}-v^{\Delta_{\phi}}}\right)= 1 
  \end{equation}
  
  Then, as we have done many times now, we take $N$ derivatives around the self-dual point (in this case we made the change of variables $z \rightarrow \frac{a +\sqrt{b}}{2}~,~ \bar{z} \rightarrow \frac{a -\sqrt{b}}{2}$ and take derivatives $\partial_{a}^{m}\partial_{b}^{n}|_{a=1,b=0}$ ), and get $N+1$ equations ($H$ was defined in \ref{eq:H}):
  \begin{align}
  & \sum_{\mathcal{O}\ne I} f_{\phi\phi\mathcal{O}}^{2} H_{\Delta,l}^{\Delta_{\phi}}(1,0) = 1 \\
  & \sum_{\mathcal{O}\ne I} f_{\phi\phi\mathcal{O}}^{2} \partial_{a}^{2m}\partial_{b}^{n}H_{\Delta,l}^{\Delta_{\phi}}(a,b)|_{a=1,b=0} = 0
  \end{align}
  The values $m,n \in \mathbb{N}$ are chosen such that we take $N$ derivatives, and they are written $(2m,n)$ because other derivatives vanish in this choice of $a,b$ variables.
  The major simplification is our assumption of truncability of the OPE, which leaves the sum with a finite number (say $M$) of terms:
  \begin{equation}
  	f_{\phi\phi\mathcal{O}_{1}}^{2} \partial_{a}^{2m}\partial_{b}^{n}H_{\Delta_{1},l_{1}}^{\Delta_{\phi}}(a,b)|_{a=1,b=0} + \dots + f_{\phi\phi\mathcal{O}_{M}}^{2} \partial_{a}^{2m}\partial_{b}^{n}H_{\Delta_{M},l_{M}}^{\Delta_{\phi}}(a,b)|_{a=1,b=0} = 0 
  \end{equation}
  
  With this, we have $N$ homogeneous linear equations for the $M$ unknown OPE coefficients ($	f_{\phi\phi\mathcal{O}_{1}}^{2},\dots,f_{\phi\phi\mathcal{O}_{M}}^{2}$), and a single non-homogeneous linear equation:
  \begin{equation}
  	f_{\phi\phi\mathcal{O}_{1}}^{2} H_{\Delta_{1},l_{1}}^{\Delta_{\phi}}(1,0) + \dots + f_{\phi\phi\mathcal{O}_{M}}^{2} H_{\Delta_{M},l_{M}}^{\Delta_{\phi}}(1,0) = 1
  \end{equation}
  
  It is interesting to notice the similarities to the EFM method, where our equations also become truncated, but by extremality, not by our knowledge of the Fusion rule, and our manifest neglecting of Operators appearing after the first $M$ we choose.
  
  Now, the other key ingredient of this method comes in. If we have $N$ linear homogeneous equation on $M$ unknowns, and $N>M$, the following statement (Rouché-Capelli theorem) holds: The system will have a solution if and only if the determinants of all $M \times M$ minors of the matrix of coefficients vanish.
  In particular, in an $N\times M$ matrix there are $ \binom{N}{M}$ determinants to require to vanish. Note that these determinants are a function of $\Delta_{\phi}$ and of up to $M$ dimensions $\Delta_{\mathcal{O}_{i}}$. 
  Then we can write down the fundamental equations of the method of determinants:
  \begin{equation}
  	D_{i}(\Delta_{\phi},\Delta_{\mathcal{O}_{i}}) \equiv \det(  \partial_{a}^{2m_{i}}\partial_{b}^{n_{i}}H_{\Delta_{i},l_{i}}^{\Delta_{\phi}}(a,b)|_{a=1,b=0}) = 0
  \end{equation}
  
  Where $m_{i},n_{i}$ label the matrix elements of the $i$th minor we consider. Now these are implicit (non-linear) equations for the dimensions of the operators, which define hyper-surfaces in $\Delta$-space. At their common intersection we will find a spectrum which is consistent with our $M$-truncation. Having found the solution for our spectrum, we can plug it into the truncated linear equations (including the non-homogeneous one), and solve it for an approximated set of OPE coefficients (even though we have more equations than unknowns, note that by setting the determinants to zero we created linear dependency between the equations, and we get a normalization from the inhomogeneous equation which allows us to completely solve the problem). 
  Therefore, this makes it possible to solve the (approximate) CFT data of a particular theory, just by knowing its Fusion rules (which can be fixed to some degree by symmetry or other physical arguments).
  \subsection{Solving a 4D free scalar model with the method of determinants}
  Our method seems reasonable, but we have had no care at all about the error introduced by truncation. Therefore we do a simple check on everybody's first passion, the free scalar field. This will allows to see how good (or bad) the method actually is, at least when compared to an exactly solved CFT.
  First, we know that the OPE of the fundamental field $\phi$ should be \cite{Gliozzi:2013ysa}:
  \begin{equation}
  	\phi \times \phi \sim I + \varepsilon[\Delta_{\phi^{2}},0] + T_{\mu\nu}[4,2] + [\Delta_{\phi^{2}}+4,4] + ...
  \end{equation}
  Where we denoted operators by their symbol, but also by their quantum numbers $[\Delta,l]$.
  Of course, we know that, in a free scalar theory, dimensional analysis is good enough to calculate operator dimensions: $\Delta_{\phi}=1 , \Delta_{\varepsilon}=2$ but this will \textbf{not} be an assumption. We will see that this follows from the Bootstrap.
  With our choice of truncation, we introduced $M=3$ unknowns to the Bootstrap equations: $f_{\phi\phi\varepsilon}$ , $ f_{\phi\phi T}$ , $f_{\phi\phi ,l=4}$
  Where $l=4$ means the only spin 4 operator included in our Truncated OPE.
  Note however that we only have 2 unknown operator dimensions: $\Delta_{\phi} , \Delta_{\varepsilon}$. For our method to work, we need more than $3$ derivatives. We will use the minimum possible value: 4 derivatives. We take $(2m,n) = (0,1);(2,0);(0,2);(2,1)$, which are the lowest possible order non-vanishing derivatives. We then have a $4\times 3$ matrix of coefficients which gives us $\binom{4}{3} = 4 $ equations for vanishing determinants. We first plot the $4$ equations $D_{i}(\Delta_{\phi},\Delta_{\varepsilon}) = 0$, which are the curves in Figure \ref{fig:Gliozzi}:
   \begin{figure}[htbp]
   	\centering
   	\includegraphics[width= 9 cm]{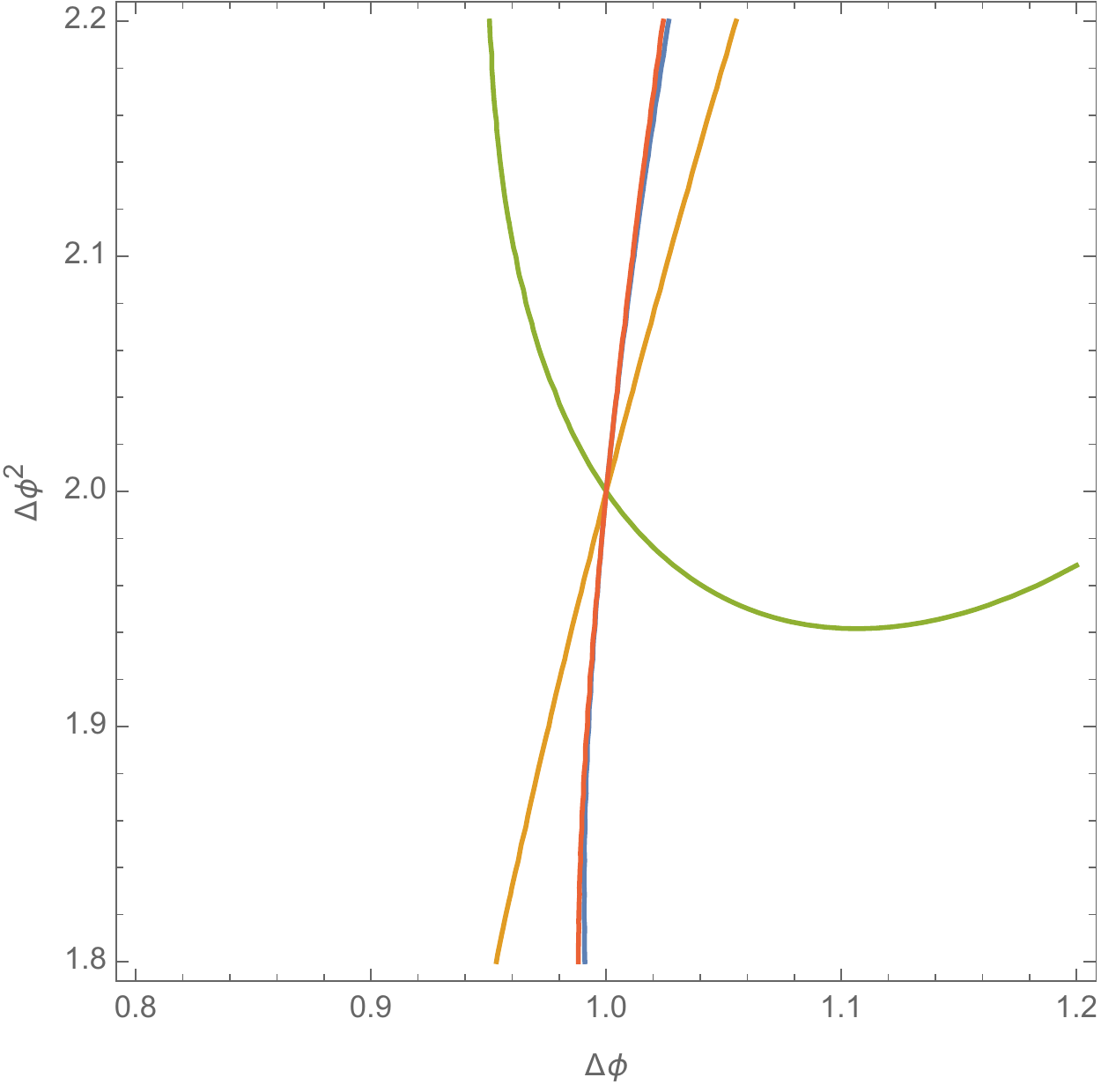}
   	
   	\caption[The vanishing determinant curves for the 4D free scalar]{The $4$ curves corresponding to the vanishing of the determinant for the 4 minors. A single intersection is manifest. }
   	\label{fig:Gliozzi}
   \end{figure}
   
   It is clear that all 4 curves (the blue and red one almost coincide in this region of $\Delta$-space) meet at a single point, close to the expected value ($\Delta_{\phi} =1, \Delta_{\varepsilon} =2 $). Indeed, if we numerically solve each pair of simultaneous equations:
   \begin{align}
   	\nonumber & D_{i}(\Delta_{\phi},\Delta_{\varepsilon})=0 \\
   	& D_{j}(\Delta_{\phi},\Delta_{\varepsilon})=0
   \end{align}
   
   Some of the curves intersect on more than one place, but all of them have the common solution:
   \begin{equation}
   	(\Delta_{\phi},\Delta_{\varepsilon})= (1.,2.)
   \end{equation}
   Where the $1.$ and $2.$ are rounded to 7 decimal places but have minor numerical differences beyond this order depending on the pair of minors chosen. 
   It is remarkable how precise this result is, but the introduction of a spin 4 operator with the dimension given in terms of $\Delta_{\varepsilon}$ is giving a lot of information about the fact that the theory is uncoupled. Any way, for such a small number of operators, the results, in what concerns the spectrum, are remarkable.
   Having fixed the dimensions, we still have a set of 4+1 linear equations which we can use to find OPE coefficients, introducing the newly found values for the dimensions we have:
   \begin{equation}
   	\begin{pmatrix}
   H_{2,0}^{\Delta_{\phi}}(1,0) & H_{4,2}^{\Delta_{\phi}}(1,0) &
   H_{4+2,4}^{\Delta_{\phi}}(1,0)\\
   	\partial_{b}^{1}H_{2,0}^{\Delta_{\phi}}(1,0) & \partial_{b}^{1}H_{4,2}^{\Delta_{\phi}}(1,0) &
   	\partial_{b}^{1}H_{4+2,4}^{\Delta_{\phi}}(1,0) \\
   	\partial_{a}^{2}H_{2,0}^{\Delta_{\phi}}(1,0) & \partial_{a}^{2}H_{4,2}^{\Delta_{\phi}}(1,0) &
   	\partial_{a}^{2}H_{4+2,4}^{\Delta_{\phi}}(1,0)\\
   	\partial_{b}^{2}H_{2,0}^{\Delta_{\phi}}(1,0) & \partial_{b}^{2}H_{4,2}^{\Delta_{\phi}}(1,0) &
   	\partial_{b}^{2}H_{4+2,4}^{\Delta_{\phi}}(1,0)\\
   	\partial_{a}^{2}\partial_{b}^{1}H_{2,0}^{\Delta_{\phi}}(1,0) & \partial_{a}^{2}\partial_{b}^{1}H_{4,2}^{\Delta_{\phi}}(1,0) &
   	\partial_{a}^{2}\partial_{b}^{1}H_{4+2,4}^{\Delta_{\phi}}(1,0)
   	\end{pmatrix} 
   	\cdot
   	\begin{pmatrix}
   	f_{\phi\phi\varepsilon}^{2}\\
   	f_{\phi\phi T}^{2}\\ 
   	f_{\phi\phi,l=4}^{2}\\
   	\end{pmatrix}
   	= \begin{pmatrix}
   	1\\
   	0\\ 
   	0\\
   	0\\
   	0
   	\end{pmatrix}
   \end{equation}
   Where $\partial_{a}^{2m}\partial_{b}^{n}H_{\Delta_{i},l_{i}}^{\Delta_{\phi}}(1,0)$ is shorthand notation for $\partial_{a}^{2m}\partial_{b}^{n}H_{\Delta_{i},l_{i}}^{\Delta_{\phi}}(a,b)|_{a=1,b=0}$. These equation do have a solution because of linear dependence of the bottom 4, which is:
   \begin{equation}
   	f_{\phi\phi\varepsilon}^{2} = 2.196 ~,~ f_{\phi\phi T}^{2} = 0.278 ~,~ f_{\phi\phi,l=4}^{2} = 0.044 
   \end{equation}
   
   These are in the same order of magnitude of the exact values (we compute them in the Appendix \ref{AppendixA}), but have errors as high as 54$\%$. The exact values are $	f_{\phi\phi\varepsilon}^{2} = 2 ~,~ f_{\phi\phi T}^{2} =\frac{1}{3}= 0.333 ~,~ f_{\phi\phi,l=4}^{2}=\frac{1}{35} = 0.029 $.
   It is clear that the truncation is not as miraculous as it seemed when we computed the dimensions. The convergence of the OPE is delicate, and of course, only including 3 operators induces errors to compensate the ones that were suppressed. Even so, it is impressive that from such a large simplification we got a really good picture of the theory underneath our fusion rules, the free scalar.
   
   These results suggest that there is something intrinsically important in the unitarity bound and the dimension for a free scalar, as the dimensions obtained are far more accurate than the OPE coefficients. Of course, increasing the number of operators in the OPE, we get more unknowns, and the OPE coefficients begin to converge, as is claimed by Gliozzi \cite{Gliozzi:2013ysa}, and is also expected from the point of view of the EFM.  
\chapter{A Twist field correlator approach to Modular Invariance}
\label{twist}

So far we have reviewed a series of well established results in the Bootstrap literature, and obtained them with minimal effort from the computational point of view, that is, without resorting to complicated semi-definite solvers or very optimized Linear Programming solvers.

Before concluding this thesis, we attempt an alternative approach to a famous result in CFT, which, as far as we know, has not been tackled in this manner.
\section{The Hellerman Bound}

In his famous paper \cite{Hellerman2011}, Simeon Hellerman obtained a Universal bound for Unitary 2D CFTs that states the following:

\textit{Every Unitary 2D CFT ($c,\tilde{c} >1$) possesses a primary operator of dimension $\Delta_{1}$, such that $0<\Delta_{1}< \frac{c + \tilde{c}}{12} + 0.473695$.}

$c$ and $\tilde{c}$ are known as the left and right moving central charges, which are a very important property of CFTs (We present a summary of the most important results in 2D CFT in Appendix \ref{appendix2d}). For our purposes, we can take $c=\tilde{c}$ and the bound is on $\frac{c}{6}$.

This has a beautiful interpretation from AdS$_{3}$/CFT$_{2}$ dictionary, which establishes that the lowest massive excitation of a 3D theory of gravity with negative cosmological constant $\Lambda$ must satisfy $m< \frac{1}{4G_{N}} + O(\Lambda^{1/2})$.

Originally this bound was obtained by performing Bootstrap techniques known as Modular Bootstrap \cite{Collier2016}, which take advantage of the so-called Modular invariance of CFTs on a torus. By this we mean CFTs with a compactified spatial dimension and imaginary time, which we can think of as turning on a temperature $\frac{1}{\beta}$. By thinking of taking the path integral in slices of constant spacial coordinate or constant temperature and requiring them to match, the modular invariance equation is obtained:
\begin{equation}
	Z(\beta) = Z(\frac{4\pi^{2}}{\beta})
\end{equation}

What we called modular Bootstrap, then, consists of assuming a spectrum and choosing linear functionals which could be of the form:
\begin{equation}
	\alpha = \sum_{n=1}^{N} (\beta\partial_{\beta})^{n}|_{\beta= 2\pi}
\end{equation}

The point $\beta=2\pi$ is a self-dual point which is analogous to the crossing symmetric point ($z=\bar{z}=\frac{1}{2}$) of the usual correlator Conformal Bootstrap.
With these functionals, we can write a similar set of positivity constraints and derive bounds, for example by imposing a gap between the ground state (which has energy $-\frac{c}{12}$) and the first excited state, and checking whether the positivity constraints are satisfied or not. The argument is by contradiction, just like we described in Chapter \ref{Boot}.
Usually, Modular Bootstrap is slightly more complicated as the partition function can be organized in terms of characters of representations of the Virasoro Algebra $\chi$. Then the action of the derivatives is not as trivial as simply differentiating the standard Boltzmann form of a partition function $Z= \sum e^{-\beta \Delta}$.

\section{Introducing topology in flat space: Twist fields}
In order to be able to use the correlator Bootstrap we have developed so for, we need to be able to translate the partition function on the Torus, to a correlation function on a flat-space CFT, which is all we know how to handle.
To do this, we use the so-called twist fields (not to be confused with the quantity $\tau = \Delta -l$ which is also called twist), which generates additional sheets for our CFT and give it non-trivial geometry, by introducing branch points and cuts. This is explained in \cite{Calabrese:2004eu} in the context of entanglement entropy of Conformal Field Theories.

Their description is made in terms of a partition function on an $n$-sheeted surface: $Z_{n}$. They also consider subparts of configuration space which are described with branch cuts with their end on a pair of branch points $(u_{j},v_{j})$, and there could be, in General, $N$ such cuts. In fact, the corresponding partition function is calculated on an $n$-sheeted Riemann Surface with $N$ branch cuts, and their prescription, is that computing this is the same as considering the following 2N-pt function:
\begin{equation}
	\langle \Phi_{n}(u_{1}) \Phi_{-n}(v_{1}) \dots  \Phi_{n}(u_{N}) \Phi_{-n}(v_{N})\rangle
\end{equation}
These $\Phi_{n}$ are very special operators, called twist-n operators, which create the geometry of the branch cuts and define how path integrals should be taken when we change between different sheets (how many twists we have to take to come back to the sheet where we started).
They have Conformal dimensions $h_{n}= \bar{h}_{n}$ that satisfy \cite{Calabrese:2004eu}:
\begin{equation}
	h_{n}= \bar{h}_{n}= \frac{c}{24}\left( 1-\frac{1}{n^{2}}\right) 
\end{equation}
As is mentioned in the Appendix B, the Conformal dimensions are related to $\Delta$ and $l$, such that these twist-n operators have:
\begin{equation}
	l_{n}=0 ~,~ \Delta_{n} =  \frac{c}{12}\left( 1-\frac{1}{n^{2}}\right)   
\end{equation}

Now, from the previous section, we saw that the Hellerman bond was obtained from the invariance of the partition function under a modular transformation of the Torus. To use the framework of twist operators and apply it to torus modular invariance, we make the following claim:

\textit{A 2-sheeted Riemann surface with 4 twist-2 insertions,i.e, with 2 branch cuts, is a Torus.}

The proof is the following series of drawings:
   \begin{figure}[htbp]
   	\centering
   	\includegraphics[width= 6 cm]{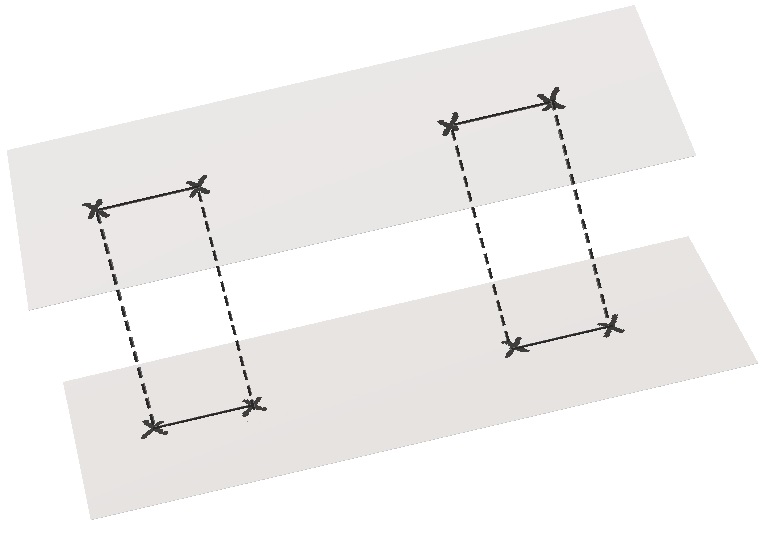}
   	
   	\caption[Two Riemann Sheets with two branch cuts]{The claimed 2-sheeted CFT with 4 operator insertions(two cuts) }
   	\label{fig:twosheet}
   \end{figure}
   
   \begin{figure}[htbp]
   	\centering
   	\includegraphics[width= 6 cm]{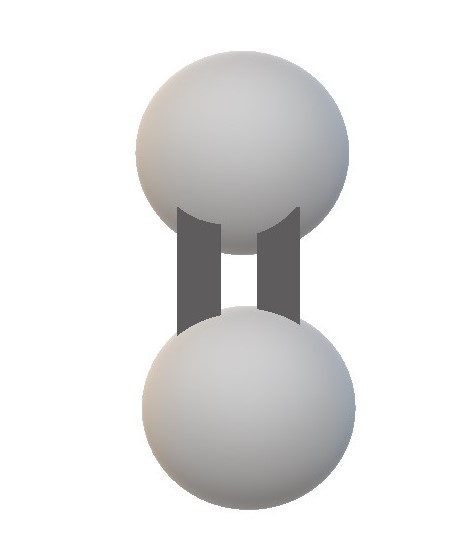}
   	
   	\caption[Two Riemann Spheres connected with two branch cuts]{The 2 sheets are seen as Riemann Spheres with the points at infinity, and are connected by the two cuts }
   	\label{fig:twoballs}
   \end{figure}
   
    \begin{figure}[htbp]
    	\centering
    	\includegraphics[width= 6 cm]{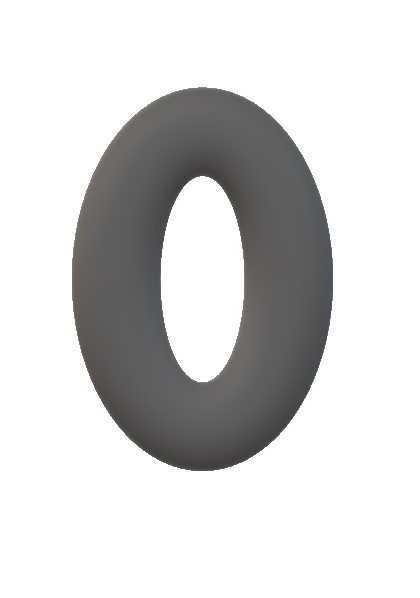}
    	
    	\caption[After a continuous homotopy, a torus]{Performing a continuous homotopy we see that the we object we constructed is a torus }
    	\label{fig:torus}
    \end{figure}
    
    Now it is clear that the relevant quantity to study modular invariance with the Bootstrap on the correlator is a $2 \times N = 4$ point function of twist-2 operators (2-sheets): \begin{equation}
    	\langle \Phi_{2}(u_{1})\Phi_{-2}(v_{1}) \Phi_{2}(u_{2})\Phi_{-2}(v_{2})\rangle
    \end{equation}
    
    Of course, because this is a four point function of operators with the same quantum numbers, we can use the OPE and impose crossing symmetry, just as we have for every other correlator in this thesis.
    From this, we expect to extract information similar to the one obtained by modular invariance methods.
    
    \section{Bootstrapping twist-2 correlators}
    The key fact we used to employ the numerical Bootstrap strategy on the crossing equation, was that it was organized in a sum over all possible primary operators, where each one had the contribution of its Conformal family embedded in a Conformal Block.
    Now we are interested in a case where there are many more descendants associated to the infinite family of Virasoro Generators. Therefore there will be a more general Conformal Block with which we will need to worry about: the Virasoro Block.
    \subsection{Virasoro Blocks}
    An explicit expression for the Virasoro Blocks is not known \cite{Perlmutter2015}. However there are a couple of Recursion relations, developed by Zamolodchikov \cite{Zamolodchikov1984,Zamolodchikov1987}, and made systematic by Perlmutter \cite{Perlmutter2015}. We have then, two possible  main strategies:
   \begin{enumerate}
   	\item{Decompose the Virasoro Representations in global representations, by summing over global Blocks which correspond to Quasi-primaries (which are descendants from the Virasoro point of view) From the tree in Appendix \ref{fig:Virtree}, this corresponds to choosing  quasi primaries and summing over the branches generated by acting only with $L_{-1}$.
   	This can be implemented with a recursion relation based on poles in the central charge, as null states that appear in MM values of the central charge will make it blow up. This makes sense, because in the $c\rightarrow \infty$ limit, the contribution of Virasoro descendants are completely suppressed by their norms (which grow with $c$ because of the commutation relations), and all that remains is the Global Conformal family of the primary. As we pick up more $c$ poles, more and more quasi-primaries begin contributing}
   
   \item{As described in Chapter \ref{Boot}, we can use poles on the Conformal dimensions of quasi-primaries directly, to recursively obtain the Block. In the Virasoro case this is much more complicated, and the recursion relation leads to Blocks being written in the Elliptic variable $q = \exp\left(-\pi \frac{K(1-z)}{K(z)}\right) $ as a power series whose order grows with the number of poles used. ($K(z)$ is the complete elliptic integral of the first kind)}
   \end{enumerate}
   
   The first strategy is better from the point of view of the large $c$ limit, as we immediately obtain back the Global Block, which makes it easier to compare to results we already had. However, the recursion depends simultaneously on values of $c$ and $h$ which makes it more inefficient, and harder to implement at high order.
   The second strategy is more commonly used in the literature, for example \cite{Chen:2017yze,Ribault:2015sxa,Lin2015} have used it, and in particular, \cite{Chen:2017yze} made their implementation available. Interestingly the Virasoro Blocks also correspond to some cases of SuperConformal Blocks, as was seen in \cite{Lin2015} and in calculations in the paradigmatic case of Liouville theory \cite{Alday2010,Ribault:2015sxa}.
   To be more explicit, the $q$-expansion reads:
   \begin{equation}
   	\mathcal{G}(c,h_{i},h_{p};z) = (16q)^{h_{p}- \frac{c-1}{24}} z^{\frac{c-1}{24}} (1-z)^{\frac{c-1}{24}-h_{2}-h_{3}} \theta_{3}(q)^{\frac{c-1}{2}-4\sum_{i} h_{i}} H(c,h_{i},h_{p},q)
   \end{equation}
   
   Where the $h_{i}$ are the external operator dimensions (in particular note that even if they are all the same, the Block will still depend on $h_{ext}$, unlike the global case), $h_{p}$ is the Conformal dimension of the primary being exchanged, $\theta_{3}(q)$ is the Jacobi elliptic theta function $\theta_{3}(q) = \sum_{n=-\infty}^{+\infty} q^{n^{2}}$, and $H$ contains all the information about the recursion and the series expansion in $q$: 
   \begin{equation}
   	H(c,h_{i},h_{p},q) = 1 + \sum_{m,n=1}^{\infty} \frac{(16q)^{mn}}{h_{p} - h_{\langle m,n \rangle}} R_{m,n} H(c,h_{i},h_{\langle m,-n\rangle}) 
   \end{equation}
   Where $\langle m,n \rangle$ labels the null states, and $R_{m,n}$ is a factor of the residue which depends only on the external dimension and the dimensions of null states.
   
   We also need to recall that the actual Conformal Block contains two factors as in Eq. \ref{eqn: 2dblock}, so we need to multiply this by the corresponding $\bar{z}$ Block.
   
   It is clear however that this form for the Blocks is not very amenable to the derivative  method around the crossing symmetric point. First, the expansion is written in a complicated variable, and, although it is a series expansion, it is centered around $q=0$ which also corresponds to $z=0$. The Blocks themselves, though, are really efficiently implemented in \cite{Chen:2017yze} and so computing their value (but not their $z$ derivatives, specially of higher order) is a rather fast task. 
   
  \subsection{A multi-point Bootstrap}
   Motivated by our paradigm of being minimalistic from the computational point of view, and using a technique in the spirit of \cite{Echeverri:2016ztu} 
 , we choose a different kind of subspace for the crossing equation \ref{eqn: 2.9} of this problem. Instead of focusing on a single point $z=\bar{z} =\frac{1}{2}$, we compute the crossing symmetrized Virasoro Block $\mathcal{F}$ (the Virasoro version of \ref{eqn: 2.9}) on a series of points on the line $\bar{z}= \frac{1}{2}, z \in [0,\frac{1}{2}]$, and compute first order derivatives on both directions (note that $\partial_{z}\ne \partial_{\bar{z}}$ because we are away from the crossing symmetric point). In particular, we built the vectors:
   
   \begin{equation}
   	\overrightarrow{\mathcal{F}_{\Delta,l,c}^{\Delta_{2}}}= \begin{pmatrix}
   	& \mathcal{F}_{\Delta,l,c}^{\Delta_{2}}(0.49,0.5)\\
   	& \mathcal{F}_{\Delta,l,c}^{\Delta_{2}}(0.48,0.5) \\
   	& \mathcal{F}_{\Delta,l,c}^{\Delta_{2}}(0.47,0.5) \\
   	& \mathcal{F}_{\Delta,l,c}^{\Delta_{2}}(0.46,0.5) \\
   	& \partial_{z} \mathcal{F}_{\Delta,l,c}^{\Delta_{2}}(z,0.5)|_{z=0.5} \\
   	& \partial_{z} \mathcal{F}_{\Delta,l,c}^{\Delta_{2}}(z,0.5)|_{z=0.4} \\
   	& \partial_{\bar{z}} \mathcal{F}_{\Delta,l,c}^{\Delta_{2}}(0.4,\bar{z})|_{\bar{z}=0.5}\\ 
    & \partial_{z} \mathcal{F}_{\Delta,l,c}^{\Delta_{2}}(z,0.5)|_{z=0.3} \\
  	& \partial_{\bar{z}} \mathcal{F}_{\Delta,l,c}^{\Delta_{2}}(0.3,\bar{z})|_{\bar{z}=0.5}\\ 
    & \partial_{z} \mathcal{F}_{\Delta,l,c}^{\Delta_{2}}(z,0.5)|_{z=0.2} \\
    & \partial_{\bar{z}} \mathcal{F}_{\Delta,l,c}^{\Delta_{2}}(0.2,\bar{z})|_{\bar{z}=0.5}\\ 
    & \partial_{z} \mathcal{F}_{\Delta,l,c}^{\Delta_{2}}(z,0.5)|_{z=0.1} \\
    & \partial_{\bar{z}} \mathcal{F}_{\Delta,l,c}^{\Delta_{2}}(0.1,\bar{z})|_{\bar{z}=0.5} 
   	\end{pmatrix}
   \end{equation}
   
   Which are 12 dimensional, and contain points which are not in the euclidean regime $\bar{z} = z^{*}$. The Bootstrap equation holds for any value of $(z,\bar{z})$, so we chose the points merely by numerical convenience.
   Also, $\Delta_{2} = 2 h_{2}$ denotes the dimension of the external twist-2 operator.
   Note that having more components does not automatically give a stronger constraint. For example in Chapter \ref{Bounds}, we mainly used 10 components, but high order derivatives depend on the value of the function on a vicinity of the point. This means intuitively that they should contain more information.
   
   Now we arrive at the point where we can perform the Bootstrap. Our assumptions for the spectrum and the correlation functions are as follows: 
   \begin{itemize}
   	\item{The external dimension of the twist operator is $\Delta_{2} = 2 \frac{c}{24}(1-\frac{1}{2^{2}}) = \frac{c}{16}$.}
   	\item{We will take $c$ as the parameter, and constrain $\Delta_{1}(c)$ with $c\gg 1$.}
   	\item{The identity Block will be exchanged, and the next family will correspond to primaries of dimension $\Delta_{1}$. All operators included satisfy $\Delta > \max(\Delta_{1},l)$.}
   \end{itemize}
   As we impose the constraint for all operators, regardless of spin, our bound is on a primary, not necessarily a scalar primary. This was clear in \cite{Hellerman2011}. In the 2D case the unitarity bound is just $\Delta \geq l$.
   In this particular problem, we just chose to minimize the zero vector (Feasibility LPP), and normalized the contribution of the Identity to one, which vastly improved numerical stability. We obtained the plot of Figure \ref{fig:Hellbound}:
   
   \begin{figure}[htbp]
   		\centering
   		\includegraphics[width= 9 cm]{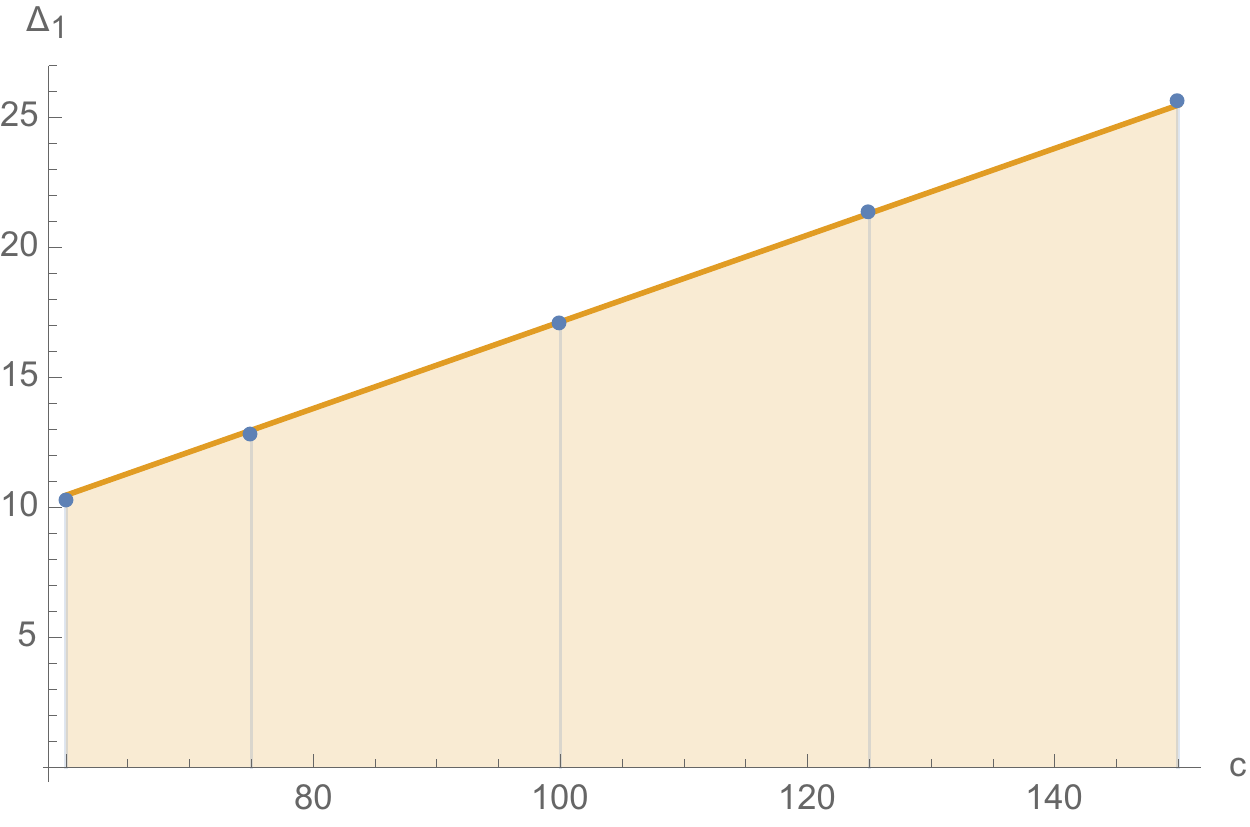}
   		
   		\caption[The Hellerman bound, reproduced from twist correlator]{The maximum allowed value for the gap on a primary $\Delta_{1}$ as a function of $c$. The blue dots are the obtained results and the yellow line is the original bound of \cite{Hellerman2011}}
   		\label{fig:Hellbound}
   \end{figure}

   Indeed, we were able to reproduce the Hellerman bound. 
   In fact, the best fit line for our data is:
   
   \begin{equation}
   	\Delta_{H}(c) = \frac{c}{5.86} + 0.014
   \end{equation}
   
   This weakens the factor in $c$ but improves the order 1 parameter.
   However, fixing the slope to be six, all the points satisfy:
   \begin{equation}
   	\Delta_{1} \le \frac{c}{6} + 0.6
   \end{equation}
   And most of them actually satisfy a stronger version with the constant being smaller than $0.473$.
   
   This is not a surprise as this bound was severely improved in \cite{Collier2016}. The technique was still modular Bootstrap, but the increasing power of cluster computations allows for derivatives of an extremely high order to be accounted for, and the bounds indeed improve dramatically.
   We also attempted to perform the Bootstrap for twist operators of larger twist, which should correspond to exploring a single modular parameter of more complicated Riemann surfaces. This still provided bounds, but all were weaker than the one for twist two. Perhaps introducing multiple correlators simultaneously as in \cite{Kos2014}, would make stronger bounds, but we leave this for future work.

   Even though our construction is essentially just another way to write a partition function on the torus, from the numerical point of view there are substantial numerical differences between our approach and the standard Modular way. The Virasoro Blocks are extremely complicated and are not known explicitly, unlike the Virasoro characters $\chi$.
   We conjecture it may be possible to improve the bound of \cite{Collier2016}, using our techniques combined with the computational power of the cutting edge numerical calculations, but leave this for future work.

\chapter{Conclusion}
\label{Conclusion}

To finish off this thesis, we recap the main results that were obtained, and give brief suggestions of alternative paths that could have been taken, and of possible future work to follow up the results presented here.

As we argued in Chapter \ref{Chapter2}, Conformal Field theory is an excellent playground to probe strongly coupled Quantum field theories. Conformal Symmetry allows us to kinematically fix many important properties, leaving only some model dependent quantities to be computed. Here enters the Bootstrap approach advocated in Chapter \ref{Boot}: We could consider a Lagrangian for a specific theory and attempt to solve it, as a one-off endeavor, but instead, we proceed to write down further general statements about every Conformal Field Theory, and use them to obtain information, without worrying about what microscopic physics is behind the field theory. We did not even need to write a Lagrangian.

Having understood the main philosophy of the Bootstrap, we proceeded to exploit the Bootstrap equation \ref{eqn: 2.9}, by considering finite dimensional subspaces of it, and using linear functionals to write down a Dual LPP, which is an approachable problem, unlike the full Bootstrap equation whose analytic properties still remain vastly unexplored.

This opened the door for Chapter \ref{Bounds}, where we applied the previously described techniques, to recover several standard results in the literature. We bounded scalar dimension $\Delta_{\varepsilon}(\Delta_{\phi})$ in 2D, where we found the famous Ising kink, and checked that many well-known theories were inside our predicted region. We extended our work to 4D, where we were also able to bound OPE coefficients, with a slight modification of the previous methods.

Having shown the power of the Bootstrap as a cartographer of the space of CFTs, we went on to more restrictive results, in Chapter \ref{Beyond}. We refined our interpretation of the bounds previously derived and took advantage of the Extremal cases, to apply the EFM and actually obtain an approximate set of operator dimensions and OPE coefficients for several operators in the 2D Ising Model. We also made a brief description of another approach to the Bootstrap, known as Gliozzi's method. We exploited the knowledge of the fusion rules to calculate Operator dimensions and OPE coefficients, in the simple case of a 4D free scalar field.

With this, we went on to attack a very well known problem in the literature, bounding operator dimensions as a function of the Central charge in a 2D CFT. Here, we took an alternative approach to the methods of Modular Bootstrap which rely on properties of the partition function, and used Twist-fields that allowed us to build a 4-pt correlation function for a flat CFT, which was the bread and butter of our Bootstrap approach. The application of this novel method allowed us to reproduce the results of \cite{Hellerman2011}.

On what regards alternative paths and future work, there are a few points to be touched.
We chose to use Linear Programming methods instead of Semi-definite ones, which in general give stronger bounds and are very widely used. This was a choice based on simplicity, but also on the will to obtain results without resorting to external software, showing that interesting Bootstrap results can be obtained in a single laptop, without being an expert in programming.

This leads to the fact that many results in the literature resort to Bootstrapping various simultaneous correlation functions as in \cite{Kos2014}. This has allowed the appearance of islands and archipelagos in the space of CFTs, that is, generating restricted regions that narrow down the operator dimensions for certain theories. It is generally said that these multiple correlator problems are only possible in the semi-definite approach, but this false, as was shown in \cite{El-Showk:2016mxr}, where multi-correlator problems are solved in the Linear programming framework. Indeed, this would be one possible path to further develop this work.

In fact, it would be interesting to use more than one correlation function for twist operators of different $n$, to improve the results obtained in chapter 5, and possibly compete with the state-of-the-art bound of \cite{Collier2016}.

As a final note, a recent paper \cite{Li:2017kck}, has been able to produce these famous islands with single-correlator Bootstrap, using stronger assumptions on the spectrum. This suggests that the actual improvement obtained in \cite{Kos2014}, is mostly due to the assumption on the number of relevant operators, and not on the multiple correlators.
Pursuing these ideas would also be a very natural continuation of this thesis.


\addtocontents{toc}{\vspace{2em}} 

\appendix 

\chapter{OPE coefficients for a 4D free scalar} 

\label{AppendixA} 

In this appendix, we will compute the OPE coefficients $  	f_{\phi\phi\varepsilon}^{2}$ , $
f_{\phi\phi T}^{2}$ , 
$f_{\phi\phi,l=4}^{2}$ for the 4D free scalar field. We recall from Chapters \ref{Chapter2} and \ref{Boot} that the 4-pt function can be written as:
\begin{equation}
	 \langle \phi(x_{1}) \phi(x_{2}) \phi(x_{3}) \phi(x_{4}) \rangle = \frac{g(u,v)}{x_{12}^{2\Delta_{\phi}} x_{34}^{2\Delta_{\phi}}}
\end{equation}
With the conformal block expansion:
\begin{equation}
		g(u,v)= \sum_{\mathcal{O}} f_{\phi\phi \mathcal{O}}^{2} g_{\Delta_{\mathcal{O}},l_{\mathcal{O}}}
\end{equation}
We will compute the 4-pt function exactly and then expand it conformal blocks to read the OPE coefficients.
For the 4D free scalar, it is really simple to compute the 4-pt function. We just use Wick contractions ($\phi_{i} \equiv \phi(x_{i})$):
	\[
	\langle \phi_{1} \phi_{2} \phi_{3} \phi_{4} \rangle =
	\contraction{\langle}{\phi_{1}}{}{\phi_{2}}
	\contraction{\langle \phi_{1} \phi_{2}}{\phi_{3}}{}{\phi_{4}}
	\langle \phi_{1} \phi_{2} \phi_{3} \phi_{4} \rangle
	+
	\contraction[2ex]{\langle}{\phi_{1}}{\phi_{2} \phi_{3}}{\phi_{4}}
	\contraction{\langle \phi_{1} }{\phi_{2}}{}{\phi_{3}}
	\langle \phi_{1} \phi_{2} \phi_{3} \phi_{4} \rangle
	+ 
	\contraction[2ex]{\langle}{\phi_{1}}{\phi_{2}}{\phi_{3}}
	\contraction{\langle \phi_{1} }{\phi_{2}}{\phi_{3}}{\phi_{4}}	
		\langle \phi_{1} \phi_{2} \phi_{3} \phi_{4} \rangle	
	\tag{A.3-I}
	\label{eq:A.3-I}
	\]
	
	And the contraction of 2-fields is replaced by their 2-pt function. But these were fixed in Chapter \ref{Boot} as well, which means that the 4-pt function is:
	\begin{equation}
		\langle \phi_{1} \phi_{2} \phi_{3} \phi_{4} \rangle =\frac{1}{|x_{12}|^{2}|x_{34}|^{2}} + \frac{1}{|x_{14}|^{2}|x_{23}|^{2}} + \frac{1}{|x_{13}|^{2}|x_{24}|^{2}}
	\end{equation}
	Where we used $\Delta_{\phi}=1$ for  a free scalar field.
	Now, relating the 4-pt function to $g(u,v)$ and recalling the definition of $u$ and $v$ we obtain:
	\begin{equation}
		g(u,v) = 1 + u +\frac{u}{v}
	\end{equation}

Of course the one is just the contribution of the identity in a conformal block expansion, and we will write it on the left hand side. In $z,\bar{z}$ variables we have then:
\begin{equation}
	g(u,v)-1= z\bar{z} + \frac{z \bar{z}}{(1-z)(1-\bar{z})} = 2z\bar{z} + z\bar{z}(z+\bar{z}+z^{2}+ \bar{z}^{2} + z\bar{z} +z^{3} + \bar{z}^{3} +z^{2}\bar{z} + z\bar{z}^{2}+ O(z^{4},\bar{z}^{4}))
\end{equation}
Where we expanded around $z=\bar{z}=0$, and all the higher order terms have coefficient one.
Now, because of our assumption of fusion rules in Chapter \ref{Beyond}, we can write the expected CB decomposition:
\begin{equation}
	g(u,v)-1 =  f_{\phi\phi\varepsilon}^{2} g_{\Delta_{\varepsilon},0} +
	f_{\phi\phi T}^{2}  g_{4,2} +
	f_{\phi\phi,l=4}^{2} g_{4+\Delta_{\varepsilon},4} + \dots
\end{equation}

We expanded the right hand-side around $z=\bar{z}=0$ which gives, using the more compact notation $  	(f_{\phi\phi\varepsilon}^{2},
f_{\phi\phi T}^{2},
f_{\phi\phi,l=4}^{2})=(a,b,c)$ for readability:
\begin{equation}
	g(u,v)-1 = a(z\bar{z} +\frac{z^{2}\bar{z}}{2} + \frac{z\bar{z}^{2}}{2} ) + (\frac{a}{3}+b)(z^{3}\bar{z} + z\bar{z}^{3}+z^{2}\bar{z}^{2}) + (\frac{a}{4} + \frac{3b}{2})P(z^{5},\bar{z}^{5}) + (\frac{a}{5}+\frac{12b}{7}+c)P(z^{5},\bar{z}^{5})
\end{equation}
Where $P(z^{n},\bar{z}^{n})$ denotes the sum of all terms of the form $z^{p}\bar{z}^{q}$ with $p+q=n$ and $p,q \ge 1$.
By matching the coefficients of our exact result and the CB decomposition order by order we obtain the result presented in Chapter \ref{Beyond}:
\begin{equation}
	f_{\phi\phi\varepsilon}^{2} =2~,~
	f_{\phi\phi T}^{2} = \frac{1}{3}~,~
	f_{\phi\phi,l=4}^{2} = \frac{1}{35}
\end{equation}
\chapter{Summary of main results in 2D CFT}
\label{appendix2d}

In this appendix we summarize and briefly motivate the necessary results about 2D CFT \cite{Blumenhagen:2009zz,DiFrancesco:1997nk,Ribault:2014hia} which are necessary to understand this thesis, mainly in Chapter \ref{twist}. 

\section{A larger set of symmetry}

In Chapter \ref{Chapter2} we saw that the defining equation for the Conformal transformations was the Conformal Killing equation \ref{eq:Ckil}. In 2D it reads:
\begin{equation}
	\partial_{1}\epsilon_{1} = \partial_{2}\epsilon_{2} ~,~\partial_{1}\epsilon_{2} = -\partial_{2}\epsilon_{1}
\end{equation}

Which are the Cauchy-Riemann equations of Complex analysis, which means that $\epsilon_{1},\epsilon_{2}$ are the real and imaginary parts of a complex (locally) holomorphic function $\epsilon(z)$ (In fact, there also exists a anti-holomorphic analogue of this story. Everything we will do has an anti-holomorphic counterpart in terms of $\bar{z}$. This is holomorphic factorization).
Therefore any 2D conformal transformation will admit a decomposition in Laurent modes:
\begin{equation}
	z' = z + \epsilon(z) =  z + \sum_{n \in \mathbb{Z}}\epsilon_{n}(-z^{n+1})
\end{equation}

Where $\epsilon_{n}$ are the Laurent coefficients, and therefore we have generators:
\begin{equation}
	\ell_{n}\equiv-z^{n+1}\partial_{z}
\end{equation}

Which obey the Witt Algebra: 
\begin{equation}
	[\ell_{n},\ell_{m}] =(m-n) \ell_{m+n}
\end{equation}

Indeed, we have infinitely many generators, and the subset $\ell_{-1}$ , $\ell_{0}$ , $\ell_{1}$ corresponds to global conformal transformations which we discussed previously.

\section{The Virasoro Algebra, central charge and the Stress Tensor}

The previous considerations are completely classical. However, when one performs radial quantization in 2D, there is anomaly, called the conformal anomaly $c$ (or central charge), which is intimately connected to the structure of the 2D CFT \cite{Ginsparg1988}. In fact the real algebra is the Virasoro Algebra, which is a central extension of the Witt Algebra, with the center being the constant operator $c$. The Virasoro Algebra, is then:
\begin{equation}
	[L_{n},L_{m}] =(m-n) L_{m+n} + \frac{c}{12}(n^{3}-n)\delta_{n,-m}
\end{equation}

These commutation relations are also consistent with another crucial part of a CFT, its stress-energy tensor. It is common to write: 
\begin{equation}
	T(z) = \sum_{n \in \mathbb{Z}} \frac{L_{n}^{z_{0}}}{(z-z_{0})^{n+2}}
\end{equation} 
Where the generators are acting at $z_{0}$. Indeed, this allows for the construction of many Ward identities, by inserting the Stress-energy tensor in correlation functions.
As we mentioned, for a consistent Virasoro Algebra, we must have the OPE:

\begin{equation}
	T(z)T(w)= \frac{c/2}{(z-w)^{4}} + O((z-w)^{-2})
\end{equation}

So the central charge also measures the behavior of the stress energy tensor.
A further role of the central charge is to measure the number of degrees of freedom of a theory, as is stated in the famous formula by Cardy:
\begin{equation}
	\log N(\Delta) \propto \sqrt{c \Delta}
\end{equation}
This formula holds in the large $\Delta$ limit, for the number of states of dimension $\Delta$: $N(\Delta)$.
In general, the central charge corresponding to the anti-holomorphic part $\tilde{c}$ is taken to be equal to $c$.

Note that under these conventions, the quantum numbers that label states/operators are the conformal dimensions $h,\bar{h}$ which are the eigenvalues of the generators $L_{0}, \bar{L}_{0}$ (recall that $\bar{L}$ generators are associated to anti-holomorphic generators). These are related to $\Delta$ and $l$ as follows:
\begin{equation}
	h = \frac{\Delta +l}{2} ~,~ \bar{h} = \frac{\Delta - l}{2}
\end{equation}

\section{A bigger family}
One might wonder what having such a large set of symmetry generators means from the point of view of the states and operators of our theory (if you want, how the representation theory changes). As we mentioned, we now have a much larger set of operators to rise and lower conformal dimensions. If we recall our discussion from Chapter \ref{Chapter2}, our definition of primary depended on an operator satisfying:
\begin{equation}
	L_{1}\mathcal{O}=0
\end{equation}
In the context of 2D CFT this condition defines a Quasi-primary, and a full Virasoro Primary is required to satisfy:
\begin{equation}
	L_{n} \mathcal{O} ~,~ n \geq 1
\end{equation}

So this means, that from the global point view, there are in general more "primaries", as a Virasoro conformal family contains descendants of the form:
\begin{equation}
	L_{-n_{1}}^{p_{1}}\dots L_{-n_{q}}^{p_{q}} \mathcal{O}_{P}
\end{equation}

Which can contain many quasi-primaries.

This is clearly what happened in Chapter \ref{Beyond}, where we studied the Ising Model which has only three families of Virasoro Primaries ($I,\phi,\varepsilon$), but when we applied the Extremal functional method, we found a lot more primary operators. Indeed these were quasi-primaries, but not Virasoro primaries.
Schematically we can visualize the whole set of Virasoro descendants in the following tree:

\begin{figure}[htbp]
	\centering
	\includegraphics[width= 9 cm]{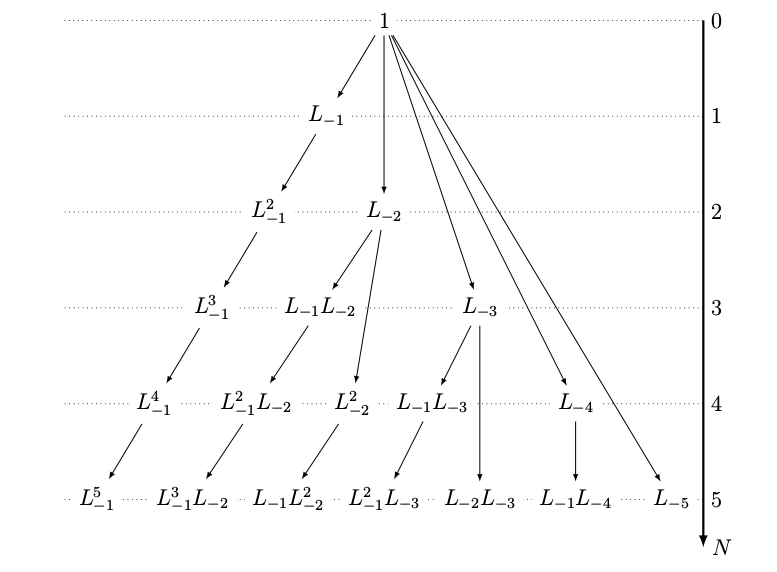}
	
	\caption[A tree of Virasoro descendants]{The set of Virasoro descendants labeled by their level $N$, obtained by acting successively with all the possible generators $L_{-n }$. Taken from \cite{Ribault:2014hia}}
	\label{fig:Virtree}
\end{figure}

The set of all Virasoro descendants of a primary of dimension $\Delta$ in known as the Verma module $\mathcal{V}_{\Delta}$ . The spectrum of a CFT is essentially the determined by the representations (and their multiplicity)  that exist in the theory. The natural representations, are in fact the Verma modules. However, there are also degenerate representations, which are crucial to define the Minimal Models we mentioned in Chapter \ref{Bounds}. These representations have special properties, and some of them even allow for theories with a finite number of Virasoro primaries, with a closed OPE.
They are obtained from the null-states, which are states which despite being descendants, are also primaries, which  means that their norms can vanish. This essentially truncates the Verma Module, or at least defines a non-trivial sub-representation.

For example at level 2 there is a null-state:

\begin{equation}
(	\frac{1}{b^{2}} L_{-1}^{2} + L_{-2})|\mathcal{O}\rangle
\end{equation}

For a primary of dimension $-\frac{1}{2} - \frac{3}{4}b^{2}$, where $b$ is related to the central charge by: 
\begin{equation}
	b = \sqrt{\frac{c-1}{24}} + \sqrt{\frac{c-1}{25}}
\end{equation}
In general, at level $N$, for each decomposition $N=rs$ for integers $r,s$ there is a null vector labeled with a dimension $\Delta_{\langle r, s \rangle}$ such that:
\begin{equation}
	\Delta_{\langle r, s \rangle} = \frac{1}{4} (Q^{2} -(rb + sb^{-1})^{2})
\end{equation}
With $Q = b + \frac{1}{b}$.

Indeed, the null-vector is obtained for a very special value of the scaling dimension which depends on the central charge, so null-vectors are not generic, but do happen in interesting physical theories such as minimal models.

\addtocontents{toc}{\vspace{2em}} 

\backmatter


\label{Bibliography}



\bibliography{library.bib} 

\end{document}